\documentclass{aa}

\usepackage{graphicx}
\usepackage{subcaption}
\usepackage{xcolor}
\usepackage{txfonts}

\usepackage{amsmath}
\usepackage{multirow}

\usepackage[version=4]{mhchem}

\newcommand{\rlight}{r_{\rm L}}

\DeclareRobustCommand{\rchi}{{\mathpalette\irchi\relax}}
\newcommand{\irchi}[2]{\raisebox{\depth}{$#1\chi$}} 

\begin{document}
                
\title{Particle motion in ultra-strong electromagnetic fields of neutron stars: The influence of radiation reaction}
\date{\today}
\titlerunning{Particle motion and radiation reaction}

\author{Ivan Tomczak \and Jérôme Pétri}

\institute{Universit\'e de Strasbourg, CNRS, Observatoire astronomique de Strasbourg, UMR 7550, F-67000 Strasbourg, France.\\
        \email{ivan.tomczak@astro.unistra.fr}         
}

\date{Received ; accepted }


\abstract
{Neutron stars are known to be efficient accelerators that produce particles with ultra-relativistic energies. As a by-product, they also emit copious amounts of photons from radio wavelengths up to gamma rays.}
{As a follow-up to our previous work on particle acceleration simulation near neutron stars, in this paper, we discuss the impact of radiation reaction on test particles injected into their magnetosphere. We therefore neglect the interaction between particles through the electromagnetic field as well as gravitation.}
{We integrate numerically the reduced Landau-Lifshitz equation for electrons and protons in the vacuum field of a rotating magnetic dipole based on analytical solutions in a constant electromagnetic field. These expressions are simple in a frame where the electric and magnetic field are parallel. Lorentz transforms are used to switch back and forth between this frame and the observer frame.}
{We found that, though due solely to the Lorentz force, electrons reach Lorentz factors up to $\gamma=10^{14}$ and protons reach them up to $\gamma=10^{10.7}$. When radiation reaction is enabled, electrons reach energies up to $\gamma=10^{10.5}$ and protons reach energies up to $\gamma=10^{8.3}$. The second set of values are more realistic since the radiation reaction feedback is predominant within the magnetosphere. Moreover, as expected, symmetrical behaviours between the north and south hemispheres are highlighted, either with respect to the location around the neutron star or with respect to particles of opposite charge to mass ratio~$(q/m)$. Consequently, it is useless to simulate the full set of geometrical parameters in an effort to obtain an overview of all possibilities.}
{The study of the influence of the magnetic dipolar moment inclination shows similar behaviours regardless of whether radiation reaction is enabled. Protons (respectively electrons) impact the surface of the neutron star less as the inclination angle increases (decreases for electrons), while if the rotation and magnetic axes are aligned, all the protons impact the neutron star, and all the electrons impact the surface if the rotation and magnetic axes are anti-aligned.
Similarly, we still find that particles are ejected away from the neutron star, in some preferred directions and Lorentz factors.}
{}

\keywords{magnetic fields -- methods: analytical -- stars: neutron -- stars: rotation -- pulsars: general}

\maketitle

\section{Introduction}
\label{sec:Intro}

Neutron stars are compact stellar remnants left after supernovae explosions. Due to their intense magnetic and electric fields, they are believed to be efficient sources of ultra-relativistic particles. They also emit photons via the synchrotron and curvature radiation mechanisms or inverse Compton, interacting with their surrounding as well as with the interstellar medium.

In this paper, we use a simple model of a neutron star described by only a few parameters, namely the inclination of the neutron star $\rchi$, corresponding to the angle between its rotation and magnetic axes, the angular rotation speed of the neutron star $\Omega$, allowing to define the light cylinder radius $\rlight=c/\Omega$, which is the distance at which an object in co-rotation with the neutron star would reach the speed of light~$c$, the radius of the neutron star~$R$ and the magnetic field strength~$B$ at the surface of the neutron star.

The extreme magnetic field of these remnant stars ranges from $B\simeq10^5~\text{T}$ to $B\simeq10^{10}~\text{T}$. Moreover, coupled with an angular speed between $\Omega \simeq 1000 ~\text{rad.s}^{-1}$ and $\Omega \simeq 0.1~\text{rad.s}^{-1}$, they generate an intense electric field~$E$, accelerating particles around the neutron star to ultra-relativistic Lorentz factors. 
For our purposes, the neutron star mass~$M$ is irrelevant because the gravitational force exerted on charged particles $F_g$ is negligible compared to the Lorentz force $F_L$. Indeed, both forces have a typical intensity of $F_g = G \, \frac{M\, m_p}{R^2} \simeq 3.17\times10^{-15} \text{N}$ for a proton at the surface of the neutron star compared to $F_L = q \, E = q \, \Omega R \, B \simeq 5.01\times10^{-7} \text{N}$. The ratio of these forces is 
$$\frac{F_g}{F_L} = \frac{G \, M \, m_p}{R^2 \, q \, E} = 6.33 \times 10^{-9},$$
so the electromagnetic force is approximately $10^{9}$ times stronger than the gravitational force for protons and even a factor $m_p/m_e \sim 2000$ larger for electrons and positrons.

Most numerical simulations of neutron star magnetospheres use the \cite{boris_relativistic_1970} scheme or better the \cite{vay_simulation_2008} scheme; however, these algorithms are not well suited for ultra-strong electromagnetic fields, and  some authors have consequently lowered the true field strengths in their simulations to unrealistically low values. 
Although scaling is sometimes applied to obtain results closer to reality, such scaling cannot be straightforwardly extrapolated, for instance, when radiation reaction is included because of the non-linearities introduced by radiative feedback \citep{vranic_classical_2016}. 
Moreover, Lorentz factors reached by the particles near pulsars hardly exceed $\gamma=10^4$ with those algorithms (see, for instance, \cite{brambilla_electronpositron_2018}, \cite{philippov_ab-initio_2018}, \cite{guepin_proton_2020}, and \cite{kalapotharakos_three-dimensional_2018}). 

This limitation arises from the huge span in timescales, starting from the ultra-high frequency gyro motion $\omega_B$ down to the stellar rotation frequency $\Omega$. These extreme values are synthesised by their ratio, which is also known as the strength parameter,
$$a = \frac{\omega_B}{\Omega} = \frac{q \, B}{m\,\Omega}\simeq10^{10}$$
for a proton near a millisecond pulsar ($\Omega=10^{3}~\text{rad.s}^{-1}$ and $B=10^{5}~\text{T}$). The situation becomes worse for electron-positron pairs and for young pulsars.
This ratio corresponds to the number of gyrations made by a particle during the timescale of evolution of the electromagnetic field due to the stellar rotation. It shows that the difference in timescales makes computing the trajectory of particles in a reasonable amount of time almost impossible since billions of time steps are needed in the pulsar period timescale. 

To tackle this issue, our aim is to propose a new technique based on analytical solutions of the Lorentz force equation in constant and ultra-strong electromagnetic fields with an acceptable computational time and, most importantly, an approach that avoids the need for scaling, allowing particles to reach high Lorentz factors with realistic fields. Several authors have worked on analytical solutions to the equation of motion including the radiation reaction, such as \cite{gordon_special_2021}, \cite{laue_acceleration_1986}, \cite{li_accurately_2021}, and \cite{heintzmann_exact_1973}. Our approach is based on these works. For plane waves in a vacuum, exact solutions are also known because of \cite{hadad_effects_2010, piazza_exact_2008}.
These solutions were applied in strong electromagnetic waves by \cite{petri_particle_2021} and around a dipole by \cite{petri_particle_2022-2}.

Neutron stars are known to act as unipolar inductors, generating huge electric potential drops between the poles and the equator of the order
\begin{equation}\label{eq:Potential}
 \Delta \phi = \Omega \, B \, R^2 \approx 10^{16}~V.
\end{equation}
As a consequence, these stars expel electrons (and maybe protons and ions), filling the magnetosphere with charged particles. The typical Lorentz factor for electrons in this static field is therefore 
\begin{equation}\label{eq:FacteurLorentz}
 \gamma = \frac{e \, \Delta\phi}{m_e\,c^2} \approx 10^{10} .
\end{equation}
If the particle injection rate is high enough, this plasma will screen the electric field, drastically mitigating the potential drop $\Delta \phi$ and the acceleration efficiency. Resistive \citep{li_resistive_2012} as well as PIC simulations \citep{cerutti_particle_2015} have indeed showed that only a small fraction of the full potential is available. However, for low particle injection rates, the plasma is unable to screen the electric field, and the full potential drop develops. In these cases, the magnetosphere is almost empty and known as an electrosphere \citep{krause-polstorff_electrosphere_1985, petri_global_2002}. Such electrospheres are the subject of the present paper. They represent inactive pulsars that are able to accelerate particles to ultra-relativistic speeds. Our aim is to accurately quantify the final Lorentz factor reached by the outflowing plasma in this large-amplitude low-frequency electromagnetic wave. A similar study was performed by \cite{michel_electrodynamics_1999}, though with a more analytical perspective.

The outline of the paper is as follows. First in Section~\ref{sec:Algo}, we summarise the principle of the algorithm. Next in Section~\ref{sec:Verif}, we show some results obtained in fields where an analytical solution is known before discussing the results of the simulations near pulsars in Section~\ref{sec:Deutsch}. Some conclusions are drawn in Section~\ref{sec:Conclusion}.

\section{Description of the numerical algorithm}
\label{sec:Algo}

In this section, we describe the algorithm developed including radiation reaction. It is similar to the one presented in \cite{petri_relativistic_2020}, which was used the basis for the work of \cite{tomczak_particle_2020}. This code successively finds analytical solutions to the equation of the motion of particles in an electromagnetic field~$(\mathbf{E}, \mathbf{B})$ assumed to remain constant within a time step integration.
During this time step, the algorithm solves the equation of motion, approximated by the reduced Landau-Lifshitz equation
\begin{equation}
\dfrac{du^{\mu}}{d\tau}=\dfrac{q}{m}{F^{\mu}}_{\nu}u^{\nu}
-\dfrac{q^4}{6\,\pi\,\varepsilon_0 \, m^3 \, c^3} \Bigg[
F^{\mu \sigma} F_{\lambda \sigma} u^{\lambda}
+\Bigg(
F^{\sigma \nu} u_{\nu} F_{\sigma \lambda} u^{\lambda}
\Bigg)u^{\mu}
\Bigg] .
\end{equation}
We note that the four-velocity of the particle (in contravariant form) is $u^{\mu}=\begin{pmatrix}
u^0 & u^1 & u^2 & u^3
\end{pmatrix}=\gamma c\begin{pmatrix}
1 & \beta^x & \beta^y & \beta^z
\end{pmatrix}=\gamma c\begin{pmatrix}
1 & \boldsymbol{\beta}
\end{pmatrix}$, with the Lorentz factor $\gamma=\dfrac{1}{\sqrt{1-\beta^2}}$; $\boldsymbol{\beta}$ is the speed of the particle normalised to the speed of light $c$, and it can be decomposed onto a Cartesian coordinate basis with the usual labels of $x$, $y$ and $z$; and 
${F^{\mu}}_{\nu}$ is the electromagnetic tensor given in components by
\begin{equation}
F{^{\mu}}_{\nu}=
\begin{pmatrix}
 0 & E_x/c & E_y/c & E_z/c \\
 E_x/c & 0 & B_z & -B_y \\
 E_y/c & -B_z & 0 & B_x \\
 E_z/c & B_y & -B_x & 0 
\end{pmatrix} .
\end{equation}
For this algorithm, we first needed to switch from the observer's reference frame denoted by $\mathcal{R}$, in which the neutron star only rotates, to a reference frame where $\mathbf{E}$ and $\mathbf{B}$ are parallel and denoted by $\mathcal{R}'$, where integrating the Lorentz force is easy. In a Cartesian coordinate system, the evolution of the four-velocity when $\mathbf{E}$ and $\mathbf{B}$ are constant and along the $z$-axis is 
\begin{subequations}
        \label{eq:4velocity}
        \begin{align}
        u^0 / c & =\gamma =a(\tau) \gamma_0 [\cosh (\omega_E \tau) + \beta_0^z \sinh (\omega_E \tau)]  \\
        u^1 / c & =\gamma \beta^x =b(\tau) \gamma_0 [\beta_0^x \cos(\omega_B \tau) + \beta_0^y \sin (\omega_B \tau)]  \\
        u^2 / c & =\gamma \beta^y =b(\tau) \gamma_0 [-\beta_0^x \sin(\omega_B \tau) + \beta_0^y \cos (\omega_B \tau)]  \\
        u^3 / c & =\gamma \beta^z =a(\tau) \gamma_0 [\sinh (\omega_E \tau) + \beta_0^z \cosh (\omega_E \tau)]  .
        \end{align}
\end{subequations}
We introduced the typical electric and magnetic frequencies as $\omega_E=q\,E/mc$ and $\omega_B=q\,B/m$.
The initial Lorentz factor is indicated with $\gamma_0$, and $\begin{pmatrix}
\beta^x_0 & \beta^y_0 & \beta^z_0
\end{pmatrix}$ is the initial velocity normalised to the speed of light~$c$. The coefficients
$a(\tau)$ and $b(\tau)$ bring corrections to the velocity induced by the radiation reaction (setting $a(\tau)=b(\tau)=1$ removes radiation reaction). The coefficients are given by
\begin{subequations}
        \label{eq:aetbtau}
        \begin{align}
        a(\tau) &= \dfrac{1}{   \gamma_0 \sqrt{(1-\beta_z^2)-(\beta_x^2+\beta_y^2)\exp(-2\tau_0 (\omega_B^2 + \omega_E^2) \tau)}   }\\
        b(\tau) &= a(\tau) \exp(-\tau_0 (\omega_B^2 + \omega_E^2) \tau) 
        \end{align}
\end{subequations} 
with $\tau_0=q^2/6\pi\varepsilon_0mc^3$, a characteristic timescale of the energy losses.
It is unfortunately not possible to integrate the four-position in an analytical manner, so we decided to use the analytical solution for the position without the radiation reaction
\begin{subequations}
        \label{eq:4position}
        \begin{align}
        c(t-t_0)=\dfrac{\gamma_0 c}{\omega_E} [\sinh (\omega_E \tau) + \beta_0^z \cosh (\omega_E \tau) - \beta_0^z]  \\
        x-x_0=\dfrac{\gamma_0 c}{\omega_B} [\beta_0^x \sin(\omega_B \tau) - \beta_0^y \cos (\omega_B \tau) +\beta_0^y]  \\
        y-y_0=\dfrac{\gamma_0 c}{\omega_B} [\beta_0^x \cos(\omega_B \tau) -\beta_0^x + \beta_0^y \sin (\omega_B \tau)]  \\
        z-z_0=\dfrac{\gamma_0 c}{\omega_E} [\cosh (\omega_E \tau) -1 + \beta_0^z \sinh (\omega_E \tau)],
        \end{align}
\end{subequations}
with the initial position and time being $(x_0,y_0,z_0)$ and $t_0$.
We integrated the position of the particle according to $\tau$, the proper time; however, as we wanted to set the observer's time step $\delta t=t-t_0$ to be constant, we resorted to finding $d\tau$, the proper time step, so as to always obtain the same observer time step~$dt$.

This scheme has some drawbacks: It returns an approximation of the four-position of the particle with radiation reaction, and it does not efficiently take into account field gradients.
However, this method is still more efficient than other approximations, such as Euler or Runge-Kutta, most of which apply a constant speed assumption. Indeed, this scheme does not assume a constant speed but only that the radiation reaction has little effect on the position of the particle. Nonetheless, it still takes into account speed variations in terms of norm and direction.
In addition, compared to other methods, it allows for longer time steps and multiple gyrations of the particle around a magnetic field line during those longer time steps. Moreover, since the gyration radius is relatively small compared to typical magnetic field scales (followed by the trajectories of the particles), the error related to the shrinking of the Larmor radius is negligible.

The light-like case, where $\mathbf{E} \cdot \mathbf{B} = 0$ and $E^2 = c^2 \, B^2$, must be treated separately because there exists no frame where $\mathbf{E}$ and $\mathbf{B}$ are parallel. Due to the low likelihood of finding such configurations, we kept solutions found in \cite{petri_relativistic_2020} as not corrected for radiation reaction. This means that if a particle were to find a light-like field, the algorithm would keep working but the radiation reaction would not be taken into account for just one time step.

\section{Tests}
\label{sec:Verif}

\begin{figure}[h]
\begin{subfigure}{.5\textwidth}
  \centering
  \includegraphics[width=\textwidth]{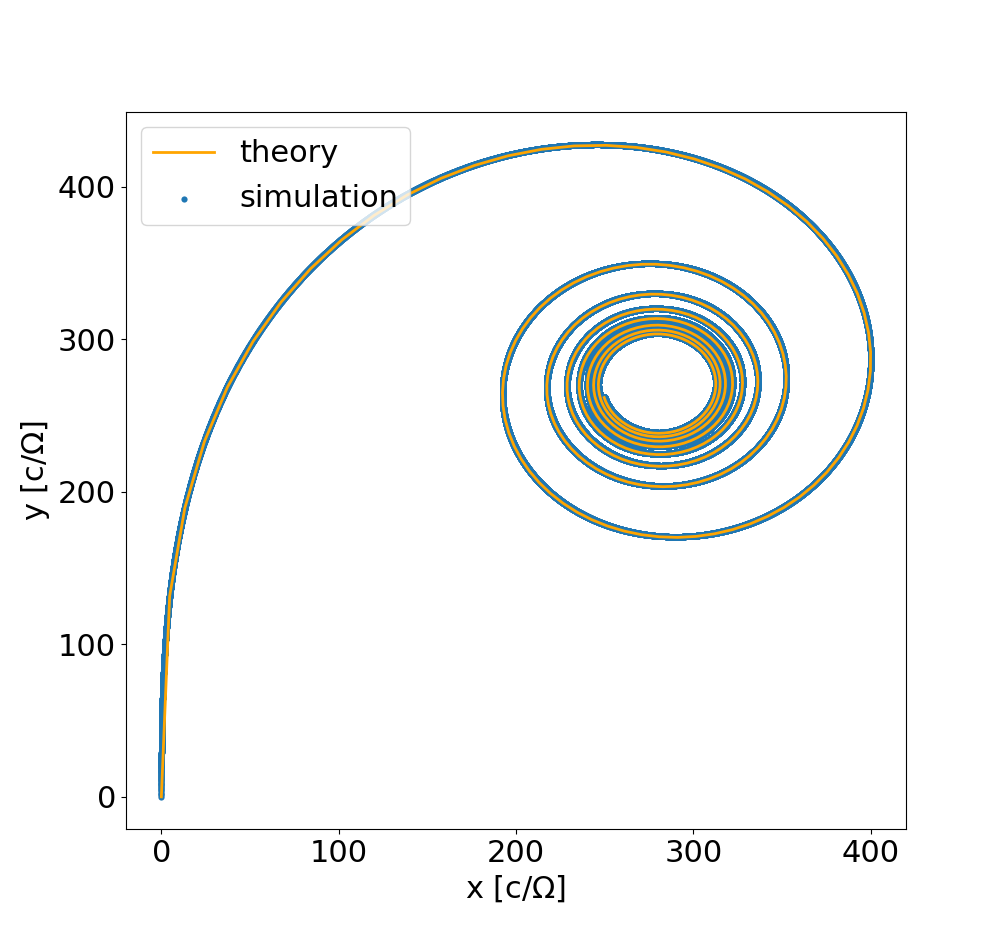}
        \put(-40,50) {\colorbox{white}{\tiny (a)}}
\end{subfigure}
\begin{subfigure}{.5\textwidth}
        \centering
        \includegraphics[width=\textwidth]{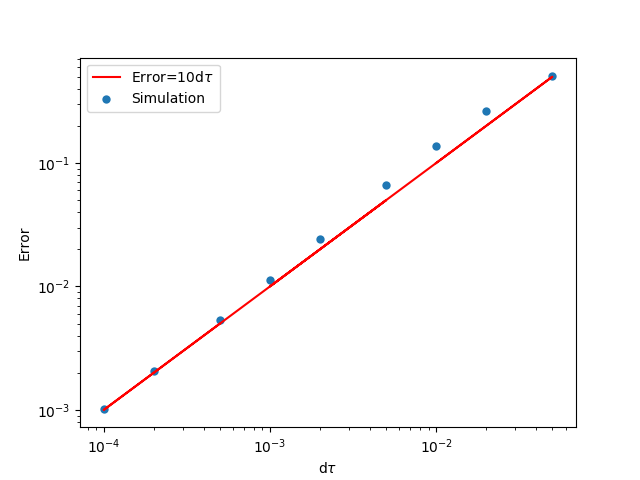}
        \put(-180,125) {\colorbox{white}{\tiny (b)}}
\end{subfigure}
\caption{Particle trajectory in a constant magnetic field with radiation reaction. Panel (a): Trajectory of the simulated particle (in blue) compared to theoretical positions (in orange) for $\tau_0\,\omega_B = 10^{-5}$, $\gamma_0=10^4$, initial speed along $\textbf{y}$. Panel (b): Relative error as a function of the proper time step $d\tau$ for the simulations in blue points, compared to the first order error expectations in red solid line.}
\label{fig:tests}
\end{figure}

In order to check the correctness and accuracy of our code implementation, we simulated the evolution of a particle in a constant and uniform magnetic field. Analytical solutions are known for the four-velocity but also for the spatial position and observer time in this case \citep{petri_particle_2022-2}.
An electron is kicked into a constant magnetic field of strength $B_z$ with an initial Lorentz factor of $\gamma_0=10^4$. The normalised damping parameter is therefore $\tau_0\,\omega_B = 10^{-5}$. 
As figure~\ref{fig:tests} highlights, the algorithm converges and finds the spiral motion described by a charged particle losing energy in a magnetic field. The exact analytical trajectory is shown in orange solid lines and the numerical simulations with blue dots. Thanks to this comparison, we observed that the algorithm is first order in proper time, according to the particle position. The first order convergence is due to the fact that the four-position was updated according to the pure Lorentz force, neglecting the radiation reaction corrections.
Being confident about the convergence and accuracy of our algorithm, we then simulated the motion of charged particles in the electromagnetic field of a rotating neutron star.

\section{Simulations in the Deutsch field}
\label{sec:Deutsch}

As an application in the astrophysical context, we explored particle acceleration and its radiation reaction in a strongly magnetised rotating magnetic dipole such as that expected around rotating neutron stars. If the star is surrounded by vacuum, simple analytical expressions are known for the electromagnetic field and given by \cite{deutsch_electromagnetic_1955}.

\subsection{Neutron star settings}

Therefore, we injected particles in the field of a rotating neutron star in a vacuum, namely, the Deutsch field \citep{deutsch_electromagnetic_1955}, decomposed in spherical coordinates and using the complex form for the magnetic field as
\begin{subequations}
        \begin{align}
        \label{eq:DeutschEM}
        B_r(\mathbf{r},t) & = 2 \, B \, \left[ \frac{R^3}{r^3} \, \cos\rchi \, \cos\vartheta + 
        \frac{R}{r} \, \frac{h^{(1)}_1(k\,r)}{h^{(1)}_1(k\,R)} \, 
        \sin\rchi \, \sin \vartheta \, e^{i\,\psi} \right] \\
        B_\vartheta(\mathbf{r},t) & = B \, \left[ \frac{R^3}{r^3} \, \cos\rchi \, \sin\vartheta + \right. \\ 
        & \left. \left( \frac{R}{r} \, 
        \frac{\frac{d}{dr} \left( r \, h^{(1)}_1(k\,r) \right)}{h^{(1)}_1(k\,R)} + \frac{R^2}{\rlight^2} \, 
        \frac{h^{(1)}_2(k\,r)}{\frac{d}{dr} \left( r \, h^{(1)}_2(k\,r) \right) |_{R}} \right) \, 
        \sin\rchi \, \cos \vartheta \, e^{i\,\psi} \right] \nonumber \\
        B_\varphi(\mathbf{r},t) & = B \, \left[ \frac{R}{r} \, 
        \frac{\frac{d}{dr} ( r \, h^{(1)}_1(k\,r) )}{h^{(1)}_1(k\,R)} \, 
        + \frac{R^2}{\rlight^2}
        \frac{h^{(1)}_2(k\,r)}{\frac{d}{dr} \left( r \, h^{(1)}_2(k\,r) \right) |_{R}}
        \, \cos 2\,\vartheta \right] \, i \, \sin\rchi \, \, e^{i\,\psi}
        \end{align}
\end{subequations}
and for the electric field as
\begin{subequations}
\begin{align}
        E_r(\mathbf{r},t) & = \Omega \, B \, R \, 
        \left[ \left( \frac{2}{3} - \frac{R^2}{r^2} ( 3 \, \cos^2\vartheta - 1 ) \right) \
        \, \frac{R^2}{r^2} \, \cos\rchi \right. \\
        & + \left. 3 \, \sin\rchi\, \sin 2\,\vartheta \, e^{i\,\psi}  \,
        \frac{R}{r} \, \frac{ h^{(1)}_2(k\,r)} \nonumber
        {\frac{d}{dr} \left( r \, h^{(1)}_2(k\,r) \right) |_{R}} \right] \\
        E_\vartheta(\mathbf{r},t) & = \Omega \, B \, R \, 
        \left[ - \frac{R^4}{r^4} \sin 2\,\vartheta \, \cos\rchi \right. \\
        & \left. + 
        \sin\rchi\, e^{i\,\psi}  \, \left(
        \frac{R}{r} \, \frac{\frac{d}{dr} \left( r \, h^{(1)}_2(k\,r) \right)}
        {\frac{d}{dr} \left( r \, h^{(1)}_2(k\,r) \right)|_{R}} \, \cos 2\,\vartheta -
        \frac{h^{(1)}_1(k\,r)}{h^{(1)}_1(k\,R)} \right) \right] \nonumber \\
        E_\varphi(\mathbf{r},t) & = \Omega \, B \, R \, \left[ \frac{R}{r} \, 
        \frac{\frac{d}{dr} \left( r \, h^{(1)}_2(k\,r) \right)}
        {\frac{d}{dr} \left( r \, h^{(1)}_2(k\,r) \right)|_{R}} -
        \frac{h^{(1)}_1(k\,r)}{h^{(1)}_1(k\,R)} \right] \, i \sin\rchi \, \cos\vartheta \, e^{i\,\psi},
        \end{align}
\end{subequations}
where $h^{(1)}_\ell$ represents the spherical Hankel functions for outgoing waves \citep{arfken_mathematical_2005}, $k\,\rlight=1$, and the phase is $\psi=\varphi-\Omega\,t$, with $\varphi$ as the phase at $t=0$.
The physical components of the electromagnetic field are the real parts of the above expressions.

In our simulations, the magnetic field at the surface of the star was set to $B=10^5$~T, which corresponds to the typical values for a millisecond pulsar. We worked in normalised units, setting the reference rotation speed of the neutron star to $\Omega$, the reference speed to the speed of light $c$, and therefore the light cylinder radius to $\rlight=c/\Omega$. By choosing the normalised neutron star radius to be $R=0.1\,\rlight$, we fixed the size of the light cylinder and the stellar rotation speed. Indeed, the neutron star radius is about $R=12~$km \citep{nattila_neutron_2017}, which means that the light cylinder radius is about $\rlight=120$~km. Moreover, since $\rlight=c/\Omega$, the angular velocity is $\Omega=2500$~rad/s. Thus, a real period of rotation of $P$ is $2.5$~ms. We needed the observer's time step to be small relative to the time of evolution of the magnetic field (which is of the order of the rotation period), so we chose $dt/P=0.0001$, meaning $dt=250$~ns. Concerning the inclination of the neutron star, the simulations were carried out with an inclination from the following set: $\rchi \in \{0^{\circ};30^{\circ};60^{\circ};90^{\circ};120^{\circ};150^{\circ};180^{\circ}\}$. 

Particle injection was made using a rejection method in order to obtain a uniform and isotropic distribution of particles around the neutron star. We generated three random numbers, each following an independent uniform distribution law in the interval $[-0.9;0.9]$ and corresponding to the Cartesian coordinates $(x,y,z)$. 
We then defined the radius at which a particle is injected, $r=\sqrt{x^2+y^2+z^2}$, and if $r/\rlight \leq 0.1$ or $r/\rlight \geq 0.9$, we removed that particle and generated it again. Otherwise, we kept the particle and injected the next one.

Particles were injected at rest and evolved in time up to a final time of $t_f/P=15$. Sometimes particles crashed onto the surface, and the integration was stopped earlier. As for particle species, we considered electrons and protons. In order to obtain reasonable statistics, we simulated 8,192 particles per configuration. In the following sections, we describe the final particle properties, including their distribution in space and energy.

\subsection{Distribution of particles in space}

In this sub-section, we discuss the particle positions at the end of the run as well as at the beginning of the run in order to link them to their final Lorentz factor.
We note, however, that the cases of aligned ($\rchi=0^{\circ}$) and anti-aligned ($\rchi=180^{\circ}$) rotators are treated separately in Sub-section~\ref{ssec:Alignes} because the electromagnetic field is static in these configurations.

The coordinates used to characterise the position of a particle were either in the Cartesian coordinate system $(x,y,z)$ or the spherical coordinate system $(r, \theta, \phi)$. In the spherical coordinate system, $r$ defines the distance of the particle relative to the centre of the neutron star, $\theta$ is its colatitude (relative to the rotation axis), and $\phi$ is its azimuth relative to the $x$ axis, knowing that at the beginning of the simulation the magnetic axis lies in the $xOz$ plane. 
We also defined three possible final states for the particles: ejected, meaning that at the end $r/\rlight\geq1$ (in our case those particles have a radial velocity); trapped, meaning that at the end $r/\rlight \in ]0.1;1[$; or crashed onto the neutron star, meaning that the particles should have reached $r/\rlight \leq 0.1$ or equivalently $r<R$ at one point.

The statistics of the particles according to their final states are summarised in Table~\ref{tab:etat_particules} for our total number of 8192~particles. It allowed us to notice that for the protons, as $\rchi$ increases from $30^\circ$ to $150^\circ$, fewer particles impact the surface of the neutron star, while, inversely, more protons are ejected away from it. We also found it interesting to notice that the maximum number of protons trapped close to the pulsar was obtained for $\rchi=90^{\circ}$ and that a symmetrical behaviour could be observed for electrons, namely, as $\rchi$ increases, more particles impact the surface of the neutron star, fewer electrons are ejected away from it, and we still find that for $\rchi=90^{\circ}$, most electrons are trapped close to the neutron star in the same proportion as the protons. We however noticed a few differences: Protons for an inclination $\rchi$ are ejected more easily than electrons for an inclination $\pi-\rchi$, which either become trapped or crash onto the surface more frequently. 

\begin{figure*}[h]
\begin{subfigure}{.5\textwidth}
  \centering
  \includegraphics[width=\textwidth]{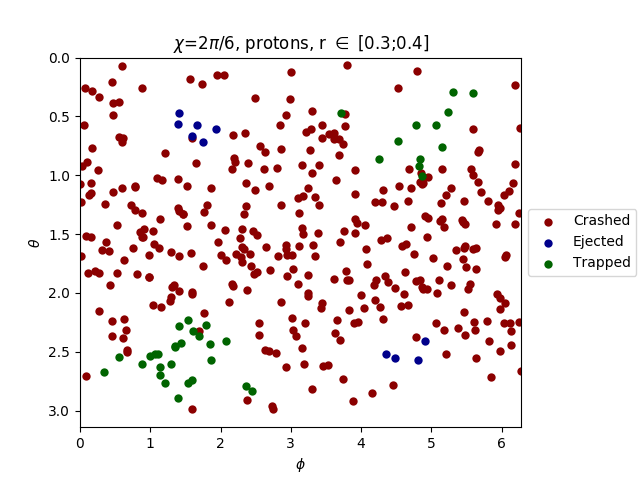}
        \put(-180,130) {\colorbox{white}{\tiny (a)}}
\label{fig:demarrage_protons_2_3}
\end{subfigure}
\begin{subfigure}{.5\textwidth}
  \centering
  \includegraphics[width=\textwidth]{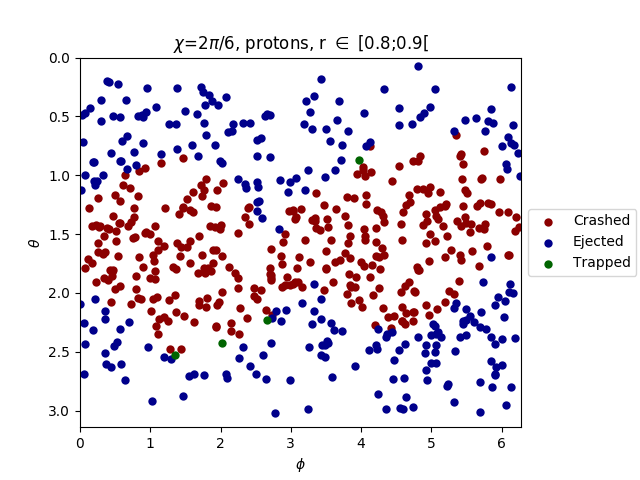}
        \put(-180,130) {\colorbox{white}{\tiny (b)}}
\label{fig:demarrage_protons_2_8}
\end{subfigure}
\begin{subfigure}{.5\textwidth}
  \centering
  \includegraphics[width=\textwidth]{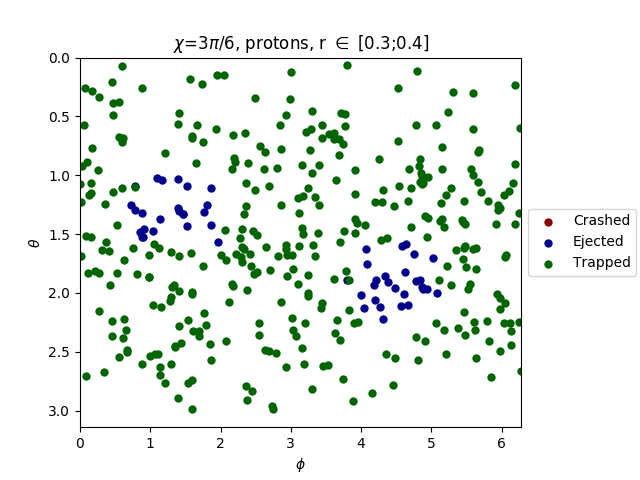}
        \put(-180,130) {\colorbox{white}{\tiny (c)}}
\label{fig:demarrage_protons_3_3}
\end{subfigure}
\begin{subfigure}{.5\textwidth}
  \centering
  \includegraphics[width=\textwidth]{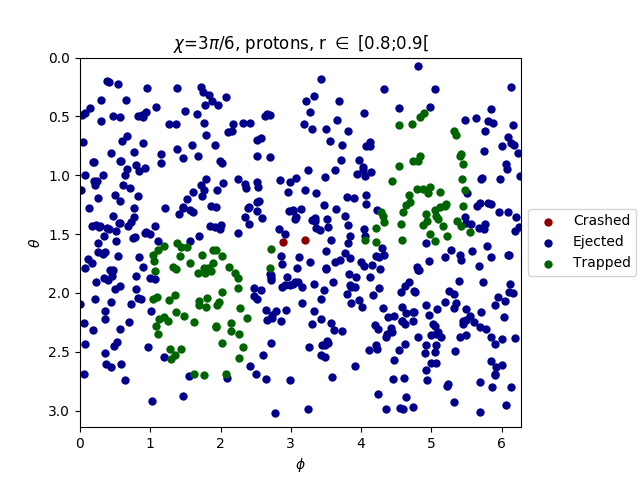}
        \put(-180,130) {\colorbox{white}{\tiny (d)}}
\label{fig:demarrage_protons_3_8}
\end{subfigure}
\caption{Final state of the protons depending on their initial positions around the pulsar. The obliquity and the initial position are $\rchi=60^{\circ}$, $r_0 \in [0.3;0.4]$ for (a) and $r_0 \in [0.8;0.9]$ for (b), $\rchi=90^{\circ}$, $r_0 \in [0.3;0.4]$ for (c) and $r_0 \in [0.8;0.9]$ for (d), radiation reaction being enabled.}
\label{fig:demarrage_devenir}
\end{figure*}

\begin{figure*}[h]
\begin{subfigure}{.5\textwidth}
  \centering
  \includegraphics[width=\textwidth]{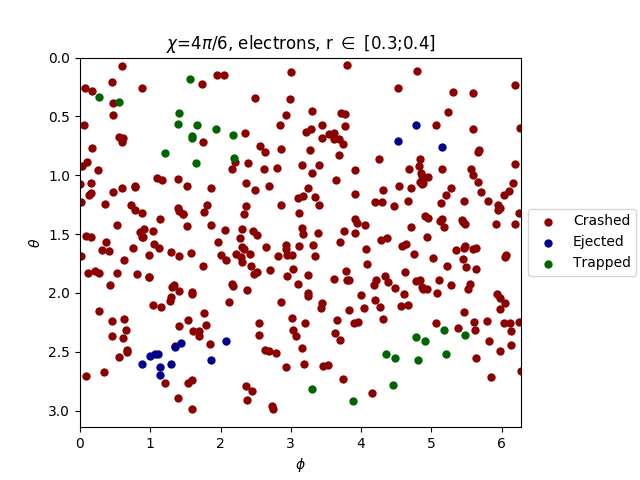}
        \put(-180,130) {\colorbox{white}{\tiny (a)}}
\label{fig:demarrage_electrons_4_3}
\end{subfigure}
\begin{subfigure}{.5\textwidth}
  \centering
  \includegraphics[width=\textwidth]{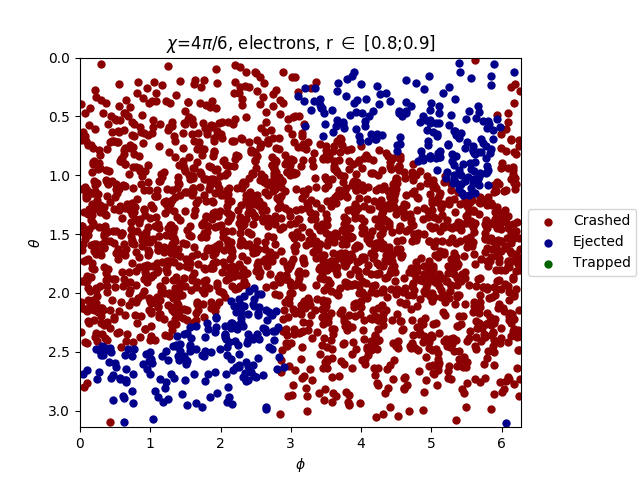}
        \put(-180,130) {\colorbox{white}{\tiny (b)}}
\label{fig:demarrage_electrons_4_8}
\end{subfigure}
\begin{subfigure}{.5\textwidth}
  \centering
  \includegraphics[width=\textwidth]{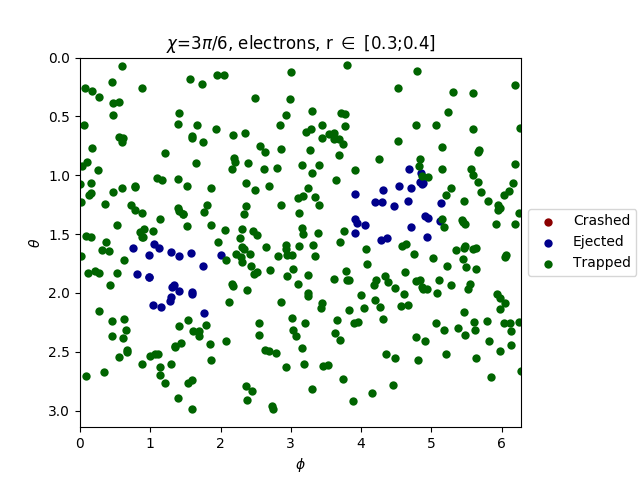}
        \put(-180,130) {\colorbox{white}{\tiny (c)}}
\label{fig:demarrage_electrons_3_3}
\end{subfigure}
\begin{subfigure}{.5\textwidth}
  \centering
  \includegraphics[width=\textwidth]{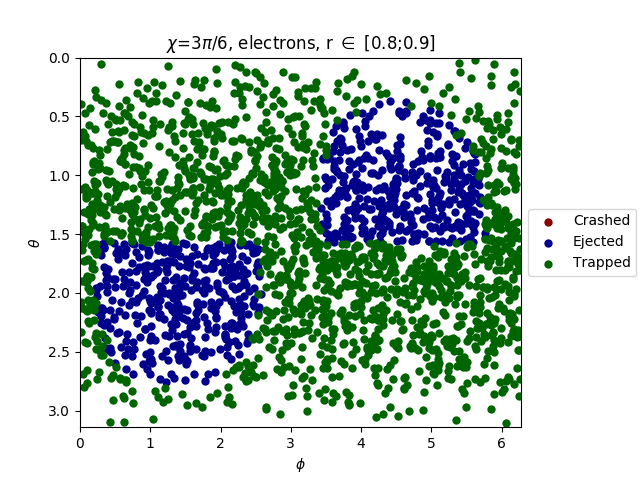}
        \put(-180,130) {\colorbox{white}{\tiny (d)}}
\label{fig:demarrage_electrons_3_8}
\end{subfigure}
\caption{Final state of the electrons depending on their initial positions around the pulsar. The obliquity and the initial position are $\rchi=120^{\circ}$, $r_0 \in [0.3;0.4]$ for (a) and $r_0 \in [0.8;0.9]$ for (b), $\rchi=90^{\circ}$, $r_0 \in [0.3;0.4]$ for (c) and $r_0 \in [0.8;0.9]$ for (d), radiation reaction being enabled.}
\label{fig:demarrage_devenir_electrons}
\end{figure*}

\begin{figure*}[h]
        \begin{subfigure}{.5\textwidth}
                \centering
                \includegraphics[width=\textwidth]{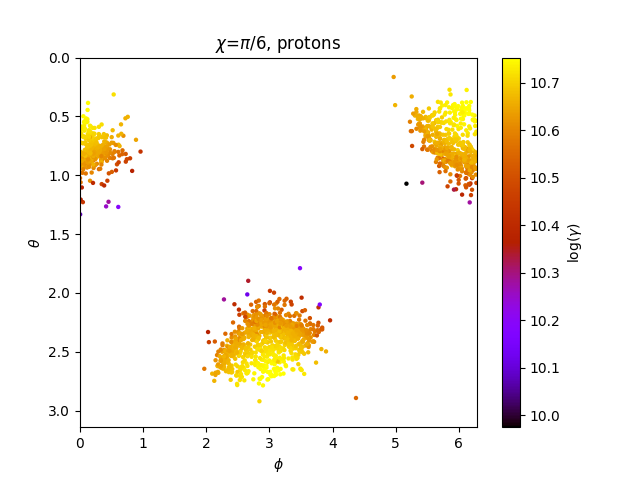}
                \put(-150,110) {\tiny (a)}
                \label{fig:crash_protons_1_rr}
        \end{subfigure}
        \begin{subfigure}{.5\textwidth}
                \centering
                \includegraphics[width=\textwidth]{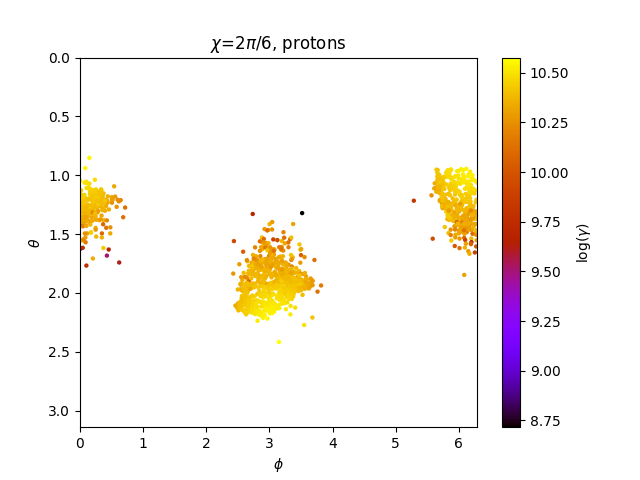}
                \put(-150,110) {\tiny (b)}
                \label{fig:crash_protons_2_rr}
        \end{subfigure}
        \begin{subfigure}{.5\textwidth}
                \centering
                \includegraphics[width=\textwidth]{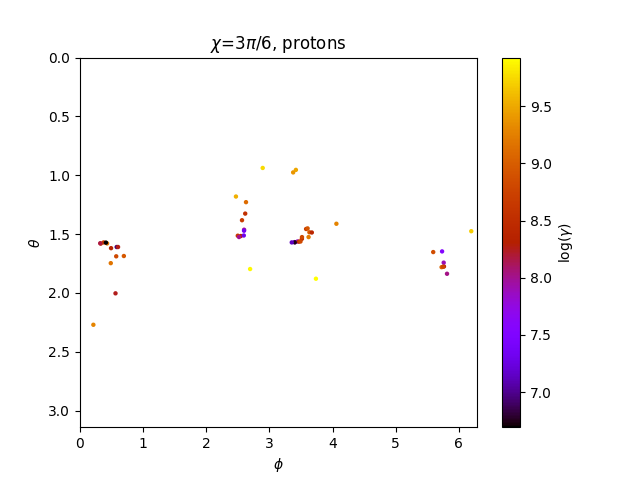}
                \put(-150,110) {\tiny (c)}
                \label{fig:crash_protons_3_rr}
        \end{subfigure}
        \begin{subfigure}{.5\textwidth}
                \centering
                \includegraphics[width=\textwidth]{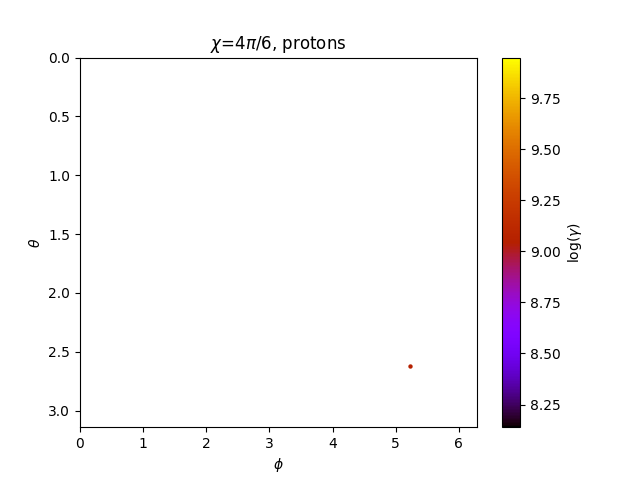}
                \put(-150,110) {\tiny (d)}
                \label{fig:crash_protons_4_rr}
        \end{subfigure}
        \caption{Map of the impact spots and Lorentz factor of protons on the surface for inclination $\rchi=30^{\circ}$ (a), $60^{\circ}$ (b), $90^{\circ}$ (c), and $120^{\circ}$ (d). Radiation reaction was enabled, and the magnetic axis is in the $\phi=0$ plane.}
        \label{fig:crash_protons_rr}
\end{figure*}

\begin{figure*}[h]
        \begin{subfigure}{.5\textwidth}
                \centering
                \includegraphics[width=\textwidth]{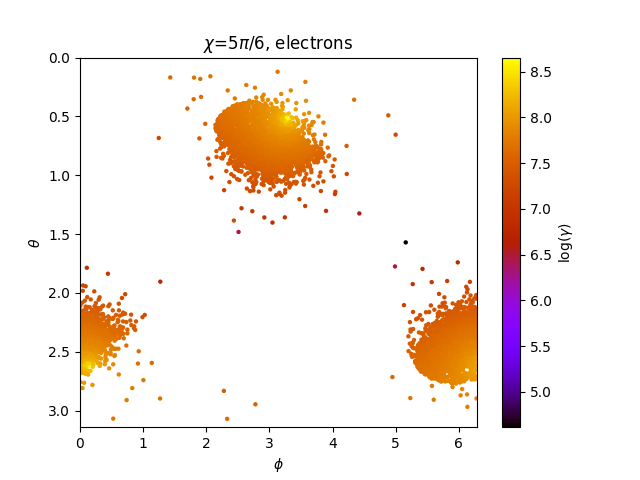}
                \put(-150,110) {\tiny (a)}
                \label{fig:crash_electrons_5_rr}
        \end{subfigure}
        \begin{subfigure}{.5\textwidth}
                \centering
                \includegraphics[width=\textwidth]{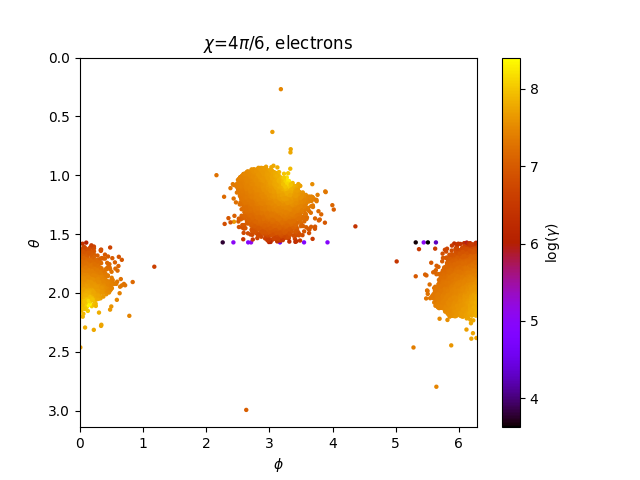}
                \put(-150,110) {\tiny (b)}
                \label{fig:crash_electrons_4_rr}
        \end{subfigure}
        \begin{subfigure}{.5\textwidth}
                \centering
                \includegraphics[width=\textwidth]{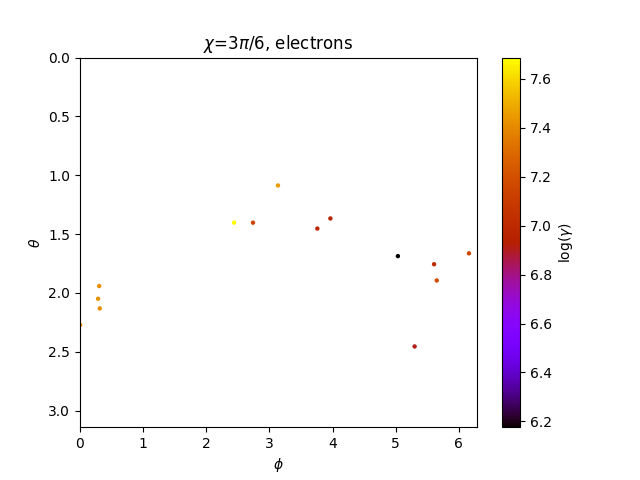}
                \put(-150,110) {\tiny (c)}
                \label{fig:crash_electrons_3_rr}
        \end{subfigure}
        \begin{subfigure}{.5\textwidth}
                \centering
                \includegraphics[width=\textwidth]{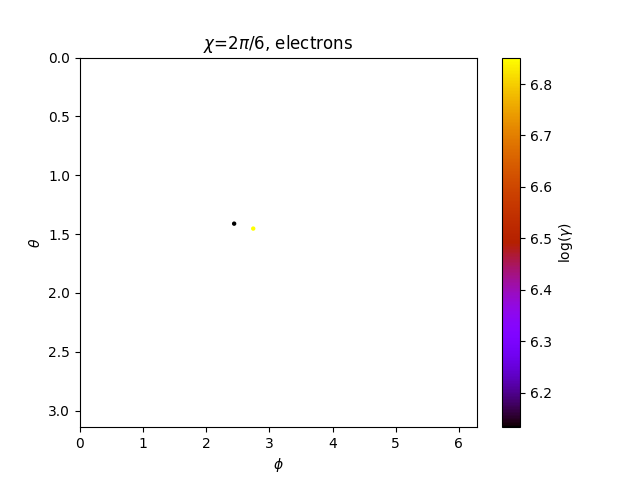}
                \put(-150,110) {\tiny (d)}
                \label{fig:crash_electronss_2_rr}
        \end{subfigure}
        \caption{Map of the impact spots and Lorentz factor of electrons on the surface for inclination $\rchi=150^{\circ}$ (a), $120^{\circ}$ (b), $90^{\circ}$ (c), and $60^{\circ}$ (d). Radiation reaction was enabled, and the magnetic axis is in the $\phi=0$ plane.}
        \label{fig:crash_electrons_rr}
\end{figure*}

\begin{figure*}[h]
        \begin{subfigure}{.5\textwidth}
                \centering
                \includegraphics[width=\textwidth]{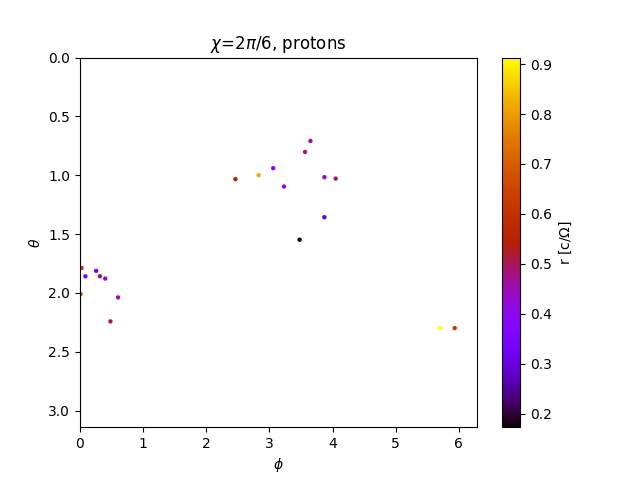}
                \put(-150,110) {\tiny (a)}
                \label{fig:trap_protons_2_rr}
        \end{subfigure}
        \begin{subfigure}{.5\textwidth}
                \centering
                \includegraphics[width=\textwidth]{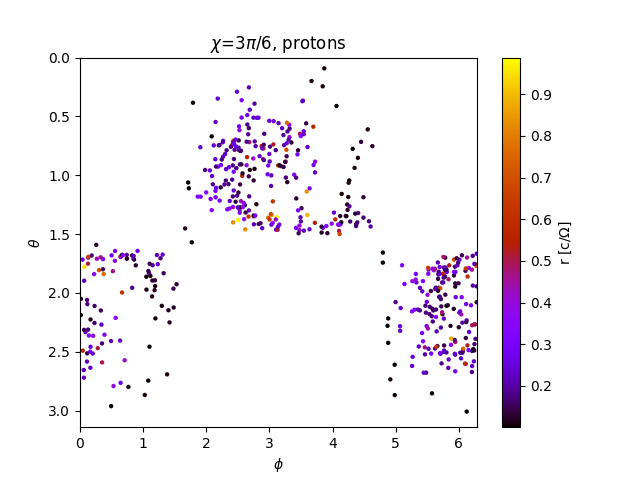}
                \put(-150,110) {\tiny (b)}
                \label{fig:trap_protons_3_rr}
        \end{subfigure}
        \begin{subfigure}{.5\textwidth}
                \centering
                \includegraphics[width=\textwidth]{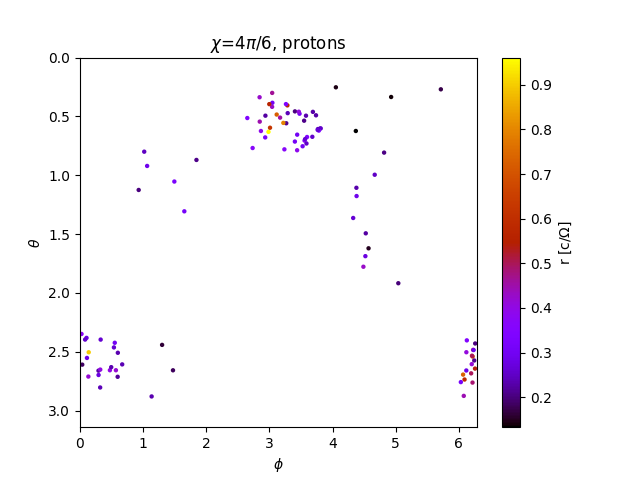}
                \put(-150,110) {\tiny (c)}
                \label{fig:trap_protons_4_rr}
        \end{subfigure}
        \begin{subfigure}{.5\textwidth}
                \centering
                \includegraphics[width=\textwidth]{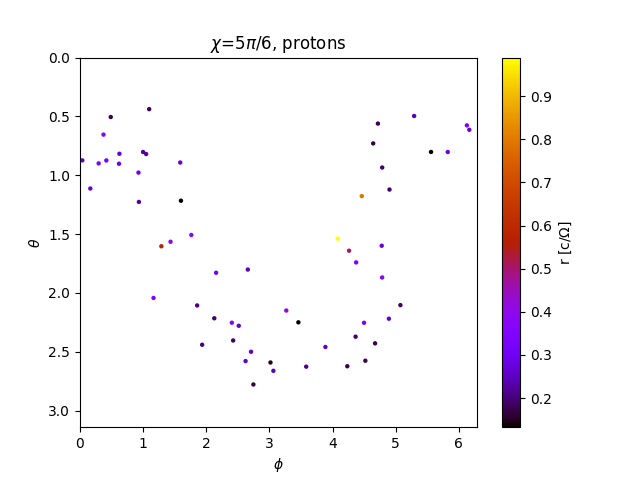}
                \put(-150,110) {\tiny (d)}
                \label{fig:trap_protons_5_rr}
        \end{subfigure}
        \caption{Map of the final colatitude, azimuth, and radius (colour) of protons around neutron stars of inclination $60^{\circ}$ (a), $90^{\circ}$ (b), $120^{\circ}$ (c), and $150^{\circ}$ (d). Radiation reaction was enabled, and the magnetic axis is in the $\phi=0$ plane.}
        \label{fig:trap_protons_rr}
\end{figure*}

\begin{figure*}[h]
        \begin{subfigure}{.5\textwidth}
                \centering
                \includegraphics[width=\textwidth]{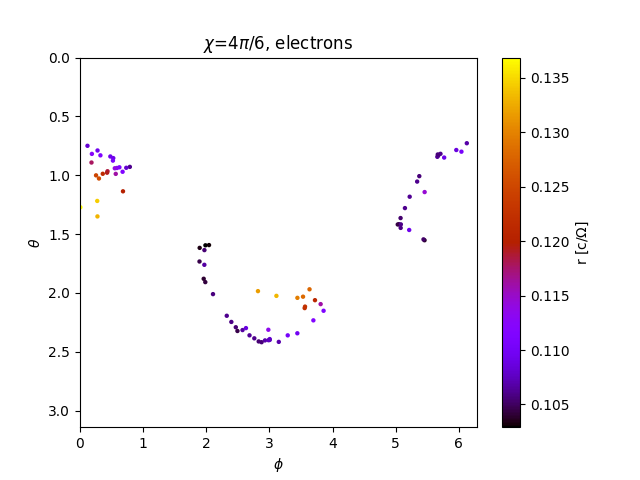}
                \put(-150,110) {\tiny (a)}
                \label{fig:trap_electron_4_rr}
        \end{subfigure}
        \begin{subfigure}{.5\textwidth}
                \centering
                \includegraphics[width=\textwidth]{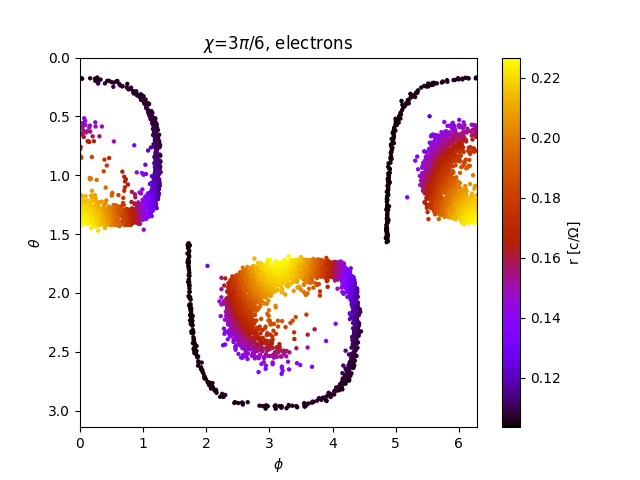}
                \put(-150,110) {\tiny (b)}
                \label{fig:trap_electron_3_rr}
        \end{subfigure}
        \begin{subfigure}{.5\textwidth}
                \centering
                \includegraphics[width=\textwidth]{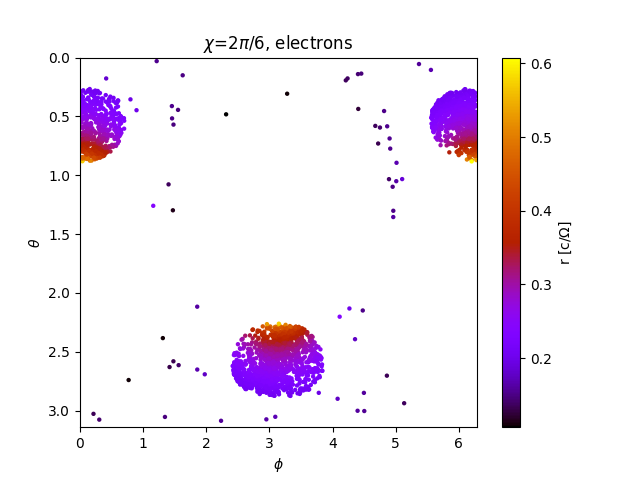}
                \put(-150,110) {\tiny (c)}
                \label{fig:trap_electron_2_rr}
        \end{subfigure}
        \begin{subfigure}{.5\textwidth}
                \centering
                \includegraphics[width=\textwidth]{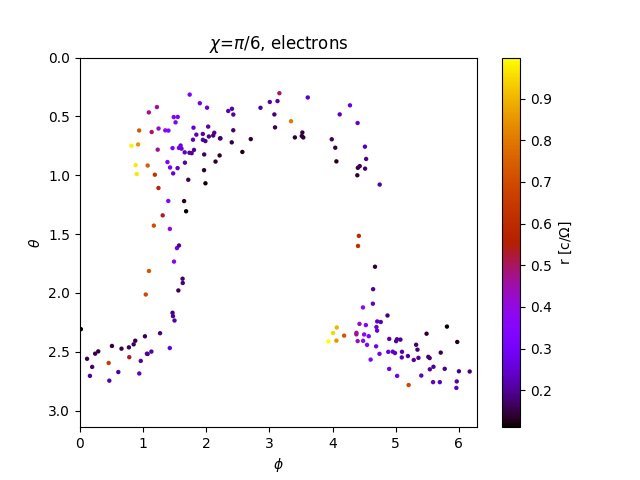}
                \put(-150,110) {\tiny (d)}
                \label{fig:trap_electron_1_rr}
        \end{subfigure}
        \caption{Map of the final colatitude, azimuth, and radius (colour) of electrons around neutron stars of inclination $60^{\circ}$ (a), $90^{\circ}$ (b), $120^{\circ}$ (c), and $150^{\circ}$ (d). Radiation reaction was enabled, and the magnetic axis is in the $\phi=0$ plane.}
        \label{fig:trap_electrons_rr}
\end{figure*}

\begin{figure*}[h]
        \begin{subfigure}{.5\textwidth}
                \centering
                \includegraphics[width=\textwidth]{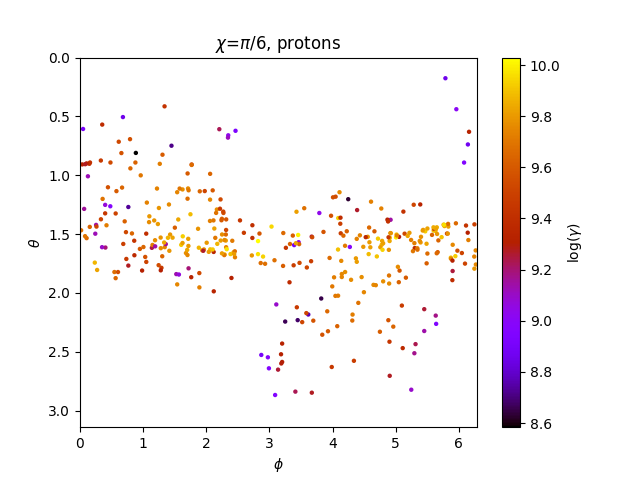}
                \put(-150,110) {\colorbox{white}{\tiny (a)}}
                \label{fig:eject_protons_1_rr}
        \end{subfigure}
        \begin{subfigure}{.5\textwidth}
                \centering
                \includegraphics[width=\textwidth]{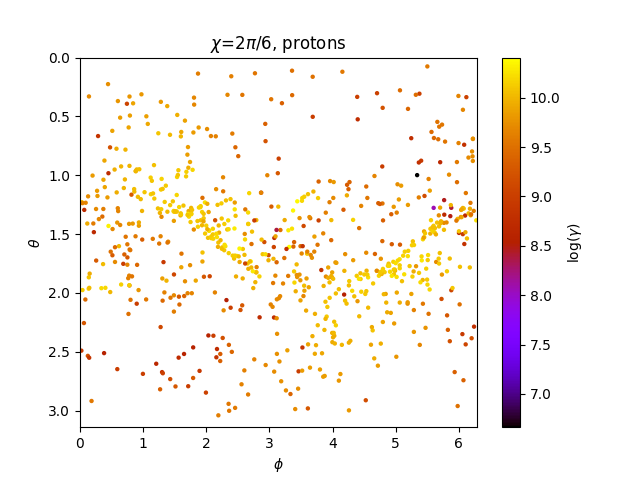}
                \put(-150,110) {\colorbox{white}{\tiny (b)}}
                \label{fig:eject_protons_2_rr}
        \end{subfigure}
        \begin{subfigure}{.5\textwidth}
                \centering
                \includegraphics[width=\textwidth]{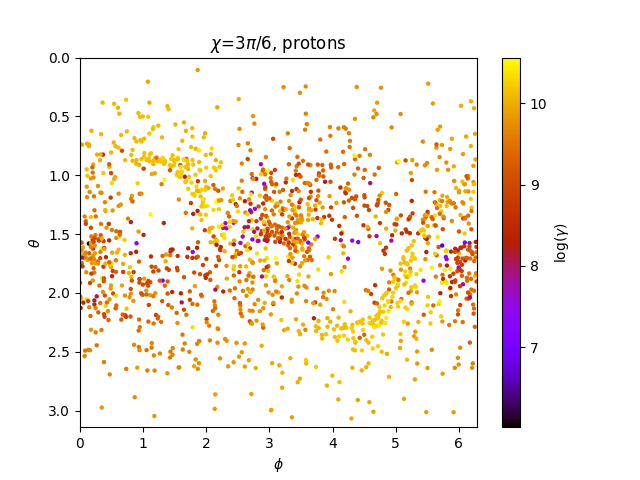}
                \put(-150,110) {\colorbox{white}{\tiny (c)}}
                \label{fig:eject_protons_3_rr}
        \end{subfigure}
        \begin{subfigure}{.5\textwidth}
                \centering
                \includegraphics[width=\textwidth]{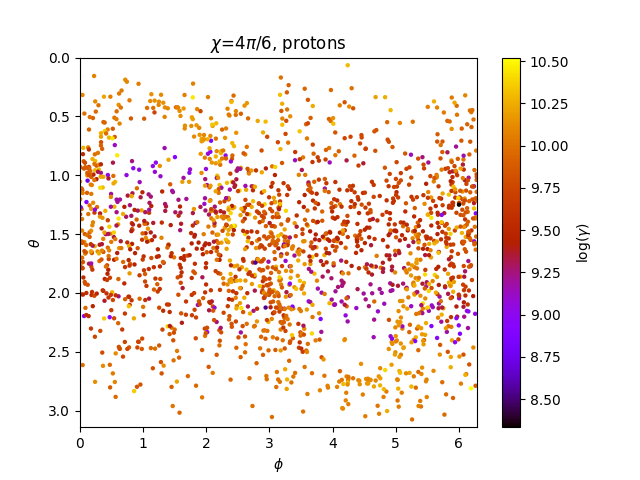}
                \put(-150,110) {\colorbox{white}{\tiny (d)}}
                \label{fig:eject_protons_4_rr}
        \end{subfigure}
        \caption{Map of the ejection colatitude and azimuth and Lorentz factor of protons on neutron stars of inclination $30^{\circ}$ (a), $60^{\circ}$ (b), $90^{\circ}$ (c), and $120^{\circ}$ (d). Radiation reaction was enabled.}
        \label{fig:eject_protons_rr}
\end{figure*}

\begin{figure*}[h]
        \begin{subfigure}{.5\textwidth}
                \centering
                \includegraphics[width=\textwidth]{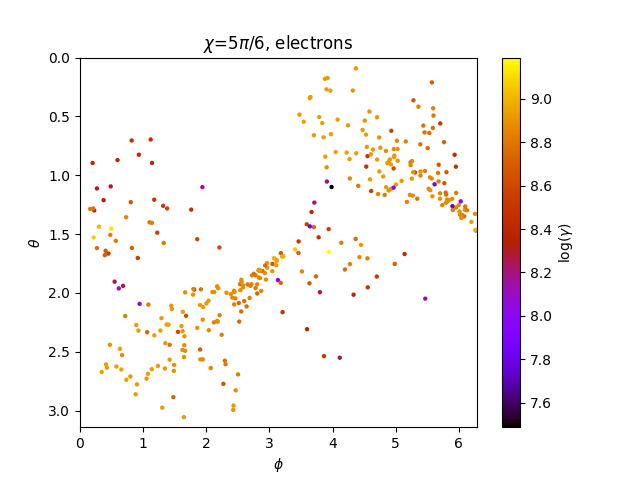}
                \put(-150,110) {\colorbox{white}{\tiny (a)}}
                \label{fig:eject_electron_5_rr}
        \end{subfigure}
        \begin{subfigure}{.5\textwidth}
                \centering
                \includegraphics[width=\textwidth]{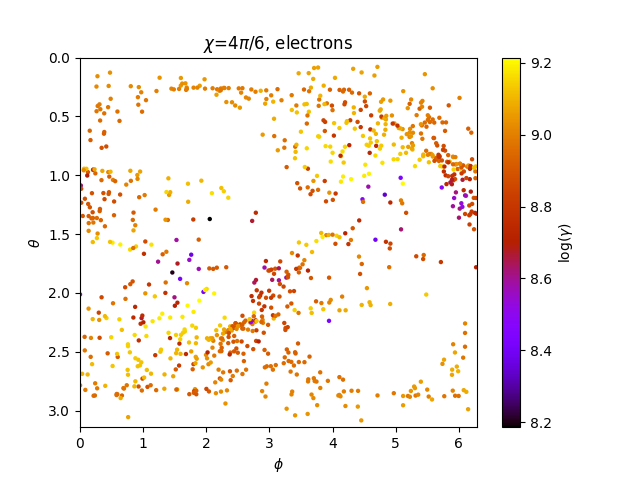}
                \put(-150,110) {\colorbox{white}{\tiny (b)}}
                \label{fig:eject_electron_4_rr}
        \end{subfigure}
        \begin{subfigure}{.5\textwidth}
                \centering
                \includegraphics[width=\textwidth]{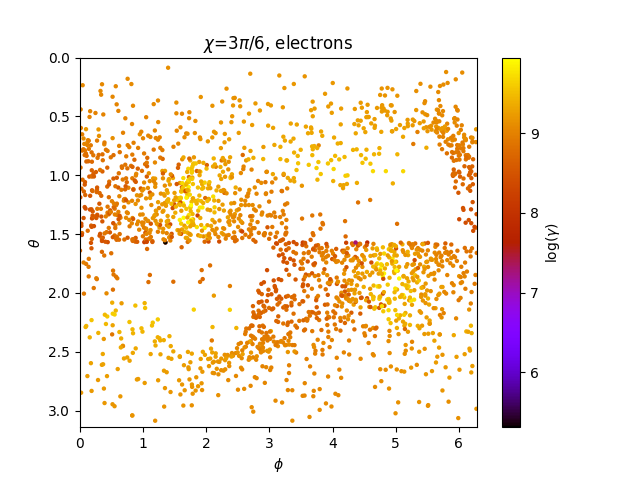}
                \put(-150,110) {\colorbox{white}{\tiny (c)}}
                \label{fig:eject_electron_3_rr}
        \end{subfigure}
        \begin{subfigure}{.5\textwidth}
                \centering
                \includegraphics[width=\textwidth]{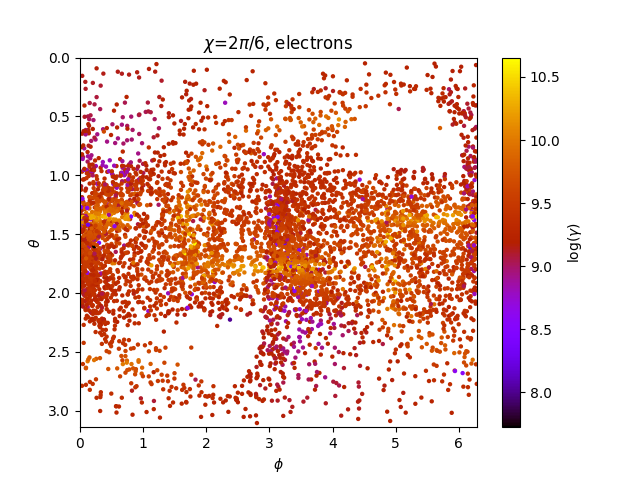}
                \put(-150,110) {\colorbox{white}{\tiny (d)}}
                \label{fig:eject_electron_2_rr}
        \end{subfigure}
        \caption{Map of the ejection colatitude and azimuth and Lorentz factor of electrons on neutron stars of inclination $150^{\circ}$ (a), $120^{\circ}$ (b), $90^{\circ}$ (c), and $60^{\circ}$ (d). Radiation reaction was enabled.}
        \label{fig:eject_electrons_rr}
\end{figure*}

Actually, we noticed that when respectively comparing the trajectories of protons and electrons in Figure~\ref{fig:demarrage_devenir} and Fig.~\ref{fig:demarrage_devenir_electrons}; Fig.\ref{fig:crash_protons_rr} and Fig.\ref{fig:crash_electrons_rr}; Fig.\ref{fig:trap_protons_rr} and Fig.\ref{fig:trap_electrons_rr}; and Fig.\ref{fig:eject_protons_rr} and Fig.\ref{fig:eject_electrons_rr}, we found the protons and electrons to possess very similar trajectories. If a proton starting at a position $(r_0;\theta_0;\phi_0)$ ends at position $(r_f;\theta_f;\phi_f)$ for a pulsar of inclination $\rchi$, an electron close to a pulsar of inclination $\pi-\rchi$ starting at position $(r_0;\pi-\theta_0;\phi_0)$ is very likely to end the simulation at position $(r_f;\pi-\theta_f;\phi_f)$. 

The symmetrical behaviour found in these simulations reflects the fact that the trajectories are relatively insensitive to the charge over mass ratio $q/m$ of the particles, except for the sign of the charge itself. Indeed, in the ultra-relativistic regime, the mass of the particles becomes negligible compared to their total kinetic energy, and they can be considered as massless particles just like photons, for instance. However, because the radiation reaction force does not scale linearly with this ratio $q/m$, it is not at all obvious that the trajectories will remain similar. Nevertheless, we found that the radiation reaction impacts the motion of protons similarly to that of electrons. However, due to the difference in mass between protons and electrons, their respective Lorentz factors, although both ultra relativistic, scale like their mass ratios $m_e/m_p$.
\begin{table}[h]
\centering
\begin{tabular}{ |c|c|c|c|c|c|c| } 
\hline
\multicolumn{2}{|c|}{$\rchi$} & $30^{\circ}$ & $60^{\circ}$ & $90^{\circ}$ & $120^{\circ}$ & $150^{\circ}$ \\
\hline
\hline
\multirow{3}{*}{Electrons} & Crashed & 1 & 2 & 14 & 7202 & 7856  \\
& Trapped & 185 & 2721 & 6158 & 83 & 0   \\
& Ejected & 8006 & 5469 & 2020 & 907 & 337   \\
\hline
\hline
\multirow{3}{*}{Protons} & Crashed & 7779 & 7069 & 10 & 1 & 0   \\
& Trapped & 2 & 108 & 6067 & 2603 & 180   \\
& Ejected & 411 & 1015 & 2115 & 5588 & 8012   \\
\hline
\end{tabular}
\caption{Final state of the particles depending on the inclination~$\rchi$ of the pulsar and on the species. Electrons are shown on the first line and protons on the second line. 8192 particles per inclination have been simulated and radiation reaction enabled.}
\label{tab:etat_particules}
\end{table}
Moreover, the statistics presented in Table~\ref{tab:etat_particules} can be linked to the initial positions of the particles.

\paragraph{Starting positions.}
Indeed, the initial position of the particles has an influence on their final state. Figure~\ref{fig:demarrage_devenir} and Figure~\ref{fig:demarrage_devenir_electrons} respectively show the map of the initial positions of protons and electrons, and their final state is indicated with a colour code. We chose two altitude intervals. The first is close to the surface, with $r/\rlight \in [0.3,0.4]$ and $\rchi = 60^\circ$, and the second is close to the light cylinder, with $r/\rlight \in [0.8,0.9]$ and $\rchi = 90^\circ$. The figures also highlight the importance of the neutron star obliquity~$\rchi$. In addition, we note the figures show regions with clear boundaries and almost no overlap that are prone to ejection, crashing, or trapping of particles, meaning that the particle's behaviour is well defined according to their starting point.

\begin{figure*}[h]
\begin{subfigure}{.5\textwidth}
  \centering
  \includegraphics[width=\textwidth]{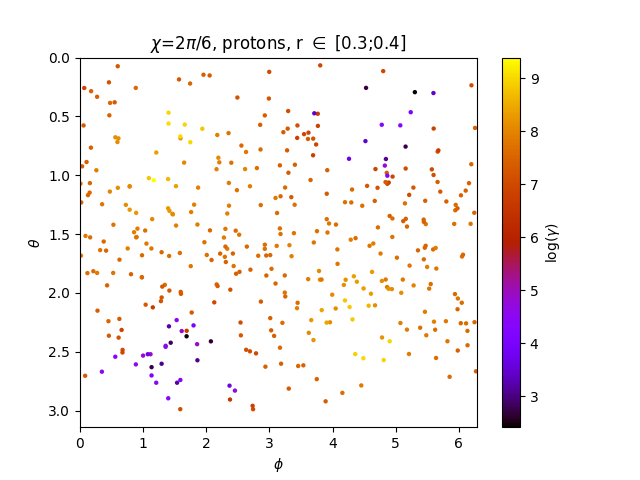}
        \put(-150,110) {\tiny (a)}
\label{fig:demarrage_protons_2_3_lorentz}
\end{subfigure}
\begin{subfigure}{.5\textwidth}
  \centering
  \includegraphics[width=\textwidth]{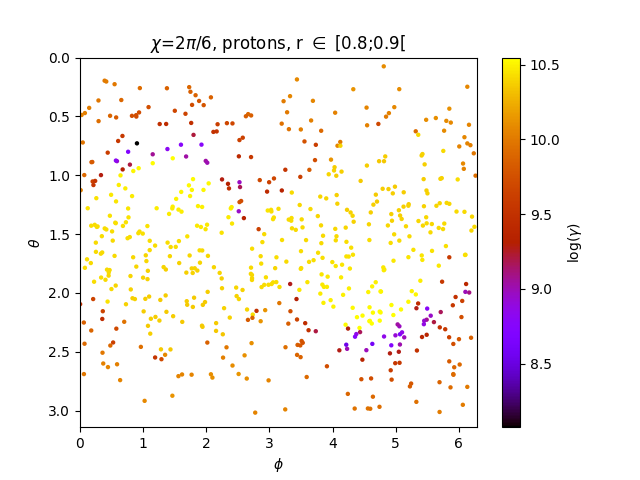}
        \put(-150,110) {\tiny (b)}
\label{fig:demarrage_protons_2_8_lorentz}
\end{subfigure}
\begin{subfigure}{.5\textwidth}
  \centering
  \includegraphics[width=\textwidth]{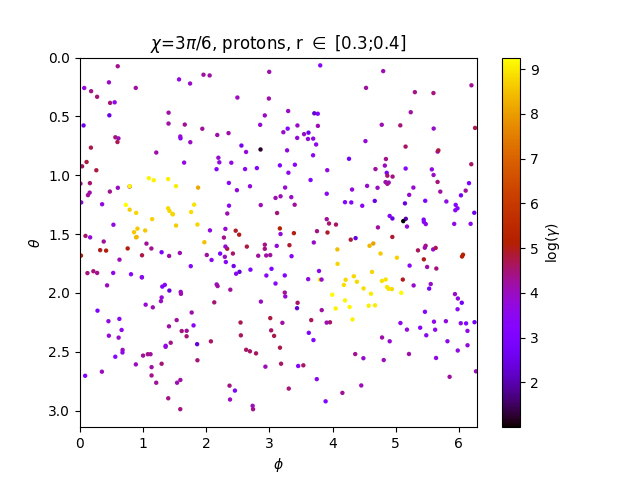}
        \put(-150,110) {\tiny (c)}
\label{fig:demarrage_protons_3_3_lorentz}
\end{subfigure}
\begin{subfigure}{.5\textwidth}
  \centering
  \includegraphics[width=\textwidth]{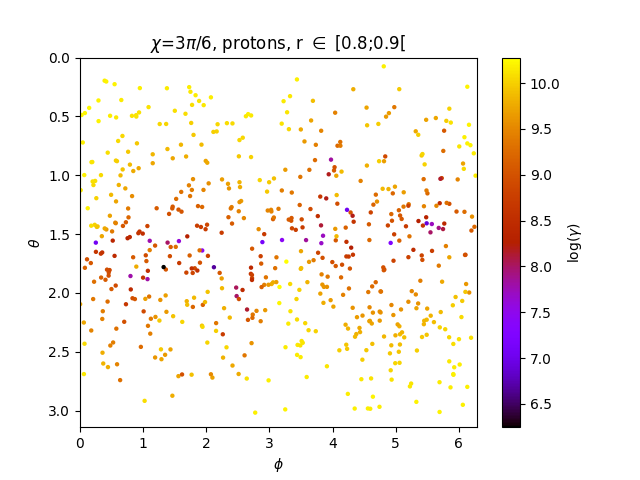}
        \put(-150,110) {\tiny (d)}
\label{fig:demarrage_protons_3_8_lorentz}
\end{subfigure}
\caption{Final Lorentz factor of the particles shown in Fig~\ref{fig:demarrage_devenir}.}
\label{fig:demarrage_Lorentz}
\end{figure*}

Comparing Figure~\ref{fig:demarrage_devenir} to Figure~\ref{fig:demarrage_Lorentz} allowed us to find either clear distinctions between the Lorentz factor of ejected particles and the other populations or that their speed would be similar no matter if they are ejected or not.
Figure~\ref{fig:demarrage_Lorentz} also shows the influence of the initial radius on the spread of energy of the particles: as the initial radius rises, proton energies become less spread for $\rchi=60^{\circ}$, spanning six orders of magnitude for low altitude ($r\in[0.3,0.4]$) and only less than two orders of magnitude for high altitude ($r\in[0.8,0.9]$). However, for some other configurations, such as the orthogonal rotator with $\rchi=90^{\circ}$, the final Lorentz factor is not affected by the initial altitude when the particles are trapped.

\begin{figure*}[h]
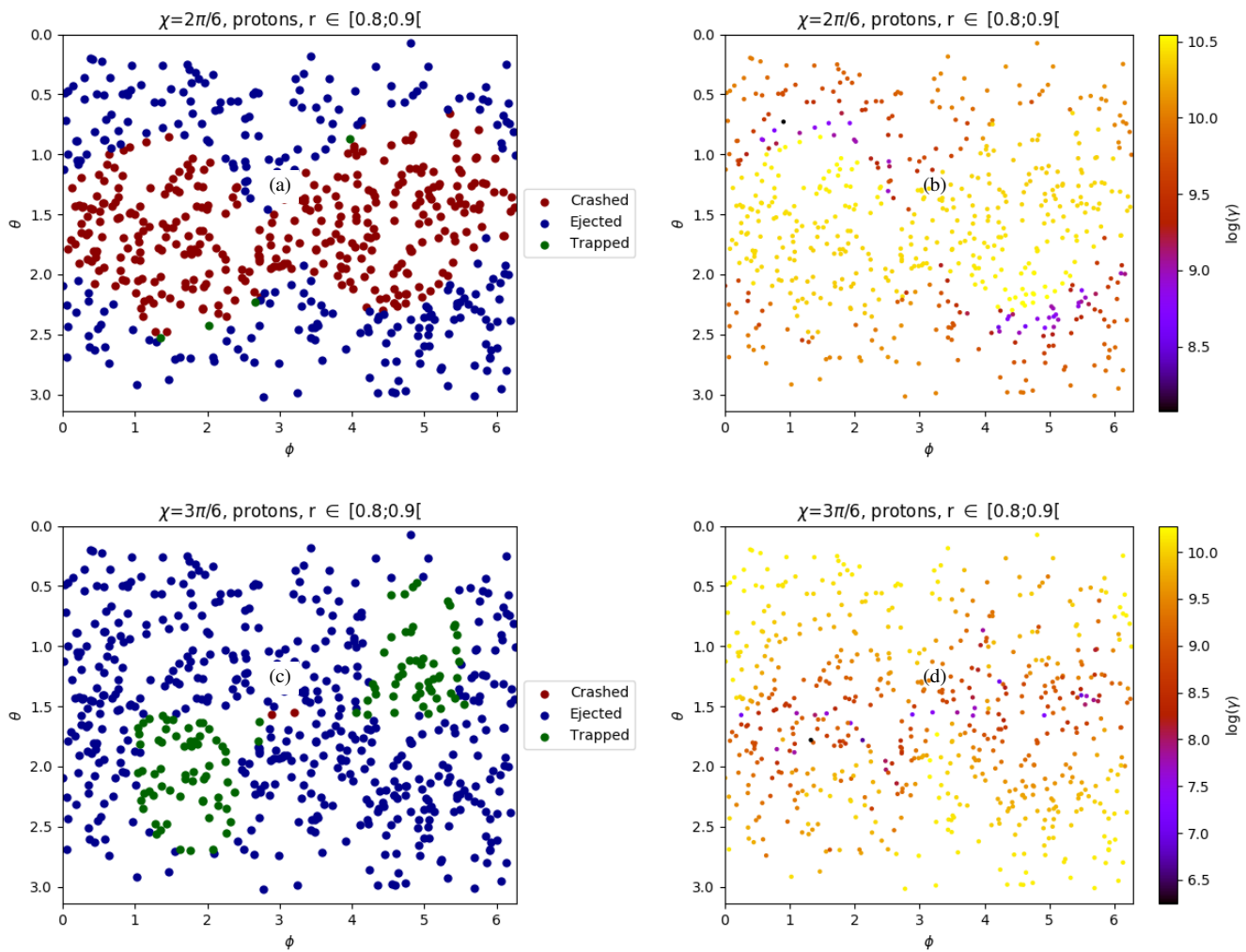

\begin{subfigure}{.5\textwidth}
  \centering
  \includegraphics[width=\textwidth]{state_proton_B5_CHI2sur6_8.png}
        \put(-150,110) {\colorbox{white}{\tiny (a)}}
\end{subfigure}
\begin{subfigure}{.5\textwidth}
  \centering
  \includegraphics[width=\textwidth]{start_proton_B5_CHI2sur6_8.png}
        \put(-150,110) {\tiny (b)}
\end{subfigure}
\begin{subfigure}{.5\textwidth}
  \centering
  \includegraphics[width=\textwidth]{state_proton_B5_CHI3sur6_8.png}
        \put(-150,110) {\colorbox{white}{\tiny (c)}}
\end{subfigure}
\begin{subfigure}{.5\textwidth}
  \centering
  \includegraphics[width=\textwidth]{start_proton_B5_CHI3sur6_8.png}
        \put(-150,110) {\tiny (d)}
\end{subfigure}
\caption{Final Lorentz factor (a) and state (b) of the particles near a neutron star of inclination $\rchi=60^{\circ}$, with $r_0 \in [0.8;0.9]$, and final Lorentz factor (a) and state (b) of the particles near a neutron star of inclination $\rchi=90^{\circ}$, with $r_0 \in [0.8;0.9]$. Radiation reaction was disabled in both cases.}
\label{fig:demarrage_no_rr}
\end{figure*}

Upon comparing the results with the radiation reaction in Figures~\ref{fig:demarrage_devenir} and \ref{fig:demarrage_Lorentz} to those without radiation reaction in Figure~\ref{fig:demarrage_no_rr}, we noticed that the radiation reaction drastically influences  the behaviour of the particles regarding their Lorentz factors. The radiation reaction decreases the final Lorentz factor by at least one order of magnitude and sometimes changes the final state of the particles, depending on their initial position.

\begin{figure*}[h]
  \centering
  \includegraphics[width=0.9\textwidth]{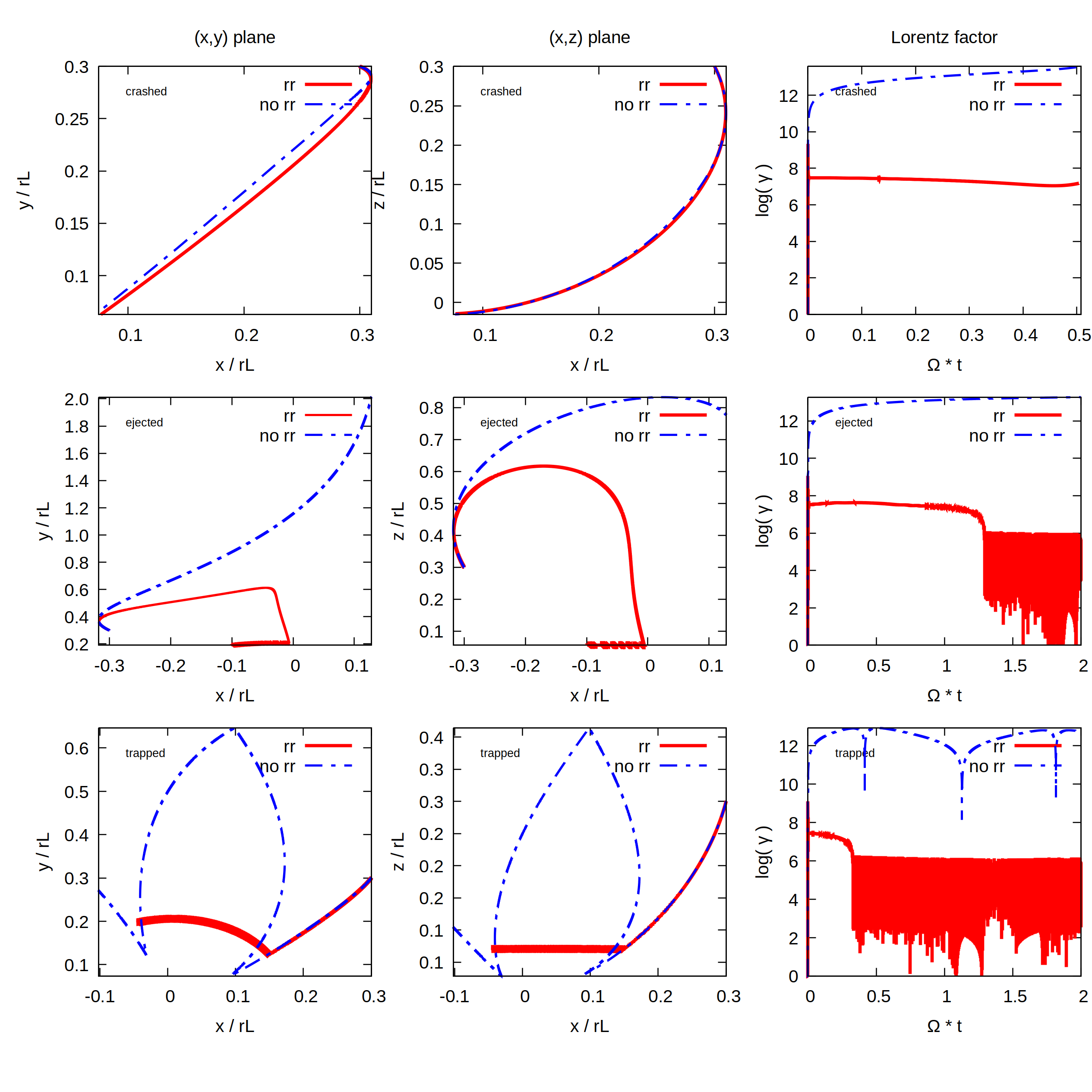}
\caption{Two trajectories of the same particle injected at the same location but with radiation reaction in one case(`rr', in red solid lines) and without it in the other (`no rr', in dash-dotted blue lines). The upper panel corresponds to a crashed particle, the middle panel to an ejected particle, and the bottom panel to a trapped particle. The left column projects the trajectory on the $(x,y)$ plane, the middle column projects it on the $(x,z)$ plane, and the right column shows the evolution of the Lorentz factor in a log scale.}
\label{fig:trajectories}
\end{figure*}

The impact of radiation reaction on the particle trajectory can be drastic. Indeed, a comparison of two trajectories of particles injected at the same location within the magnetosphere where one has radiation reaction enabled, `rr', and the other does not, `no rr', is shown in Figure~\ref{fig:trajectories}. The upper row shows a crashed particle, the middle panel shows an ejected particle, and the bottom panel shows a trapped particle. The left column of the figure projects the trajectory on the $(x,y)$ plane, and the middle column projects the trajectory on the $(x,z)$ plane. The particle with radiation reaction enabled is shown with red and the particle without it is blue. The right column shows the time evolution of the Lorentz factor in both cases. The labels crashed, trapped, or ejected refer to the trajectory as observed in the `no rr' case. Simulations were performed for electrons. For the trapped case, the particles start by following a similar path until the one losing energy because of radiation reaction brakes and only drifts, whereas the particle without radiation reaction decelerates and then accelerates again in another direction (see lower panel). Its Lorentz factor oscillates between $10^8$ and $10^{13}$ in a periodic fashion associated with a bouncing motion during the trapped stage of the `no rr' case. The particle with `rr' loses a large fraction of its initial energy quickly, decreasing the Lorentz factor from $10^8$ tp $10^6$. It then experiences violent oscillations to a point it almost rests, and then it accelerates again. For the crashed case, both particles follow similar tracks, although they are slightly different in the $(x,y)$ plane (top panel of Figure~\ref{fig:trajectories}). Nevertheless, the Lorentz factor is four to five orders of magnitude lower in the radiation reaction case. While moving towards the star, the electron efficiently radiates its energy gained in the increasingly stronger electromagnetic field. The largest difference in particle trajectories was observed between a particle ejected without radiation reaction that actually crashes when radiation is enabled (middle panel). The asymptotic Lorentz factor in the ejected case is about $10^{13}$, whereas the Lorentz factor in the crashed case  behaves as in the trapped and crashed case shown in the bottom row of Figure~\ref{fig:trajectories}.

The properties of the particles can also be viewed according to their final state: crashed, ejected, or trapped. This point of view is explored more deeply in the following sub-section.

\paragraph{Particles impacting the neutron star.}

\begin{figure*}[h]
\begin{subfigure}{.5\textwidth}
  \centering
  \includegraphics[width=\textwidth]{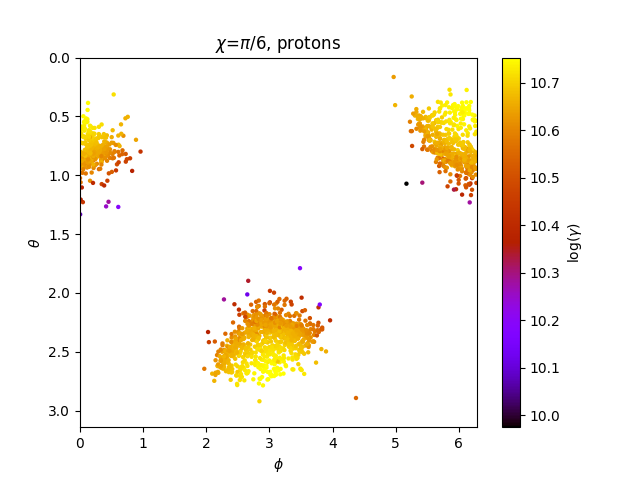}
        \put(-150,110) {\tiny (a)}
\label{fig:crash_protons_1_no_rr}
\end{subfigure}
\begin{subfigure}{.5\textwidth}
  \centering
  \includegraphics[width=\textwidth]{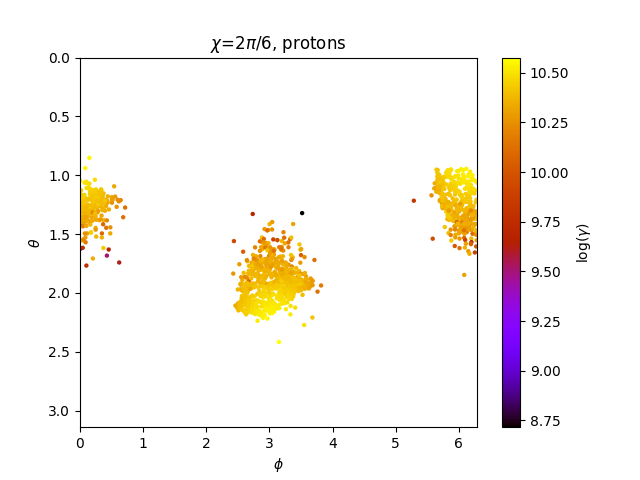}
        \put(-150,110) {\tiny (b)}
\label{fig:crash_protons_2_no_rr}
\end{subfigure}
\begin{subfigure}{0.5\textwidth}
  \centering
  \includegraphics[width=\textwidth]{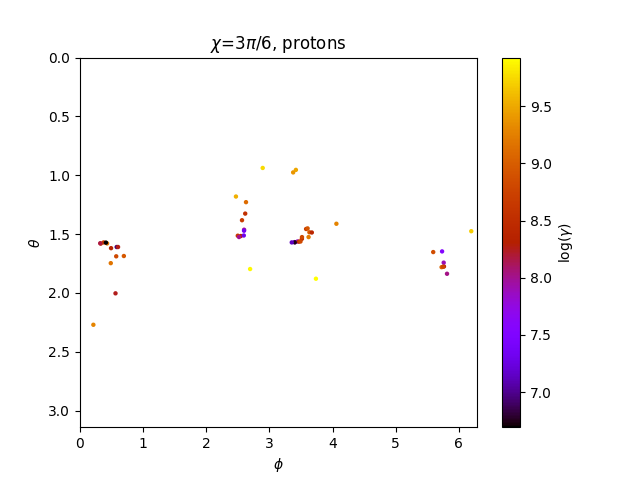}
        \put(-150,110) {\tiny (c)}
\label{fig:crash_protons_3_no_rr}
\end{subfigure}
\begin{subfigure}{.5\textwidth}
  \centering
  \includegraphics[width=\textwidth]{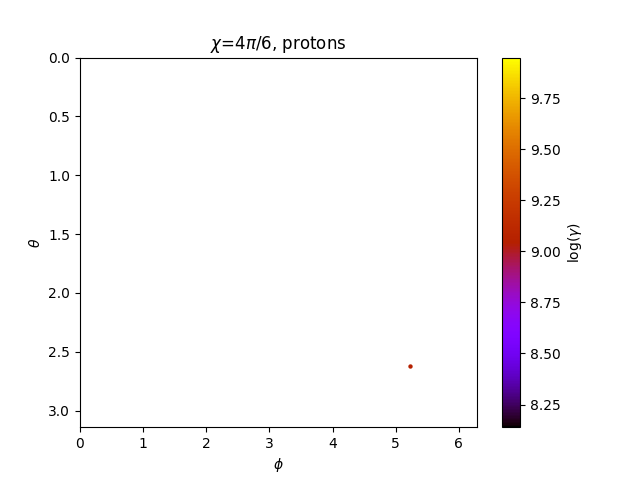}
        \put(-150,110) {\tiny (d)}
\label{fig:crash_protons_4_no_rr}
\end{subfigure}
\caption{Same as Figure~\ref{fig:crash_protons_rr} but with radiation reaction disabled.}
\label{fig:crash_protons_no_rr}
\end{figure*}

We also investigated the final positions of particles when crashing onto the stellar surface. Thus, as shown in Figure~\ref{fig:crash_protons_rr}, we report the hotspots on the polar caps, which highlights the fact that the protons impact the star in very localised areas concentrated around the magnetic axis. Indeed, those areas are always located around the azimuth $\phi=0^{\circ}$ and $\phi=180^{\circ}$, and they respectively stay in the northern and southern hemispheres as $\rchi$ increases. The spot at $\phi=0^{\circ}$ is always found between $\theta=0^{\circ}$ and $\theta=90^{\circ}$, while the spot at $\phi=180^{\circ}$ is found between $\theta=90^{\circ}$ and $\theta=180^{\circ}$. By comparison, if the radiation reaction is disabled, as in Figure~\ref{fig:crash_protons_no_rr}, the hotspots move slightly along the meridian as $\rchi$ changes. When $\rchi=30^{\circ}$, the hotspots are respectively at $\theta=45^{\circ}$ and $\theta=135^{\circ}$, while for $\rchi=90^{\circ}$, they are respectively centred around $\theta=108^{\circ}$ and $\theta=72^{\circ}$. We emphasise that there exists no south-north hemisphere symmetry in these impact maps, neither for protons nor for electrons. This is due to the nature of the electromagnetic field, which is a vector field, and it does not produce the same pattern when $\rchi>90^\circ$ compared to $\rchi<90^\circ$.

\paragraph{Trapped particles.}

The structures formed by trapped particles remained close to the neutron star. These  are shown in  Figure~\ref{fig:trap_protons_rr} and Figure~\ref{fig:trap_electrons_rr}. Again, we noticed that these particles tend to avoid some regions while populating other well-defined areas. We also note that the order of the figure is for increasing $\rchi$ for protons but decreasing $\rchi$ for electrons. This highlights the symmetry between positive and negative charges when switching from $\rchi$ to $\pi-\rchi$.

\begin{figure*}[h]
\begin{subfigure}{.5\textwidth}
  \centering
  \includegraphics[width=\textwidth]{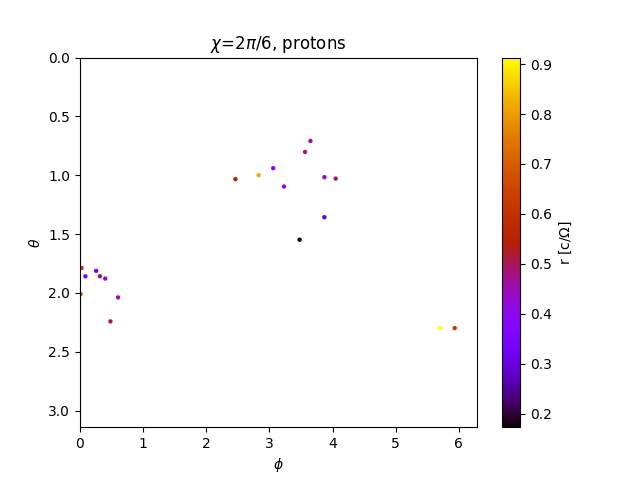}
        \put(-150,110) {\tiny (a)}
\label{fig:trap_protons_2_no_rr}
\end{subfigure}
\begin{subfigure}{.5\textwidth}
  \centering
  \includegraphics[width=\textwidth]{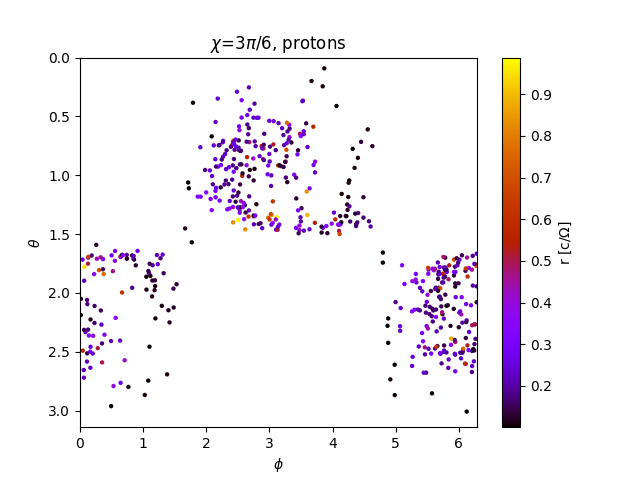}
        \put(-150,110) {\tiny (b)}
\label{fig:trap_protons_3_no_rr}
\end{subfigure}
\begin{subfigure}{.5\textwidth}
  \centering
  \includegraphics[width=\textwidth]{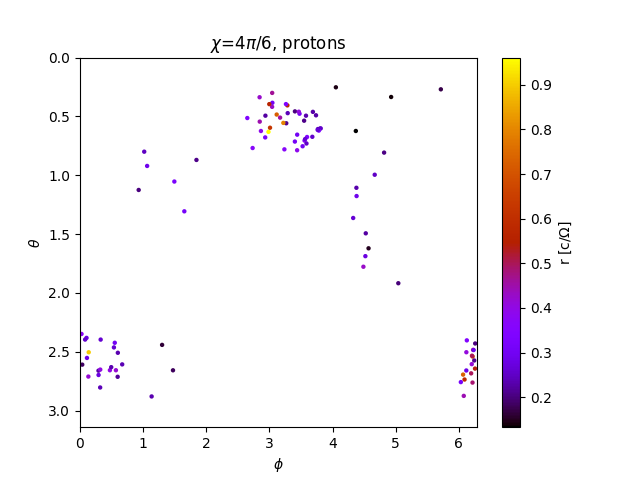}
        \put(-150,110) {\tiny (c)}
\label{fig:trap_protons_4_no_rr}
\end{subfigure}
\begin{subfigure}{.5\textwidth}
  \centering
  \includegraphics[width=\textwidth]{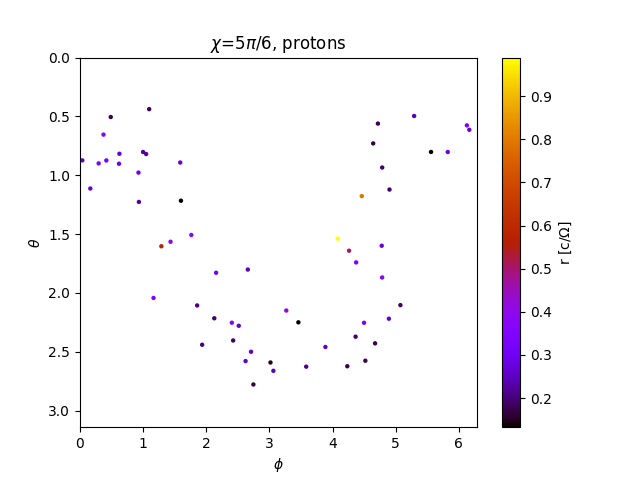}
        \put(-150,110) {\tiny (d)}
\label{fig:trap_protons_5_no_rr}
\end{subfigure}
\caption{Map of the final colatitude, azimuth, and radius (colour) of protons around neutron stars of inclination $60^{\circ}$ (a), $90^{\circ}$ (b), $120^{\circ}$ (c), and $150^{\circ}$ (d). Radiation reaction was disabled, and the magnetic axis is in the $\phi=0$ plane.}
\label{fig:trap_protons_no_rr}
\end{figure*}

As shown in Figure~\ref{fig:trap_protons_no_rr}, particles have quite similar positions whether radiation reaction is enabled or not, with the same regions being populated but some particles being farther away from the neutron star when radiation reaction is disabled. We also noticed that despite having fewer particles for the simulations, the particles cover a wider area when radiation reaction is not enabled.

Actually, Figure~\ref{fig:trap_protons_rr} shows that as $\rchi$ increases, the trapped particles are on average repelled farther away from the surface of the neutron star. 
For $\rchi=60^{\circ}$, all these particles are extremely close to the surface of the neutron star ($r \in [0.105;0.135]$), and it is probable that given more time, these particles would eventually crash onto the surface because of the energy losses . We noticed that the particles formed over densities around $\phi=0$, $\theta=2\pi/3$ and $\phi=\pi$, $\theta=\pi/3$.
For $\rchi=90^{\circ}$, the structure has a spiralling tail of particles that seems to trail towards the surface of the neutron star, while most particles are in a more densely populated region at slightly higher altitudes. Again, these particles are concentrated around $\phi=0$, $\theta=3\pi/4$ and $\phi=\pi$, $\theta=\pi/4$. 

\begin{figure*}[h]
\begin{subfigure}{.5\textwidth}
  \centering
  \includegraphics[width=\textwidth]{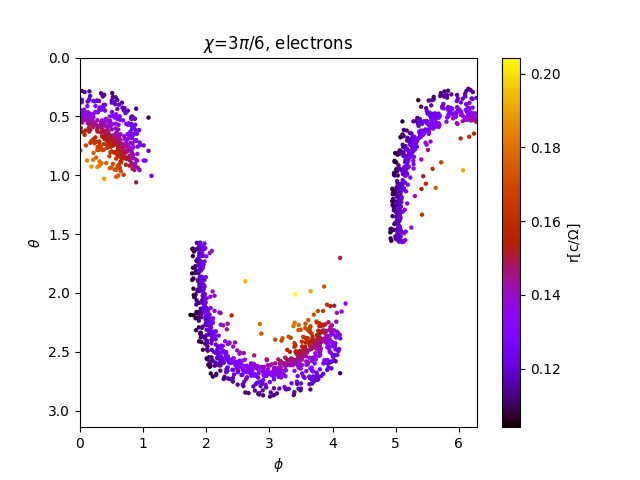}
        \put(-150,110) {\tiny (a)}
\label{fig:demis_duree}
\end{subfigure}
\begin{subfigure}{.5\textwidth}
  \centering
  \includegraphics[width=\textwidth]{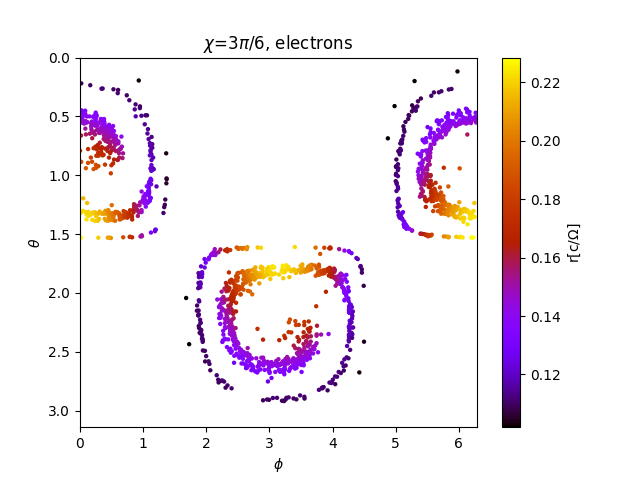}
        \put(-150,110) {\tiny (b)}
\label{fig:double_duree}
\end{subfigure}
\caption{Map of the final colatitude, azimuth, and radius (colour) of electrons around neutron stars with an inclination of $90^{\circ}$. The simulations lasted half (a) and twice (b) the time of other simulations. There were 2048 particles, radiation reaction was enabled, and the magnetic axis is in the $\phi=0$ plane.}
\label{fig:influence_duree}
\end{figure*}

Particles are reputed to be trapped for the time of the simulation. What happens later could not be guessed. However, this final state could depend on the duration of the trapping before being ejected or crashing onto the surface. In order to check the `stability' of the trapping state, we performed new simulations with half the total time and twice the total time of the fiducial run, respectively $t_f/P=7.5$ and $t_f/P=30$. In this further analysis, we injected 2048~electrons and let them evolve in an orthogonal rotator ($\rchi=90\degr$) for half and twice the time of all other simulations. It showed that as time grows, the particles spiral even more, as Figure~\ref{fig:influence_duree} highlights. However, despite their proximity to the neutron star surface, the fraction of particles impacting the surface does not vary. The statistics of the evolution of the particles are as follows: Out of the 2048~particles, 6 impact the surface, 1577 are trapped, and 465 are ejected. We did not notice any significant difference between these runs and concluded that the trapping state lasts for a significant time, at least several neutron star rotation periods, which is long enough to have an impact on the magnetosphere electrodynamics if there is any.

For $\rchi=120^{\circ}$, the structures are still around $\phi=0$ and $\phi=\pi$, but at $\theta=4\pi/5$ and $\theta=\pi/5$ respectively, and around a dense area, one can notice a more diffuse region with fewer particles at low altitude. 
For $\rchi=150^{\circ}$, the structure takes the most different form and corresponds to the part of striped wind below the light cylinder radius. The above discussion for protons also applies to electrons if the values for $\rchi$ are replaced by $\pi - \rchi$.

\begin{figure*}[h]
        \begin{subfigure}{.5\textwidth}
                \centering
                \includegraphics[width=\textwidth]{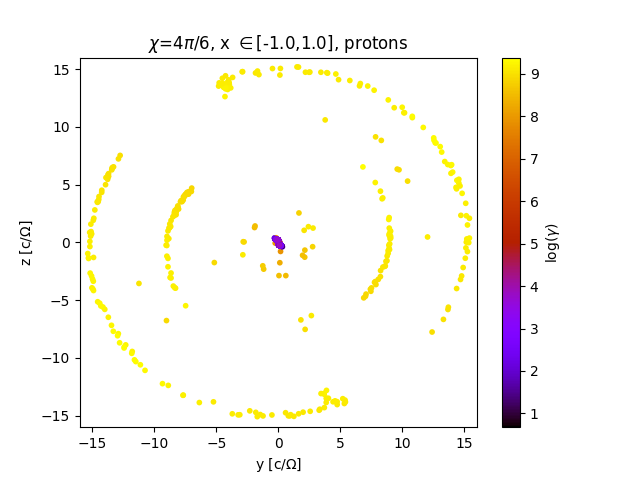}
                \put(-150,110) {\tiny (a)}
        \end{subfigure}
        \begin{subfigure}{.5\textwidth}
                \centering
                \includegraphics[width=\textwidth]{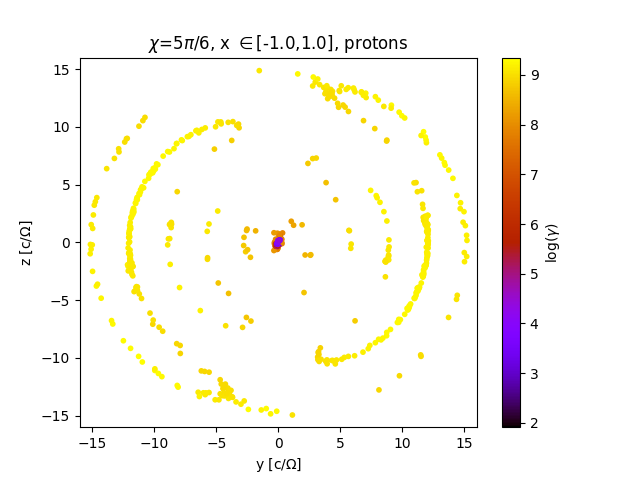}
                \put(-150,110) {\tiny (b)}
        \end{subfigure}
        \begin{subfigure}{.5\textwidth}
                \centering
                \includegraphics[width=\textwidth]{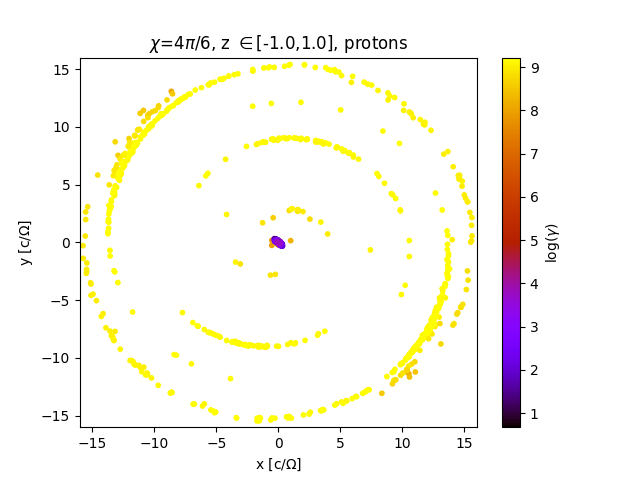}
                \put(-150,110) {\tiny (c)}
        \end{subfigure}
        \begin{subfigure}{.5\textwidth}
                \centering
                \includegraphics[width=\textwidth]{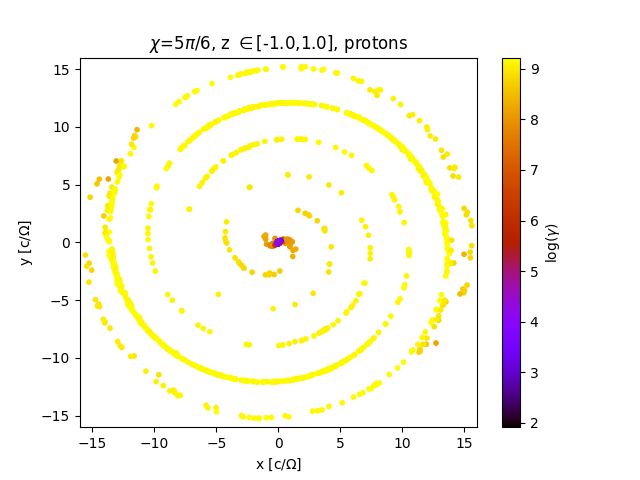}
                \put(-150,110) {\tiny (d)}
        \end{subfigure}
        \caption{Map of the final positions of protons in the $y,z$ plane for $x \in [-1;1]$
                with $\rchi=120^{\circ}$ (a) and $\rchi=150^{\circ}$ (b), and  in the $x,y$ plane for $z \in [-1;1]$ with $\rchi=120^{\circ}$ (c) and $\rchi=150^{\circ}$ (d). Radiation reaction was enabled.}
        \label{fig:winds}
\end{figure*}

\paragraph{Ejected particles} 
Since the striped wind has been mentioned, we checked if the simulations manage to produce such a structure by plotting the final positions of particles outside the light cylinder radius, as done in Figure~\ref{fig:winds}.

The distribution of the particles in space is very reminiscent to the striped wind geometry, especially due to the spiral. However, this wind geometry was retrieved only for $\rchi=120^{\circ}$ and $\rchi=150^{\circ}$ for protons ($\rchi=30^{\circ}$ and $\rchi=60^{\circ}$ for electrons). 
The theoretical equation describing the striped wind is given by \cite{bogovalov_physics_1999} and reads
\begin{equation}
r_s(t,\theta,\phi) = \rlight \, \Big[ \pm \arccos(\cot \theta   \cot \rchi)   + \dfrac{ct}{\rlight}     -\phi + 2 \, \ell \pi  \Big]  
,\end{equation}
where $\ell$ is an integer and $r_s$ the radius of the distance of the particles composing the striped wind.
We observed that the striped wind is found only in the angle $\theta \in [\pi/2 - \rchi ; \pi/2 + \rchi ]$ if $\rchi \leq \pi/2$ and $\theta \in [\pi - \rchi ; \rchi]$ if $\rchi > \pi/2$.

\begin{figure*}[h]
\begin{subfigure}{.5\textwidth}
  \centering
  \includegraphics[width=\textwidth]{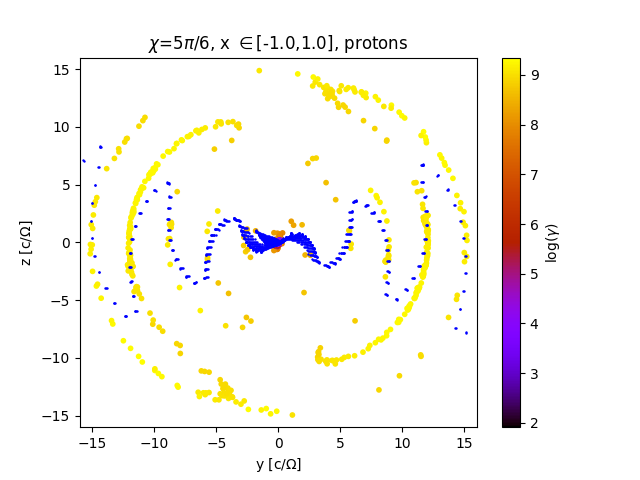}
        \put(-150,110) {\tiny (c)}
\label{fig:winds_x_comparison}
\end{subfigure}
\begin{subfigure}{.5\textwidth}
  \centering
  \includegraphics[width=\textwidth]{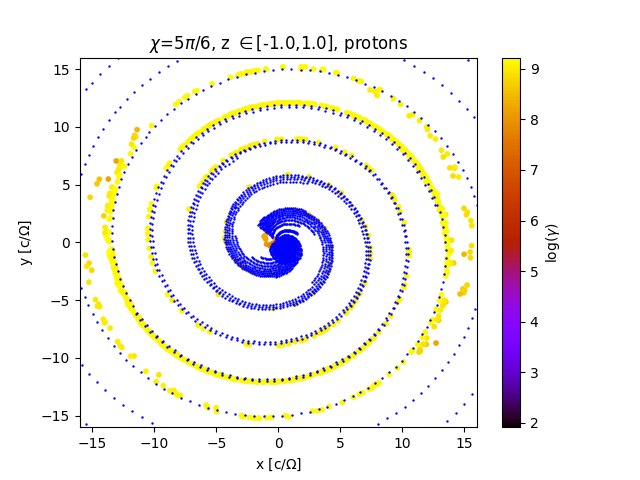}
        \put(-150,110) {\tiny (d)}
\label{fig:winds_z_comparison}
\end{subfigure}
\caption{Comparison of final proton positions to the expected striped winds in the $y,z$ plane (a) and in the $x,y$ plane (b). Radiation reaction was enabled.}
\label{fig:winds_comparison}
\end{figure*}
 
When comparing the proton positions to the expected theoretical wind structure, as in Figure~\ref{fig:winds_comparison}, we found that the spiral fits quite well. However, the opening angle at which the wind is spread is not the one expected. We believe that adding the particle interactions may improve the results by giving more realistic winds.

\begin{figure*}[h]
\begin{subfigure}{.5\textwidth}
  \centering
  \includegraphics[width=\textwidth]{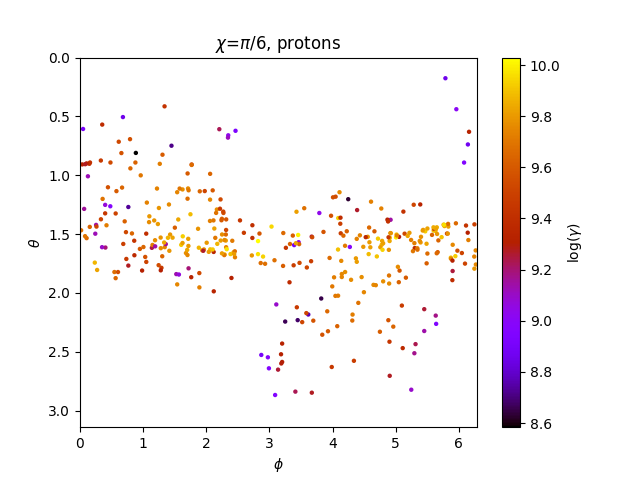}
        \put(-150,110) {\colorbox{white}{\tiny (a)}}
\label{fig:eject_protons_1_no_rr}
\end{subfigure}
\begin{subfigure}{.5\textwidth}
  \centering
  \includegraphics[width=\textwidth]{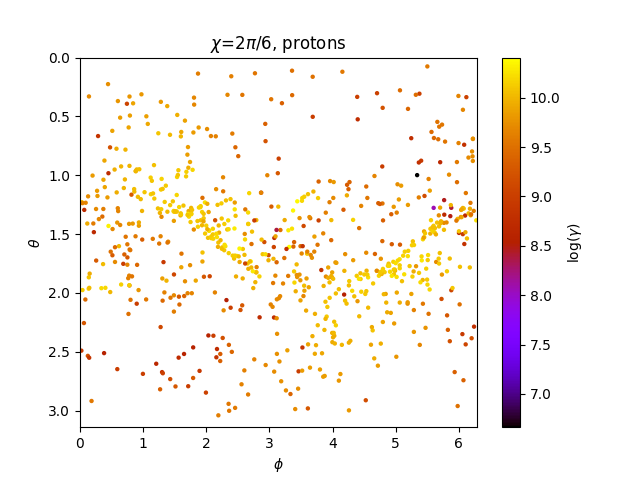}
        \put(-150,110) {\colorbox{white}{\tiny (b)}}
\label{fig:eject_protons_2_no_rr}
\end{subfigure}
\begin{subfigure}{.5\textwidth}
  \centering
  \includegraphics[width=\textwidth]{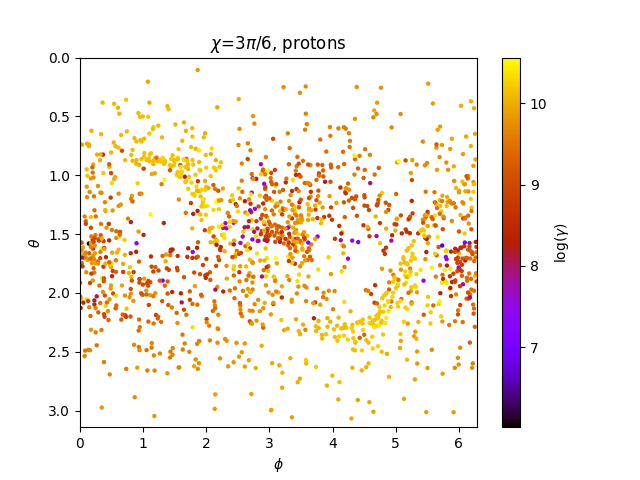}
        \put(-150,110) {\colorbox{white}{\tiny (c)}}
\label{figeject_protons_3_no_rr}
\end{subfigure}
\begin{subfigure}{.5\textwidth}
  \centering
  \includegraphics[width=\textwidth]{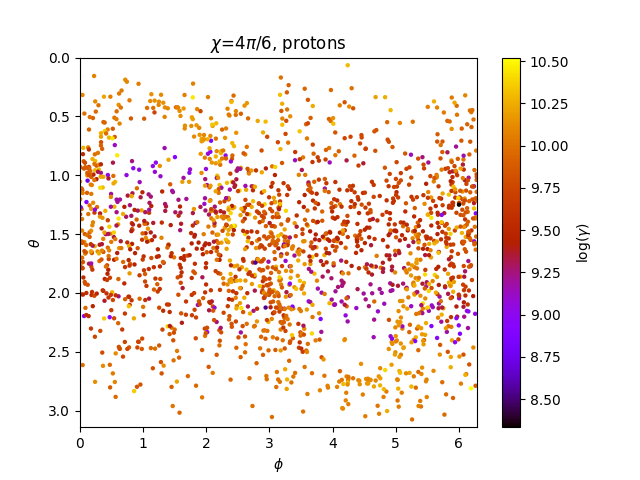}
        \put(-150,110) {\colorbox{white}{\tiny (d)}}
\label{fig:eject_protons_4_no_rr}
\end{subfigure}
\caption{Map of the ejection colatitude and azimuth and Lorentz factor of protons on neutron stars of inclination $30^{\circ}$ (a), $60^{\circ}$ (b), $90^{\circ}$ (c) and $120^{\circ}$ (d), radiation reaction disabled.}
\label{fig:eject_no_rr}
\end{figure*}

Upon looking at the final colatitude and azimuth of protons in Figure~\ref{fig:eject_protons_rr},
we noticed that the most energetic particles prefer some directions but also that a given direction of ejection can be more densely populated or, conversely, almost void of particles. Comparing Figure~\ref{fig:eject_protons_rr} to Figure~\ref{fig:eject_no_rr}, we noticed some similarities regarding the more or less densely populated directions of ejection. However, the positions of the particles and their Lorentz factors are not the same any more than when radiation reaction is enabled.

Taking a look at Figures~\ref{fig:demarrage_devenir}, \ref{fig:crash_protons_rr}, \ref{fig:trap_protons_rr}, \ref{fig:eject_protons_rr}, \ref{fig:demarrage_Lorentz}, and \ref{fig:winds}, we found a symmetrical behaviour for particles injected on one side or the opposite side of the neutron star. The centre of the neutron star is a point of symmetry for the electric field and a point of anti-symmetry for the magnetic field, meaning that $\textbf{E}(x,y,z,t)=-\textbf{E}(-x,-y,-z,t)$ and $\textbf{B}(x,y,z,t)=\textbf{B}(-x,-y,-z,t)$. This property led us to find that particles injected symmetrically relative to the centre of the neutron star have symmetrical trajectories. 

We considered two particles injected at $\textbf{x}=(x,y,z)$ and $\textbf{x}'=(-x,-y,-z)=-\textbf{x}$ with speeds of $\textbf{v}(t=0)=-\textbf{v}'(t=0)$. 
We first find the force acting on these particles at the initial time step: the Lorentz force and the radiation reaction (here in classical formulation), which gives the following for the first particle:
\begin{equation}
\begin{aligned}
\textbf{F} &=q(\textbf{E}(x,y,z)+\textbf{v}\times\textbf{B}(x,y,z)) + \dfrac{\mu_0 q^2}{6 \pi c} \dfrac{d^3 \textbf{x}}{d t^3}  
\end{aligned}
,\end{equation}
 and the following for the second particle:
\begin{equation}
\begin{aligned}
\textbf{F}' &=q(\textbf{E}(-x,-y,-z)+\textbf{v}'\times\textbf{B}(-x,-y,-z)) + \dfrac{\mu_0 q^2}{6 \pi c} \dfrac{d^3 \textbf{x}'}{d t^3}  \\
&= -q(\textbf{E}(x,y,z)+\textbf{v}\times\textbf{B}(x,y,z)) - \dfrac{\mu_0 q^2}{6 \pi c} \dfrac{d^3 \textbf{x}}{d t^3}  .
\end{aligned}
\end{equation}
Since the Lorentz force acting on the first particle $\textbf{F}_l$ is the opposite of that acting on the second particle $\textbf{F}'_l=-\textbf{F}_l$, we found that $\dfrac{d^3 \textbf{x}}{d t^3} = -\dfrac{d^3 \textbf{x}'}{d t^3}$, meaning in the end that $\textbf{F}=-\textbf{F}'$.
When we integrated the force, we found that the particles have symmetrical speeds relative to the centre of the neutron star: $\textbf{v}=-\textbf{v}'$, and if we integrated this speed, we found that $\textbf{x}'=-\textbf{x}$ holds regardless of the time of integration.

\subsection{Lorentz factor distribution function}

In addition to the spatial particle distribution, in order to better understand the effects of radiation reaction on their dynamics, we compared our results to those previously obtained by \cite{tomczak_particle_2020}, knowing that simulations with the radiation reaction enabled yield more realistic results. 
One way to find a conservative upper limit to the Lorentz factor reached by particles around neutron star has been given in the introduction (Section \ref{sec:Intro}). Taking the potential drop estimate $\Delta \Phi = \Omega B R^2 = 10^{16}$~V and multiplying it by $e/m c^2$, we found the Lorentz factor $\gamma_e=10^{10}$ for electrons and $\gamma_p=10^{6.7}$ for protons.

\begin{figure}
        \centering
        \includegraphics[width=\linewidth]{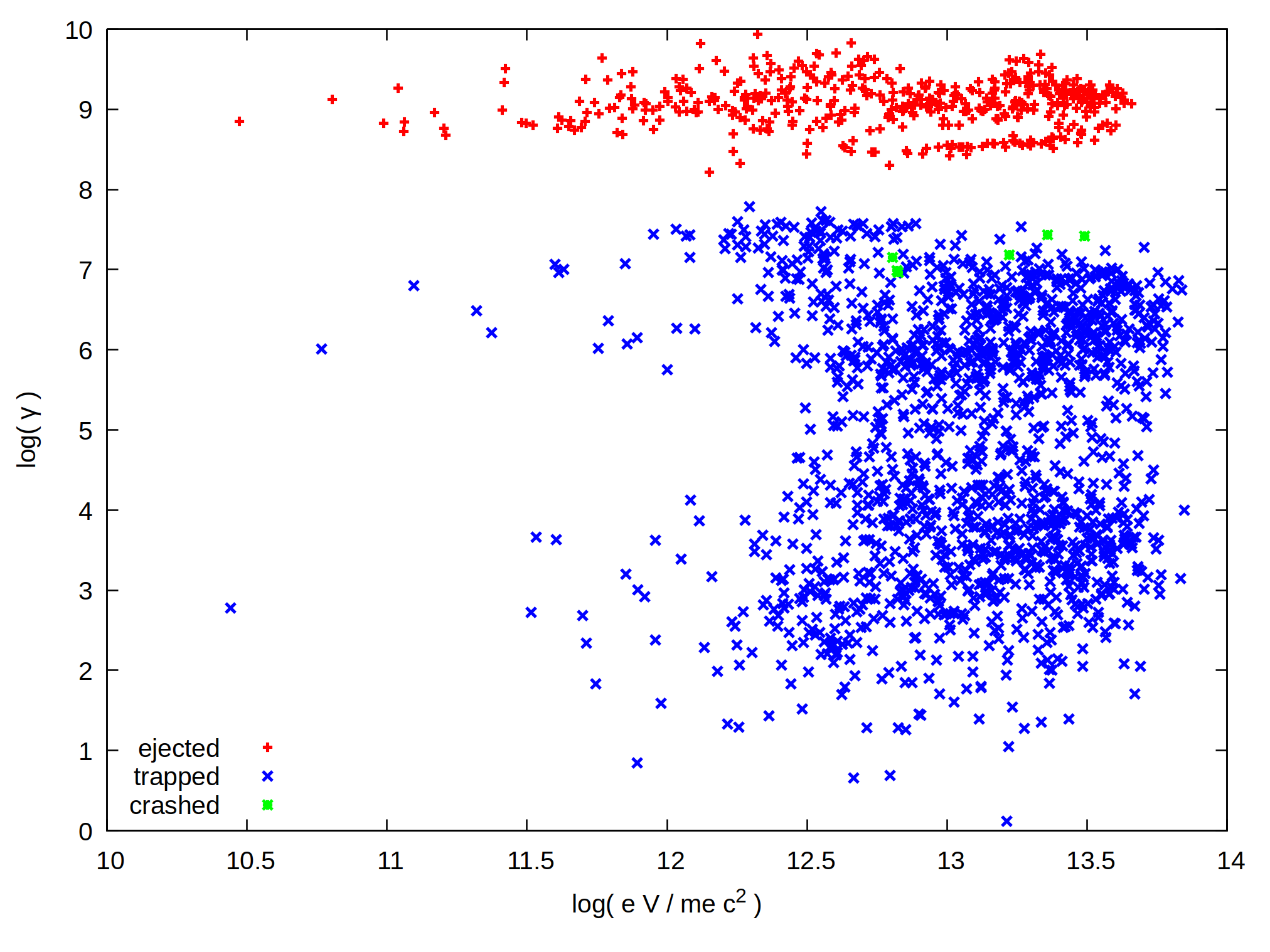}
        \caption{Final Lorentz factor of electrons as a function of the potential drop (normalised to $e\,V/m_e\,c^2$) along their path. 2048~particles were simulated, in green crosses for crashed particles, in blue crosses for trapped particles and in red plus symbols for ejected particles. The magnetic field is for an orthogonal rotator~$\rchi=90^{\circ}$. }
        \label{fig:chute_potentiel}
\end{figure}

\begin{figure}
        \centering
        \includegraphics[width=\linewidth]{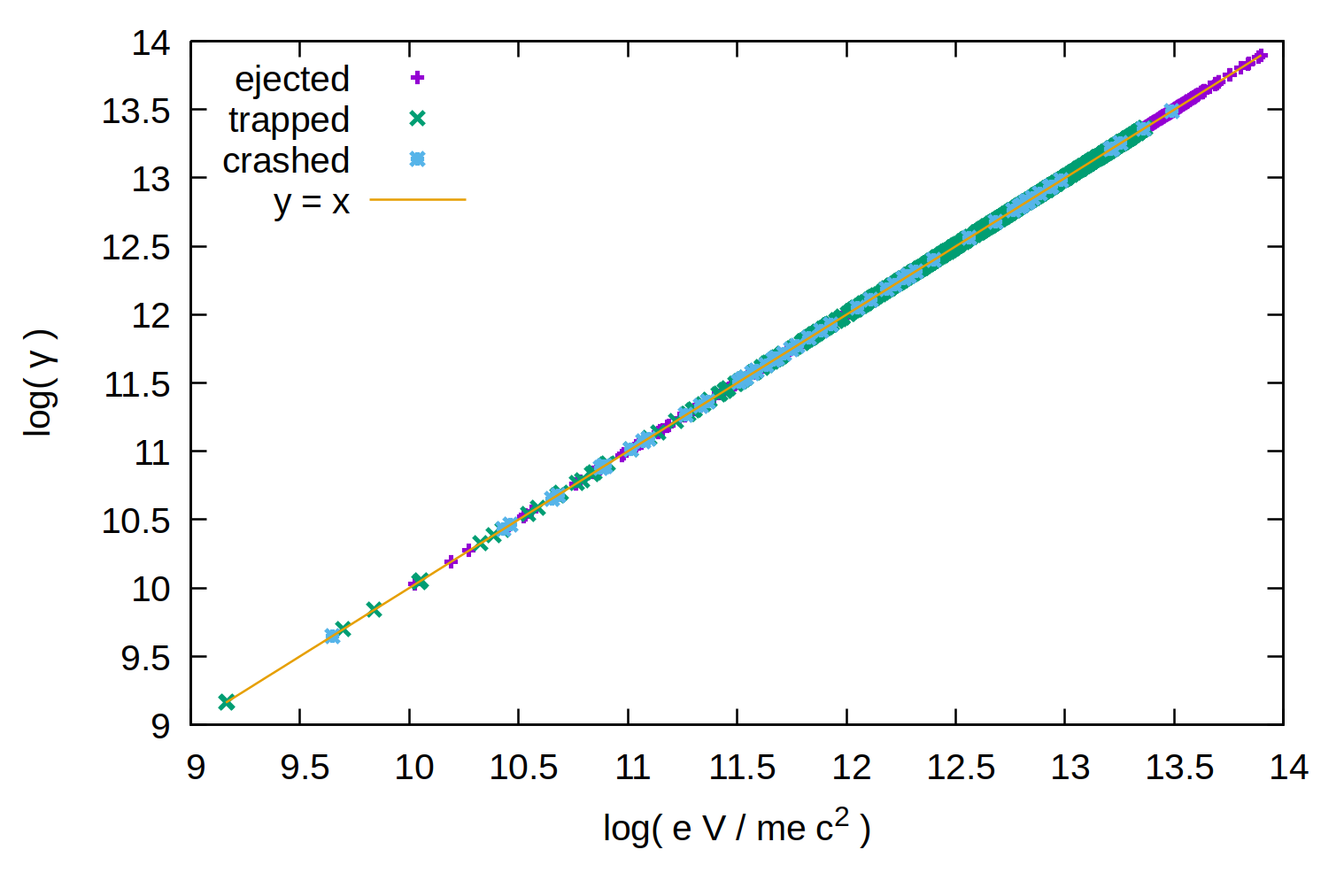}
        \caption{Same as Fig.\ref{fig:chute_potentiel} but without radiation reaction. The line $y=x$ in orange depicts the Lorentz factor if the full potential drop is used. The correlation is clearly visible for all particles.}
        \label{fig:chute_potentiel_no_rr}
\end{figure}

This estimate is however not accurate because it assumes a constant static electric field and no radiation reaction. This limit is therefore rather conservative. Our simulations are very different because the electric field varies in time and radiation reaction is taken into account. When looking for a correlation between the potential drop and the final Lorentz factor reached by the electrons in an orthogonal rotator, we obtained the plot shown in Figure~\ref{fig:chute_potentiel}, where the Lorentz factor is shown against the potential along the particle trajectory in log-log scale. The populations of crashed, trapped, and ejected particles are shown with green, blue, and red symbols, respectively. The number of crashed particles, less than ten, shown in green symbols, is too limited to perform any significant statistical analysis. For the two other populations, we found no evidence for a correlation between the final Lorentz factor of the particle and the potential drop, as Pearson's correlation coefficient is $r=0.055$ for trapped electrons and $r=-0.017$ for ejected electrons. This demonstrates that the Lorentz factor does not significantly depend on the particle motion history but is rather controlled by the local conditions (i.e. parallel accelerating electric field and curvature radius).
Without radiation reaction, the correlation is very strong. We found an excellent agreement between the potential drop and the final Lorentz factor, as shown in Fig.~\ref{fig:chute_potentiel_no_rr}, for all three kinds of particles: trapped, ejected, and crashed.

\begin{figure}
        \centering
        \includegraphics[width=\linewidth]{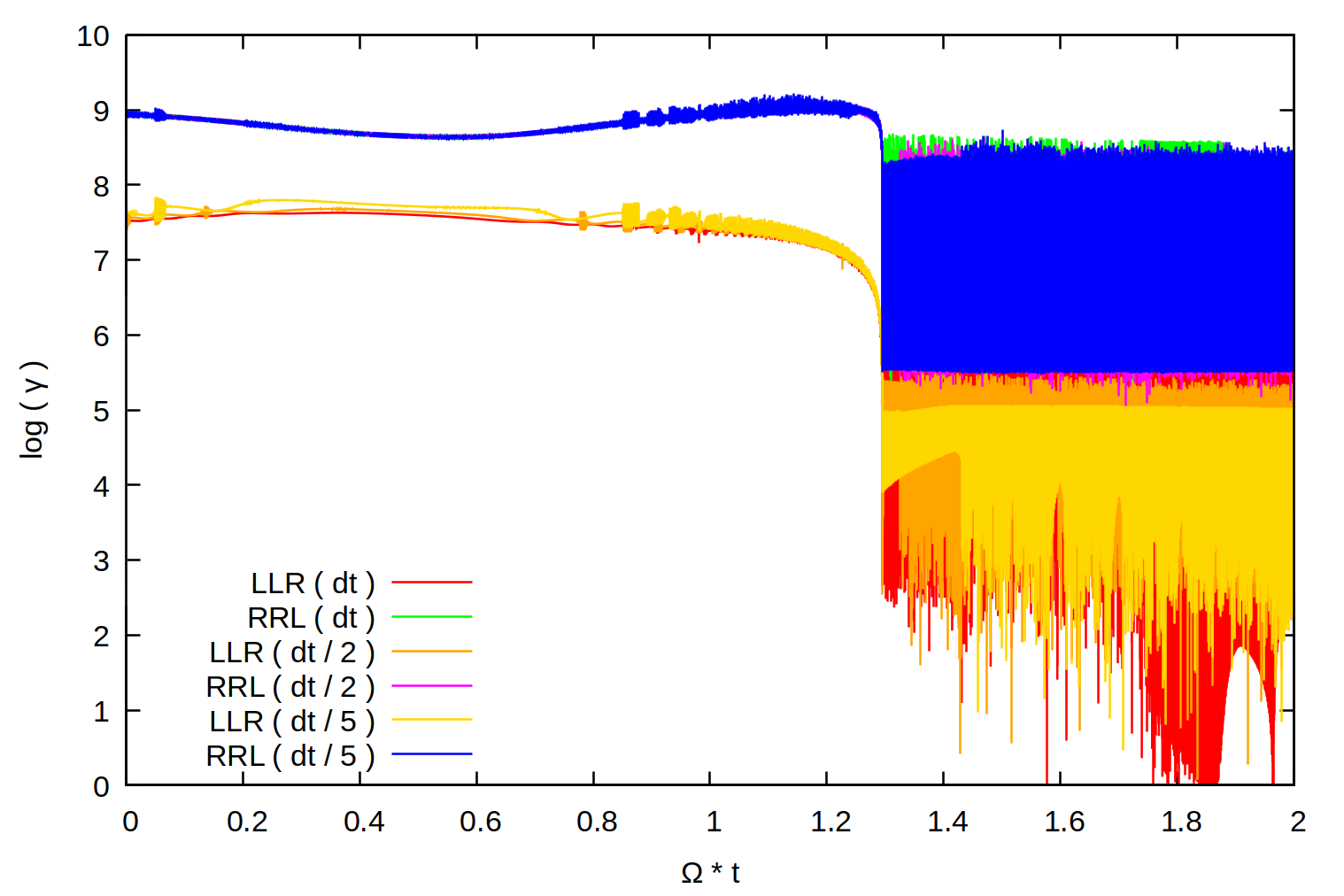}
        \caption{Lorentz factor from the Landau-Lifshitz approximation (LLR), in red and orange colours for different time steps, compared to the radiation reaction limit (RRL) guess as given by \eqref{eq:gamma_curvature}, in green and blue colours for different time steps.}
        \label{fig:comparaisongammarrlvsllr}
\end{figure}

Another way to estimate the true Lorentz factor consists of equating the power of the electric force and that of the energy losses due to the curvature radiation. Thus, we get:
\begin{equation}
q \, \textbf{E} \cdot \textbf{v} = \dfrac{q^2}{6 \, \pi \,\epsilon_0} \, \gamma^4 \dfrac{c}{\rho^2}, 
\end{equation} 
where $\rho=\sqrt{R \, \rlight}$ is a typical radius of the curvature of the trajectory of the particle (here, that of a field line close to the stellar surface); $\textbf{v}$ is the speed of the particle ($\sim c$); and $\epsilon_0$ is the vacuum permittivity. By solving for the Lorentz factor, we found that 
\begin{equation}\label{eq:gamma_curvature}
        \gamma = \left( \dfrac{6\, \pi\, \epsilon_0 \, E \, \rho^2 }{q} \right)^{1/4} = \left( \frac{q\,E_\parallel}{m\,c\,\Omega} \, \frac{\rlight}{r_e} \, \tilde{\rho}^2 \right)^{1/4}.
\end{equation}
This last expression uses quantities without dimensions, such as $\tilde{\rho} = \rho/\rlight$, and the electric strength parameter, with $r_e$ being the electron classical radius and $E_\parallel = \vec{\beta} \cdot \vec{E}$ as the accelerating electric field.
In our case, we applied the expression to millisecond pulsars and got $\gamma=10^{7.5}$, which is only a guess because the curvature radius can be very different from that on the stellar surface. The true curvature~$\kappa$ is found from the velocity vector derivative such that
\begin{equation}\label{key}
 \kappa = \frac{1}{\rho} = \left\| \frac{d\vec{v}}{c^2\,dt} \right\| .
\end{equation}
This expression accurately captures the local curvature radius~$\rho$ along the trajectory. Therefore, by following the Lorentz factor from the Landau-Lifshitz approximation and comparing it to the radiation reaction limit estimate as given by Eq.\eqref{eq:gamma_curvature}, we show that the latter always finds higher Lorentz factors, see Fig.\ref{fig:comparaisongammarrlvsllr}. To check that the results converged, several different time step integration parameters were used, two times as well as five times smaller without noticeable changes. Thus, our results have converged and are robust.

\begin{figure}
        \centering
        \includegraphics[width=\linewidth]{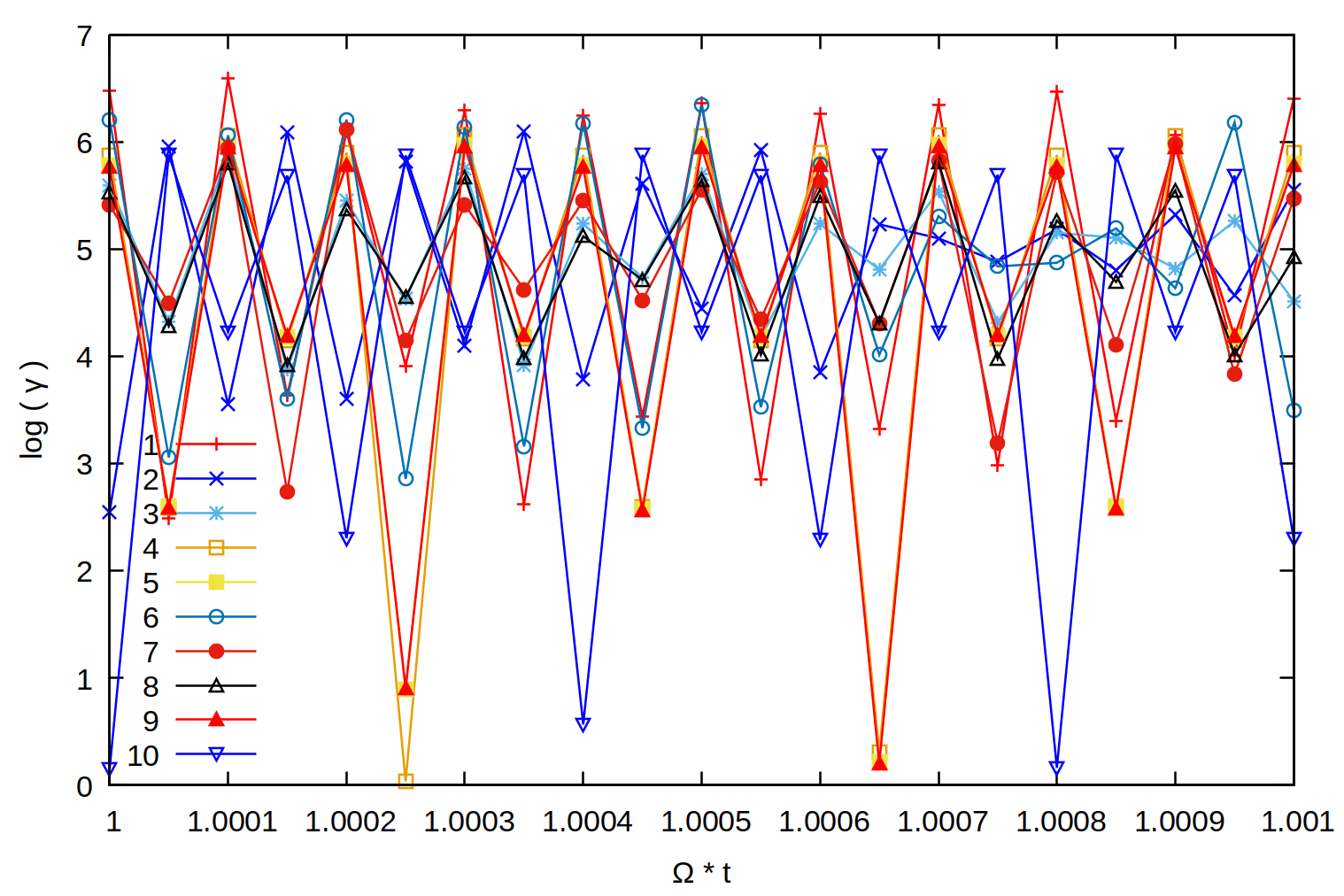}
        \caption{Fluctuation in the Lorentz factor for a sample of ten particles zoomed around the time $\Omega\,t=1$.}
        \label{fig:stroboscopique}
\end{figure}

We note that the Lorentz factors of trapped particles span a large range from almost rest $\gamma \sim 10$ to $\gamma \lesssim 10^8$. This does not necessarily mean that they always experience strong radiation damping. Indeed, the Lorentz factor variation is two-fold in this electromagnetic field environment. First, the radiation reaction decelerates the particles from a very high Lorentz factor of $\gamma \sim 10^{12-13}$ to $\gamma \sim 10^{7-8}$, decreasing by several orders of magnitude their initial Lorentz factor. Second, at a low to moderate Lorentz factor, the electric field component~$E_\parallel$ parallel to the magnetic field can accelerate but can also decelerate particles depending on the sign of $E_\parallel$. Thus, moderate energies do not necessarily mean strong radiation reaction but rather efficient electric field deceleration. Fluctuation was observed in the Lorentz factor on short timescales for trapped particles, as seen in Fig.~\ref{fig:comparaisongammarrlvsllr}. Depending on the final time, we picked out a $\gamma$ factor value between the minimum and maximum of the possible interval $\gamma \sim [10, 10^{7-8}]$. This fluctuation is a kind of stroboscopic effect, giving a sample of the Lorentz factor spreading at this interval. We checked this effect by looking at a sample of ten trapped particles and found that the Lorentz factor drastically fluctuates on very short timescales in this energy range, see Fig.\ref{fig:stroboscopique}.

\begin{figure*}[h]
\begin{subfigure}{.5\textwidth}
  \centering
  \includegraphics[width=\textwidth]{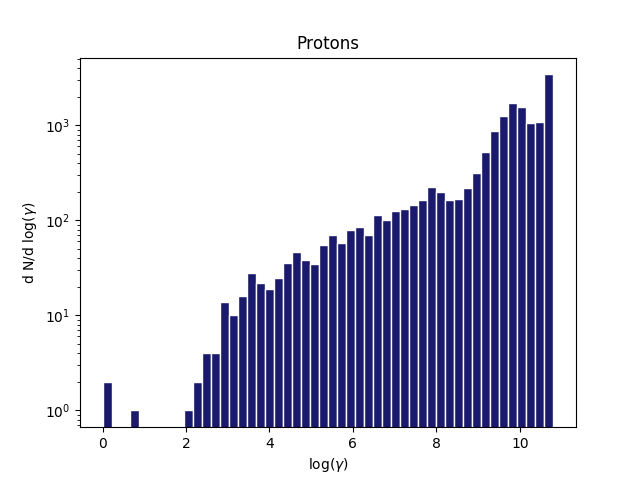}
        \put(-150,110) {\tiny (a)}
\end{subfigure}
\begin{subfigure}{.5\textwidth}
  \centering
  \includegraphics[width=\textwidth]{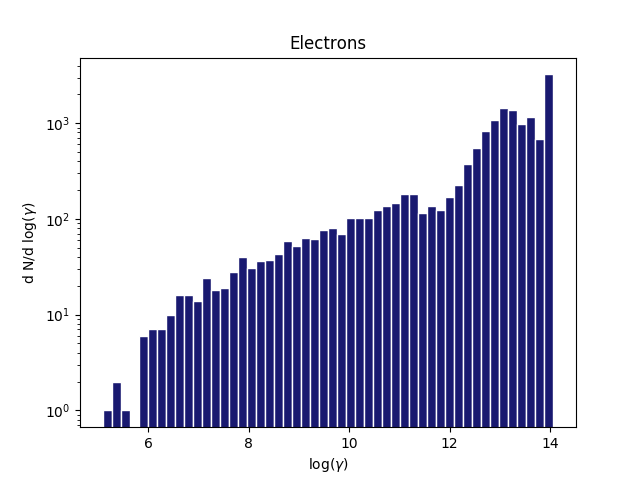}
        \put(-150,110) {\tiny (b)}
\end{subfigure}
\begin{subfigure}{.5\textwidth}
  \centering
  \includegraphics[width=\textwidth]{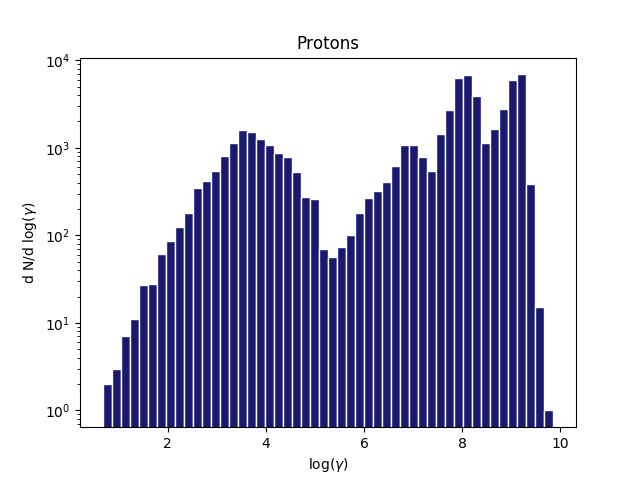}
        \put(-150,110) {\tiny (c)}
\end{subfigure}
\begin{subfigure}{.5\textwidth}
  \centering
  \includegraphics[width=\textwidth]{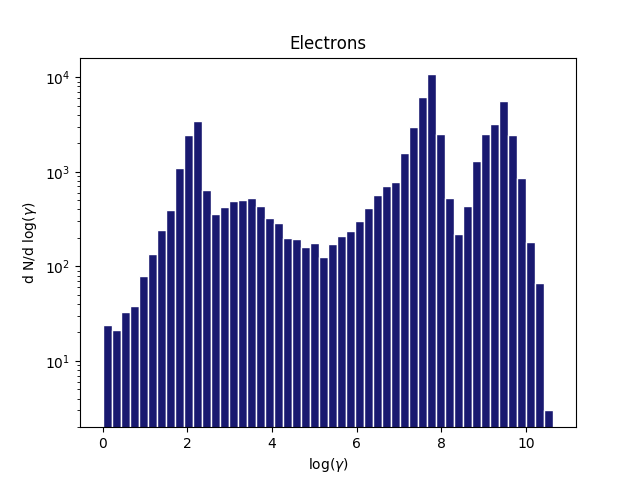}
        \put(-150,110) {\tiny (d)}
\end{subfigure}
\caption{Comparison of the Lorentz factor distributions for simulations carried out with radiation reaction disabled for protons (a) and electrons (b), and with radiation reaction enabled for protons (c) and electrons (d). The number of particles is $14,336$ in (a) and (b) and $229,376$ in (c) and (d).}
\label{fig:Comparison_rr}
\end{figure*}

As shown in Figure \ref{fig:Comparison_rr}, protons reach $\gamma\simeq10^{11}$ and electrons reach $\gamma\simeq10^{14}$ without radiation reaction, while with radiation reaction, they respectively reach $\gamma\simeq10^{10}$ and $\gamma\simeq10^{10.5}$,
which is quite different from the curvature radiation approximation. We note, however, that the particles reaching the highest energies follow the field lines with the largest curvature radii, while for our calculation, we took an averaged radius of the curvature. 

\begin{figure*}[h]
\begin{subfigure}{.5\textwidth}
  \centering
  \includegraphics[width=\textwidth]{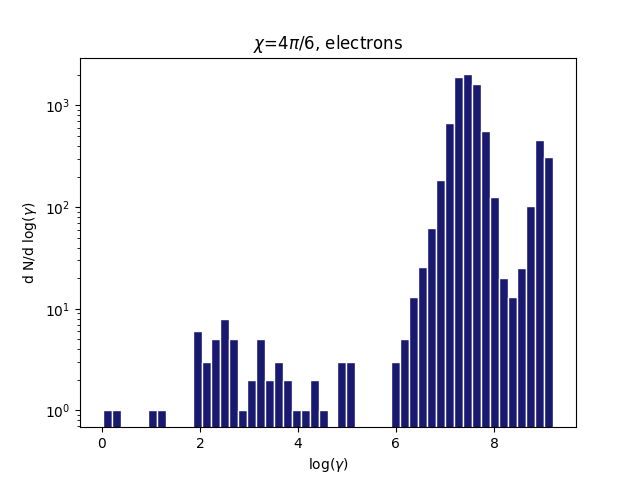}
        \put(-150,110) {\tiny (a)}
\end{subfigure}
\begin{subfigure}{.5\textwidth}
  \centering
  \includegraphics[width=\textwidth]{spect_electron_B5_CHI4sur6.png}
        \put(-150,110) {\tiny (b)}
\end{subfigure}
\begin{subfigure}{.5\textwidth}
  \centering
  \includegraphics[width=\textwidth]{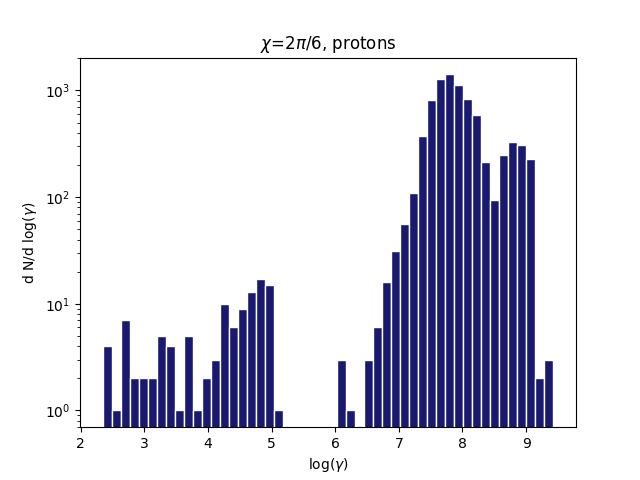}
        \put(-150,110) {\tiny (c)}
\end{subfigure}
\begin{subfigure}{.5\textwidth}
  \centering
  \includegraphics[width=\textwidth]{spect_proton_B5_CHI2sur6.png}
        \put(-150,110) {\tiny (d)}
\end{subfigure}
\caption{Comparison of the Lorentz factor distributions for electrons near a pulsar with an inclination $\rchi=120^{\circ}$, without (a) and with (b) radiation reaction enabled, and for protons near a pulsar with an inclination $\rchi=60^{\circ}$, without (c) and with (d) radiation reaction enabled.}
\label{fig:Comparison_masses}
\end{figure*}

When comparing the influence of the particle species, we noticed that particles with a high mass are less affected than those with a low mass for a given charge. Indeed, as Figure \ref{fig:Comparison_masses} shows, the loss of energy is higher for electrons than for protons since the highest energy electrons lost $\sim 4.8$ orders of magnitude for their Lorentz factor, whereas the highest energy protons lost only approximately  one order of magnitude. Since the energy of a particle is $E=\gamma m c^2$, we found that without radiation reaction, the proton energy is about
$E_p = 10^{10.5} \times m_p \, c^2 = 4,75$~J and the electron energy $E_e= 10^{13.8} \times m_e \, c^2 = 5,17$~J.  With radiation reaction enabled, we got $E_p= 10^{9.5} \times m_p \, c^2 = 0.48$~J and $E_e= 10^{10} \times m_e \,  c^2 = 0.000082$~J.
This means that even if the particles had approximately the same energy without radiation reaction, the fastest electrons have only $0.017\%$ of the energy of the fastest protons after radiation reaction has been enabled. 
We nonetheless note that with radiation reaction, the Lorentz factor distribution for protons near a pulsar of inclination $\rchi$ is similar in shape to that of electrons near a pulsar of inclination $\pi-\rchi$. Two modes, one at low energy with low statistics and another one at higher energy with up to $N \sim 10000$ at the peak for the cases, are shown in Figure \ref{fig:Comparison_masses}. The main difference is the positions of the extrema. For protons near a pulsar with an inclination of $60^{\circ}$, the low energy peak is at $\gamma=10^{7.5}$ and the high energy one is at $\gamma=10^{8.9}$, while for electrons near a pulsar with an inclination of $120^{\circ}$, the low energy and high energy peaks are reached at $\gamma=10^{7.8}$ and $\gamma=10^{8.8}$, respectively.

\paragraph{Crashed particles.}

\begin{figure*}[h]
\begin{subfigure}{.5\textwidth}
  \centering
  \includegraphics[width=\textwidth]{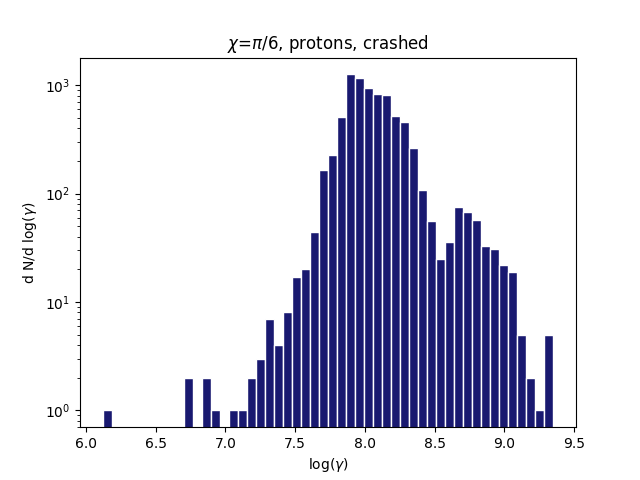}
        \put(-150,110) {\tiny (a)}
\end{subfigure}
\begin{subfigure}{.5\textwidth}
  \centering
  \includegraphics[width=\textwidth]{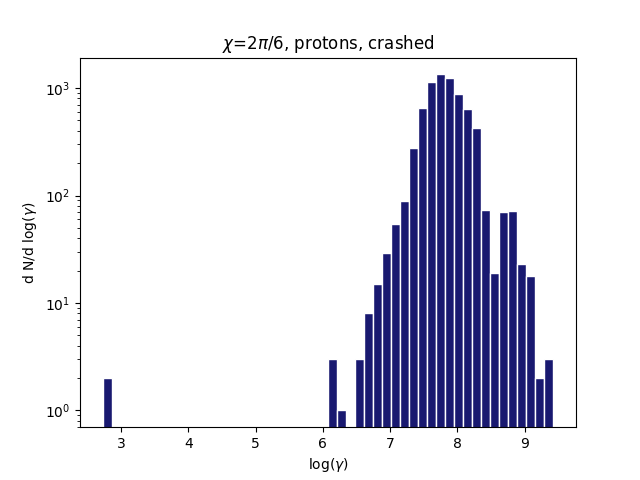}
        \put(-150,110) {\tiny (b)}
\end{subfigure}
\begin{subfigure}{.5\textwidth}
  \centering
  \includegraphics[width=\textwidth]{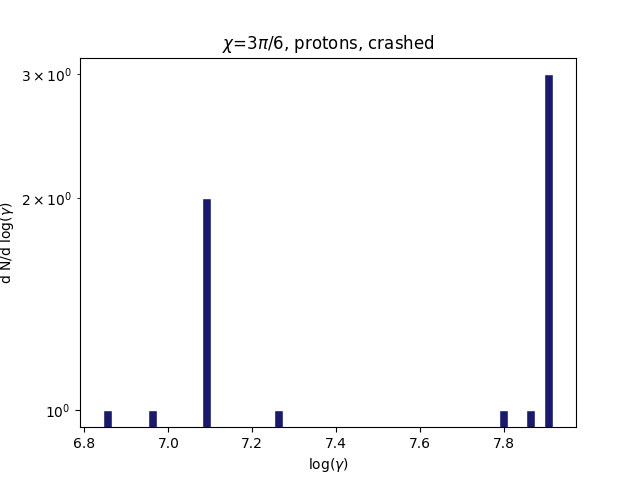}
        \put(-150,110) {\tiny (c)}
\end{subfigure}
\begin{subfigure}{.5\textwidth}
  \centering
  \includegraphics[width=\textwidth]{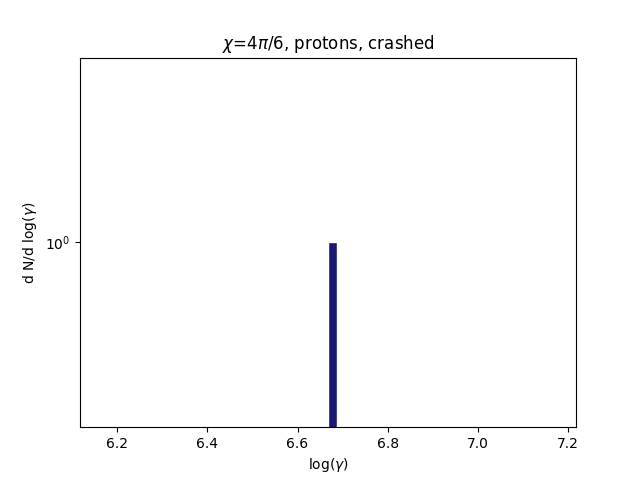}
        \put(-150,110) {\tiny (d)}
\end{subfigure}
\caption{Lorentz factor distribution of protons impacting neutron stars with an inclination $\rchi=30^{\circ}$ (a), $\rchi=60^{\circ}$ (b), $\rchi=90^{\circ}$ (c), and $\rchi=120^{\circ}$ (d). Radiation reaction was enabled.}
\label{fig:crash_spectres_protons}
\end{figure*}

Figure~\ref{fig:crash_spectres_protons} presents a closer look at particles impacting the neutron stars.
Since the number of protons impacting the surface is lower than ten when $\rchi \geq 90^{\circ}$, we could only interpret the Lorentz factor distribution for $\rchi=30^{\circ}$ and $\rchi= 60^{\circ}$. For both inclinations, the protons reach at most $\gamma=10^{9.4}$, the peak of the distributions is $\gamma=10^{8}$, and the local maximum is at $\gamma=10^{8.7}$. The main difference between these spectral distributions is at low energy. For $\rchi=60^{\circ}$, the distribution starts at $\gamma=10^{6.5}$, while for $\rchi=30^{\circ}$, the distribution starts at $\gamma=10^{7}$. 
Even if the statistics are too low, protons reach at most $\gamma=10^{7.8}$ and at least $\gamma=10^{6.8}$ for $\rchi=90^{\circ}$. However, for this case as well as the $\rchi=150^{\circ}$ case, it is certain that if given more particles, the overall shape of the distributions would drastically change, according to the initial positions.

\begin{figure*}[h]
\begin{subfigure}{.5\textwidth}
  \centering
  \includegraphics[width=\textwidth]{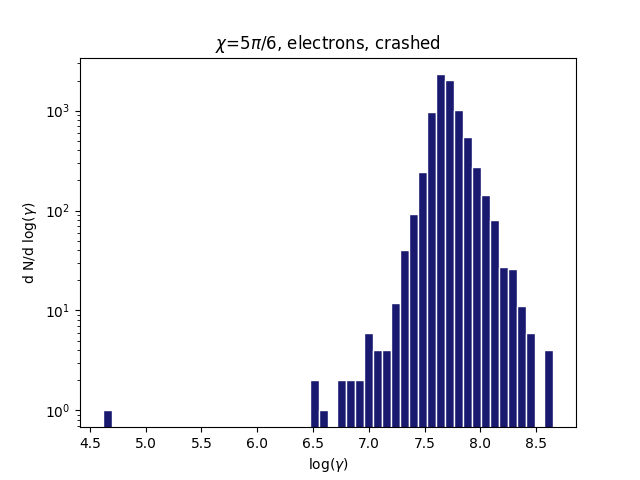}
        \put(-150,110) {\tiny (a)}
\end{subfigure}
\begin{subfigure}{.5\textwidth}
  \centering
  \includegraphics[width=\textwidth]{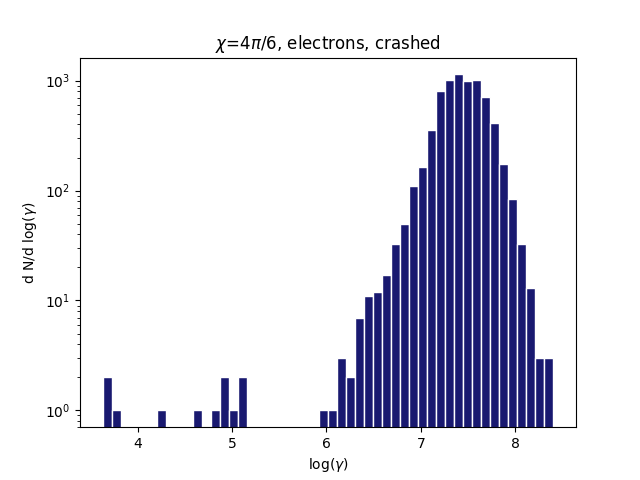}
        \put(-150,110) {\tiny (b)}
\end{subfigure}
\begin{subfigure}{.5\textwidth}
  \centering
  \includegraphics[width=\textwidth]{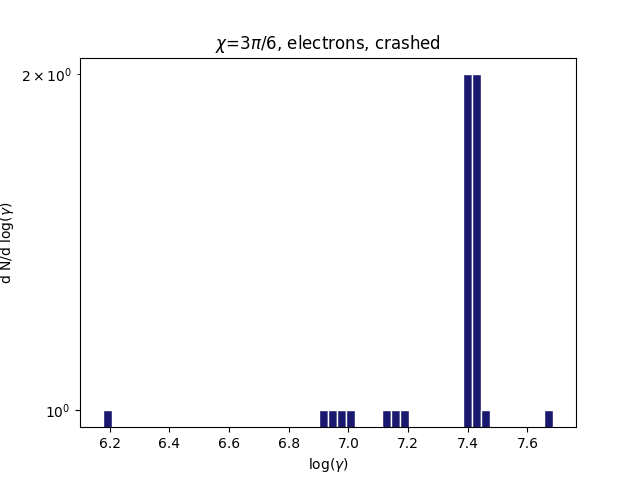}
        \put(-150,110) {\tiny (c)}
\end{subfigure}
\begin{subfigure}{.5\textwidth}
  \centering
  \includegraphics[width=\textwidth]{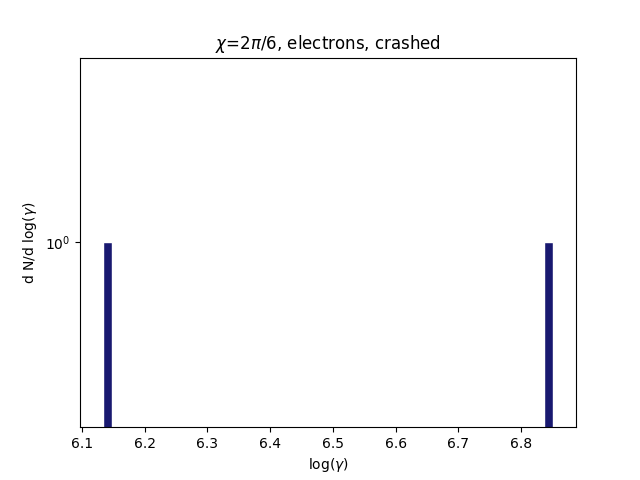}
        \put(-150,110) {\tiny (d)}
\end{subfigure}
\caption{Lorentz factor distribution of protons impacting neutron stars with an inclination $\rchi=150^{\circ}$ (a), $\rchi=120^{\circ}$ (b), $\rchi=90^{\circ}$ (c), and $\rchi=60^{\circ}$ (d). Radiation reaction was enabled.}
\label{fig:crash_spectres_electrons}
\end{figure*}

When taking a look at Figure~\ref{fig:crash_spectres_electrons}, we found that the overall shape of the spectra is slightly different for protons and electrons. In particular, the electrons do not have the high energy local maximum. The values of the Lorentz factors reached by electrons are lower than those reached by protons. For $\rchi=30^{\circ}$ and  $\rchi=60^{\circ}$, the peak of the Lorentz factor distribution is reached at $\gamma=10^{7.8},$ while for $\rchi=150^{\circ}$ and $\rchi=120^{\circ}$, the distributions of electrons peak at $\gamma=10^{7.7}$ and $\gamma=10^{7.5}$, respectively. 

\paragraph{Trapped particles.}

\begin{figure*}[h]
\begin{subfigure}{.5\textwidth}
  \centering
  \includegraphics[width=\textwidth]{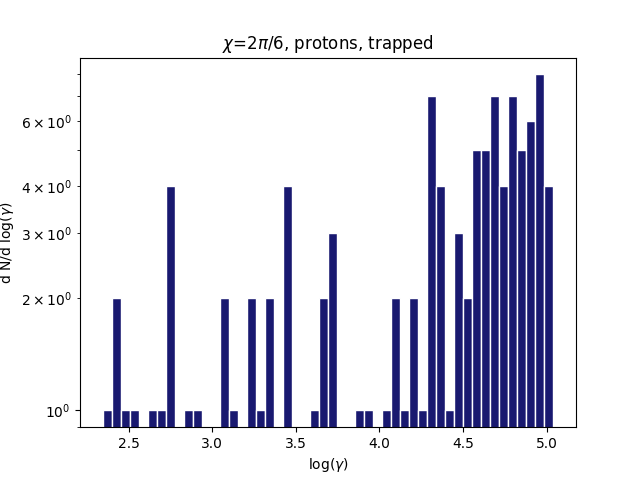}
        \put(-150,110) {\tiny (a)}
\end{subfigure}
\begin{subfigure}{.5\textwidth}
  \centering
  \includegraphics[width=\textwidth]{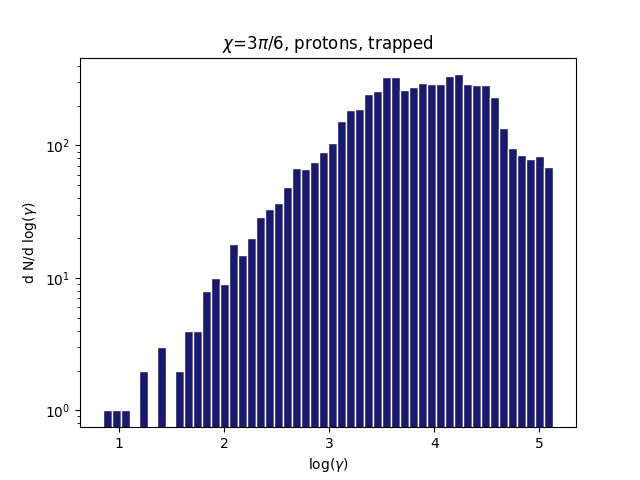}
        \put(-150,110) {\tiny (b)}
\end{subfigure}
\begin{subfigure}{.5\textwidth}
  \centering
  \includegraphics[width=\textwidth]{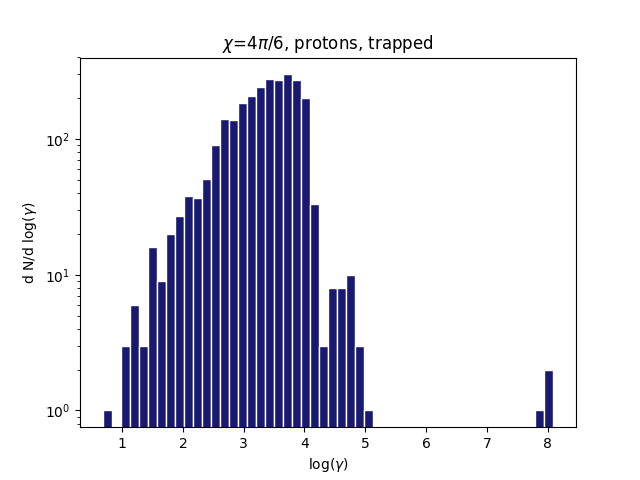}
        \put(-150,110) {\tiny (c)}
\end{subfigure}
\begin{subfigure}{.5\textwidth}
  \centering
  \includegraphics[width=\textwidth]{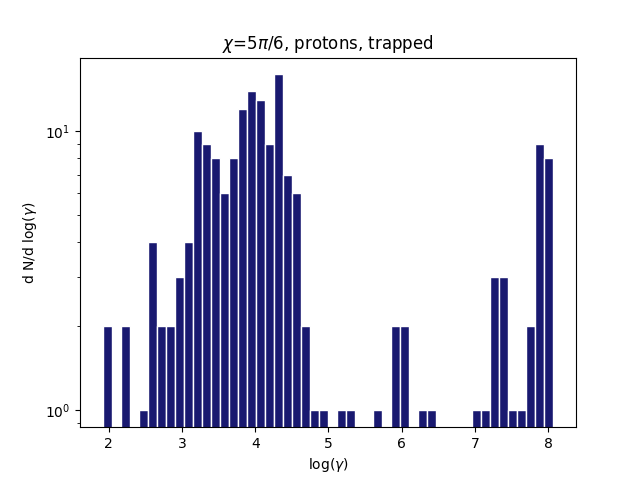}
        \put(-150,110) {\tiny (d)}
\end{subfigure}
\caption{Lorentz factor distribution of protons trapped around neutron stars with an inclination of $\rchi=60^{\circ}$ (a), $\rchi=90^{\circ}$ (b), $\rchi=120^{\circ}$ (c), and $\rchi=150^{\circ}$ (d). Radiation reaction was enabled.}
\label{fig:trap_spectres}
\end{figure*}

The distribution of the Lorentz factors of protons trapped around neutron stars is shown in Figure~\ref{fig:trap_spectres}. It always produces a mode ending at $\gamma=10^5$, but for $\rchi=120^{\circ},$ a few protons reached $\gamma=10^8$, and for $\rchi=150^{\circ}$, more particles were distributed between $\gamma=10^5$ and $\gamma=10^8$. For $\rchi=60^{\circ}$, the total number of particles is quite low and analysing the distribution function becomes problematic. 
With $\rchi=90^{\circ}$, the maximum of the distribution is a plateau between $\gamma=10^{3.5}$ and $\gamma=10^{4.5}$, with $N=300$ particles per bin. The distribution begins with a power law with a slope of approximately one between $\gamma=10$ and $\gamma=10^{3.5}$. 
With $\rchi=120^{\circ}$, the distribution also starts at $\gamma=10$, but the maximum is located at $\gamma=10^{3.9}$, with $N \sim 300$, and a local maximum is reached at $\gamma=10^{4.7}$, with $N=10$ particles per bin. 
Finally, with $\rchi=150^{\circ}$, most bins contain between $N=1$ and $N=20$ particles, meaning that the distribution is highly sensitive to noise. The number of trapped particles becomes very low in this configuration. However, protons with an energy of about $\gamma=10^8$ reach up to $N=10$ particles per bin, meaning that this part of the distribution is not simply a random event and that it  may become significant with a higher number of simulated particles.

\begin{figure*}[h]
\begin{subfigure}{.5\textwidth}
  \centering
  \includegraphics[width=\textwidth]{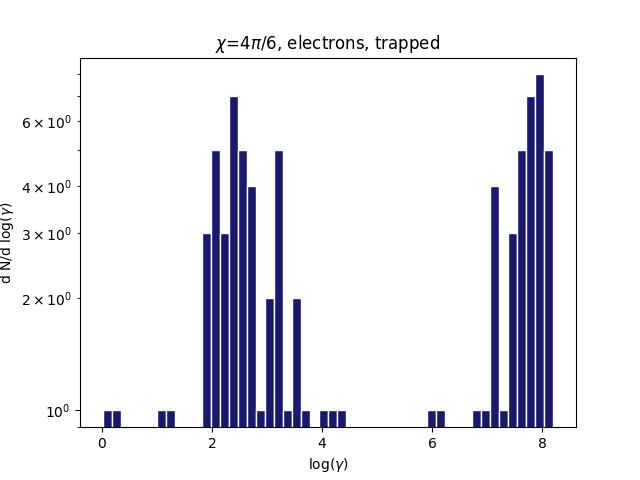}
        \put(-150,110) {\tiny (a)}
\end{subfigure}
\begin{subfigure}{.5\textwidth}
  \centering
  \includegraphics[width=\textwidth]{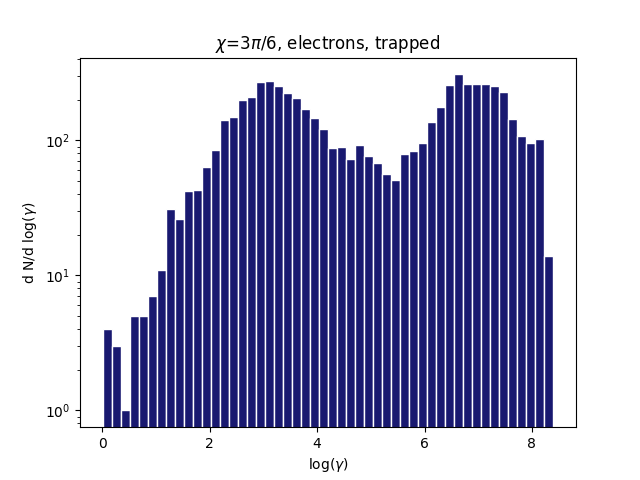}
        \put(-150,110) {\tiny (b)}
\end{subfigure}
\begin{subfigure}{.5\textwidth}
  \centering
  \includegraphics[width=\textwidth]{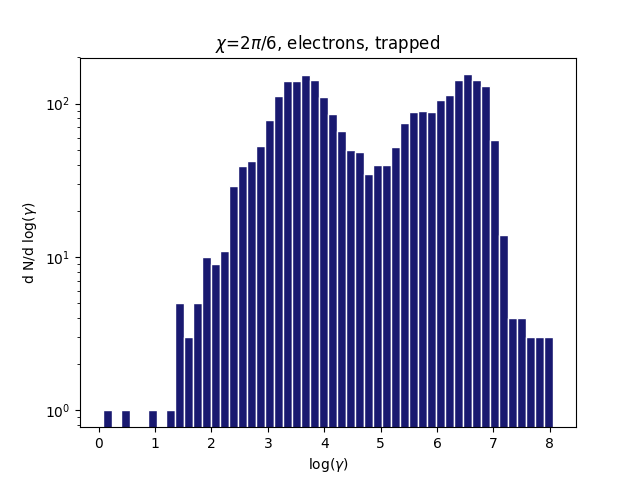}
        \put(-150,110) {\tiny (c)}
\end{subfigure}
\begin{subfigure}{.5\textwidth}
  \centering
  \includegraphics[width=\textwidth]{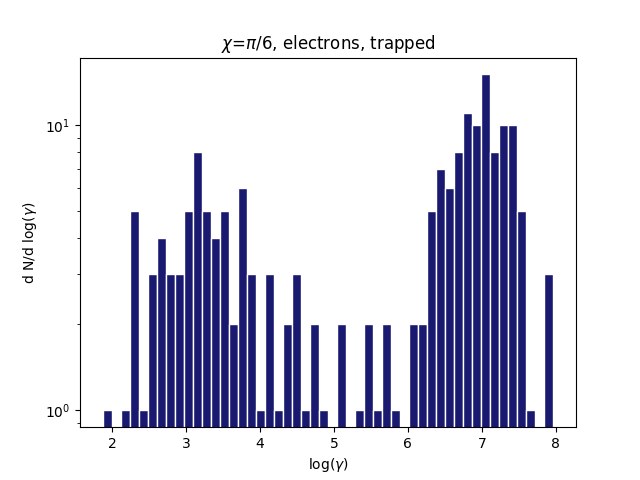}
        \put(-150,110) {\tiny (d)}
\end{subfigure}
\caption{Lorentz factor distribution of electrons trapped around neutron stars of inclination $\rchi=120^{\circ}$ (a), $\rchi=90^{\circ}$ (b), $\rchi=60^{\circ}$ (c), and $\rchi=30^{\circ}$ (d). Radiation reaction was enabled.}
\label{fig:trap_spectres_electrons}
\end{figure*}

Again, looking at Figure~\ref{fig:trap_spectres_electrons}, proton and electron spectra are different for mirror inclinations (i.e. $\rchi$ and $\pi-\rchi$). The electron distributions are always bimodal, but due to low statistics, the cases with $\rchi=120^{\circ}$ and $\rchi=30^{\circ}$ are hard to interpret. Nevertheless, the protons hardly exceed $\gamma=10^8$ in any inclination.
When $\rchi=90^{\circ}$, the two maxima are located at $\gamma=10^{3}$ and $\gamma=10^{6.5}$, and both are close to $N \sim 300$ particles per bin. The minimum between the two modes is located at $\gamma=10^{5.5},$ with $N\sim 50$ particles per bin.
For $\rchi=60$, the two maxima are also at the same statistical level of $N\sim 130$ particles per bin, but at $\gamma=10^{3.5}$ and $\gamma=10^{6.5}$,  the minimum between the modes is found at $\gamma=10^{5}$ for $N\sim 30$ particles per bin.

\paragraph{Ejected particles.} 

\begin{figure*}[h]
\begin{subfigure}{.5\textwidth}
  \centering
  \includegraphics[width=\textwidth]{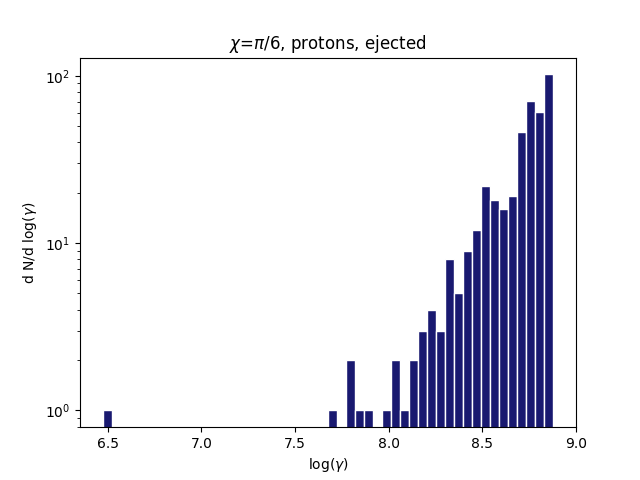}
        \put(-150,110) {\tiny (a)}
\end{subfigure}
\begin{subfigure}{.5\textwidth}
  \centering
  \includegraphics[width=\textwidth]{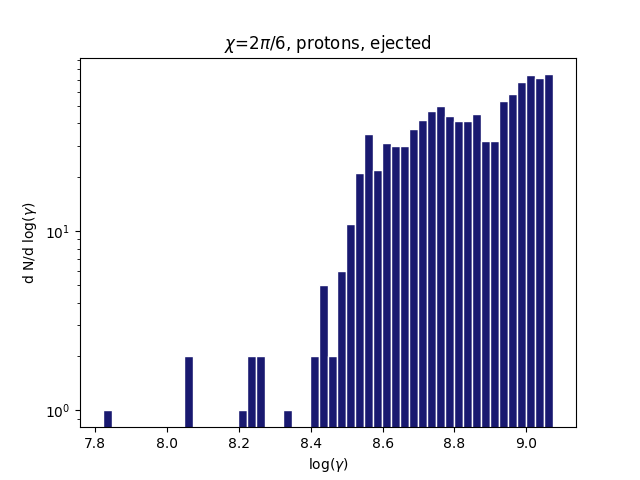}
        \put(-150,110) {\tiny (b)}
\end{subfigure}
\begin{subfigure}{.5\textwidth}
  \centering
  \includegraphics[width=\textwidth]{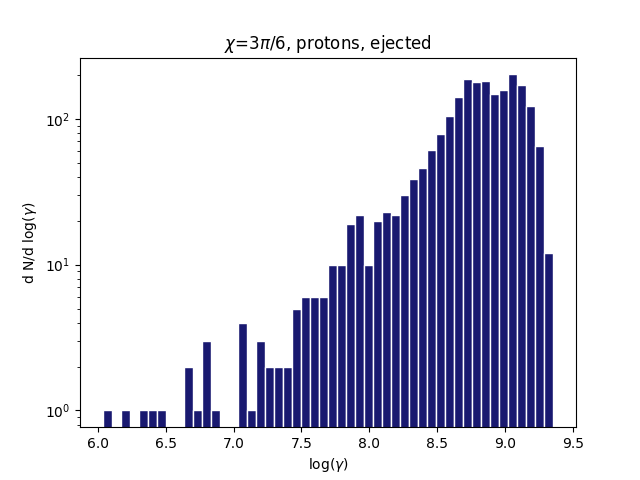}
        \put(-150,110) {\tiny (c)}
\end{subfigure}
\begin{subfigure}{.5\textwidth}
  \centering
  \includegraphics[width=\textwidth]{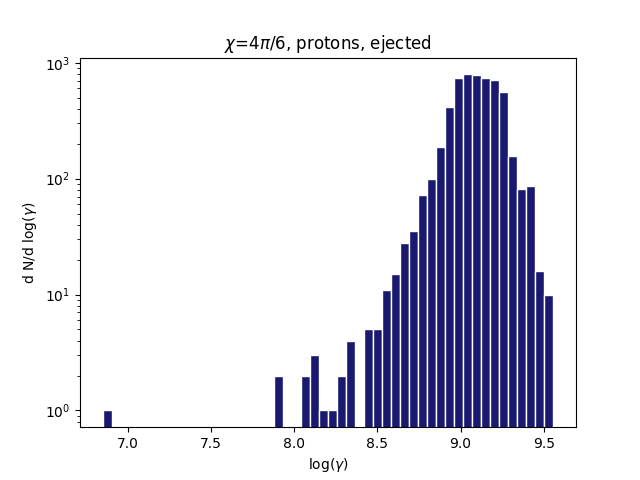}
        \put(-150,110) {\tiny (d)}
\end{subfigure}
\begin{subfigure}{.5\textwidth}
  \centering
  \includegraphics[width=\textwidth]{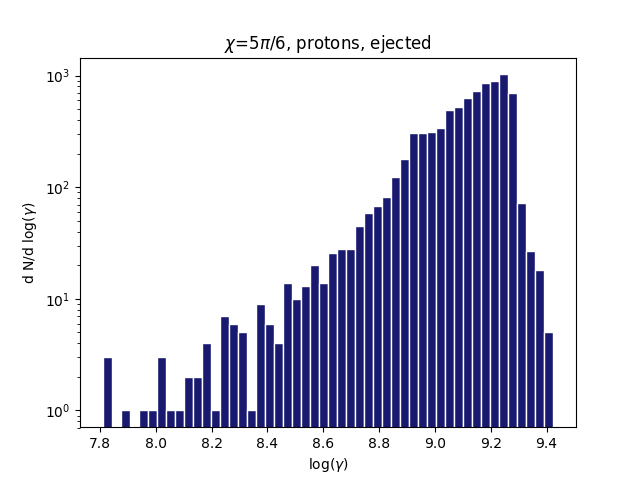}
        \put(-150,110) {\tiny (e)}
\end{subfigure}
\caption{Lorentz factor distribution of protons ejected by neutron stars with an inclination of $\rchi=30^{\circ}$ (a), $\rchi=60^{\circ}$ (b), $\rchi=90^{\circ}$ (c), $\rchi=120^{\circ}$ (d), and $\rchi=150^{\circ}$ (e). Radiation reaction was enabled.}
\label{fig:eject_spectres}
\end{figure*}

Figure~\ref{fig:eject_spectres} shows the spectral distribution of the Lorentz factor of protons ejected away from neutron stars. 
The case $\rchi=30^{\circ}$ shows a power law with a slope of approximately two between $\gamma=10^{7.9}$ ($N\sim1$) and $\gamma=10^{8.9}$ ($N\sim100$) followed by an abrupt cut-off.
The inclination $\rchi=60^{\circ}$, however, starts at $\gamma=10^{8.4}$ (apart from a few particles below this Lorentz factor) and grows fast until $\gamma=10^{8.6}$ at $N\sim 30,$ where the growth is slower until $\gamma=10^{9.1}$ at $N \sim 70$, which is where the distribution ends.
Taking a look at the orthogonal rotator, the distribution starts with some particles at $\gamma=10^{6}$, but then at $\gamma=10^{7}$, the distribution follows a power law until $\gamma=10^{8.7}$ ($N=200$), giving a slope of $\sim 1.35$. Then the distribution forms a plateau before a cut-off, ending at $\gamma=10^{9.3}$.
Regarding the $\rchi=120^{\circ}$ inclination, we noticed a rapid growth from $\gamma=10^{8}$ up to $\gamma=10^{9}$ followed by a short plateau that starts decreasing at $\gamma=10^{9.2}$ and ending  at $\gamma=10^{9.5}$.
Finally, for $\rchi=150^{\circ}$, the distribution starts as a power law of slope $\sim2.3$ until $\gamma=10^{9.3}$ and ends in $\gamma=10^{9.4}$ with a sharp cut-off.

\begin{figure*}[h]
\begin{subfigure}{.5\textwidth}
  \centering
  \includegraphics[width=\textwidth]{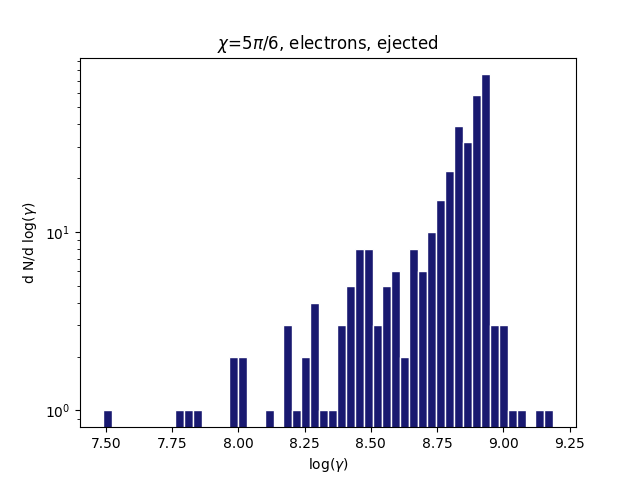}
        \put(-150,110) {\tiny (a)}
\end{subfigure}
\begin{subfigure}{.5\textwidth}
  \centering
  \includegraphics[width=\textwidth]{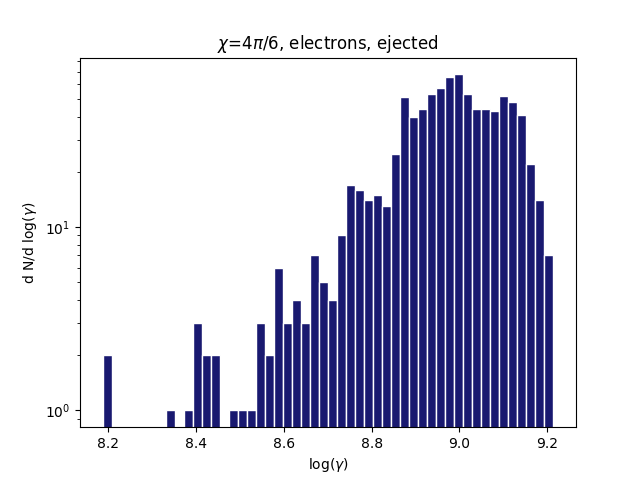}
        \put(-150,110) {\tiny (b)}
\end{subfigure}
\begin{subfigure}{.5\textwidth}
  \centering
  \includegraphics[width=\textwidth]{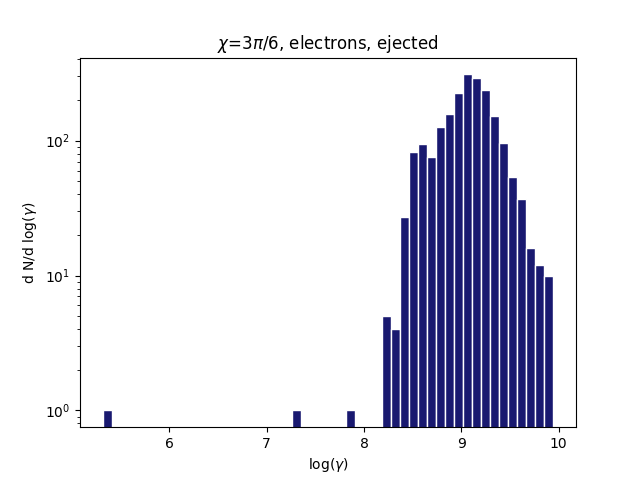}
        \put(-150,110) {\tiny (c)}
\end{subfigure}
\begin{subfigure}{.5\textwidth}
  \centering
  \includegraphics[width=\textwidth]{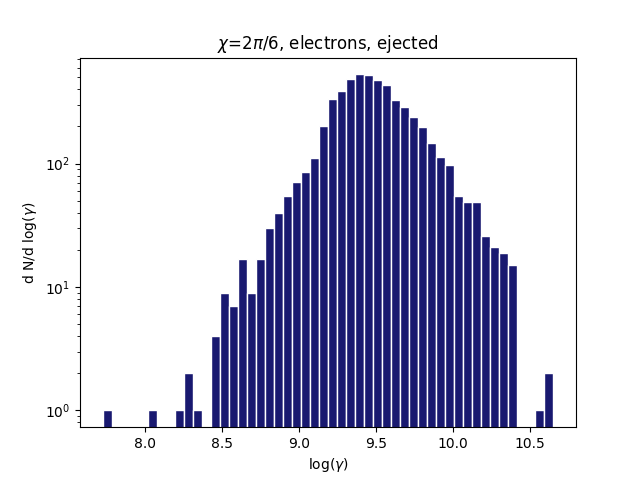}
        \put(-150,110) {\tiny (d)}
\end{subfigure}
\begin{subfigure}{.5\textwidth}
  \centering
  \includegraphics[width=\textwidth]{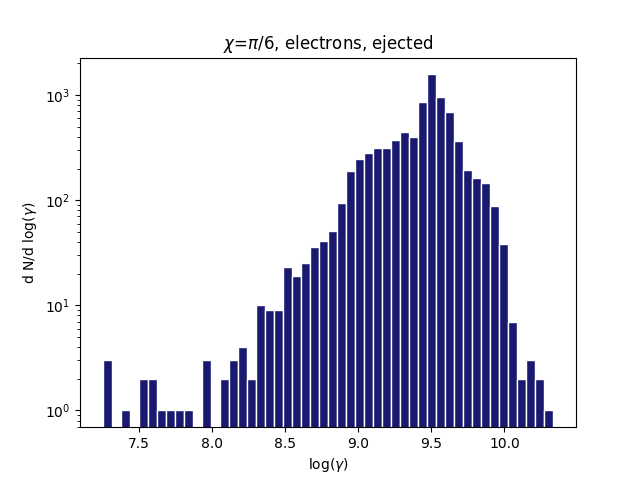}
        \put(-150,110) {\tiny (e)}
\end{subfigure}
\caption{Lorentz factor distribution of electrons ejected by neutron stars with an inclination of $\rchi=150^{\circ}$ (a), $\rchi=120^{\circ}$ (b), $\rchi=90^{\circ}$ (c), $\rchi=60^{\circ}$ (d), and $\rchi=30^{\circ}$ (e). Radiation reaction was enabled.}
\label{fig:eject_spectres_electrons}
\end{figure*}

Figure~\ref{fig:eject_spectres_electrons} shows the Lorentz factor of ejected electrons.
If $\rchi=150^{\circ}$, the distribution grows between $\gamma=10^{7.5}$ and $\gamma=10^{8.9}$ at $N \sim 60,$ and it then drops to zero at $\gamma=10^{9.2}$.
The distribution of the Lorentz factors for $\rchi=120^{\circ}$ starts at $\gamma=10^{8.2}$ with low statistics and some empty bins and grows to $N \sim 60$ at $\gamma=10^{9}$, and then the distribution drops, first slowly until $\gamma=10^{9.1}$ and then faster to zero in $\gamma=10^{9.2}$.
In the case of the orthogonal rotator, three particles were found at $\gamma<10^{8}$, but most of the distribution starts at $\gamma=10^{8.2}$. It then peaks at $\gamma=10^{9.1}$ with $N\sim 300$ and ends in $\gamma=10^{9.9}$.
If $\rchi=60^{\circ}$, the distribution starts with a growth between $\gamma=10^{8}$ ($N\sim 1$) and $\gamma=10^{9.5}$ ($N\sim 500$). It then decreases following a power law of slope $\sim1.9$ until $N \sim 15$ at $\gamma=10^{10.4}$. After that, the distribution stops, except for three particles close to $\gamma=10^{10.6}$.
Finally, if $\rchi=30^{\circ}$, the distribution grows irregularly between $\gamma=10^{7.3}$ and the peak at $\gamma=10^{9.5}$, with $N\sim 1500,$ and it then decreases irregularly until $\gamma=10^{10.3}$.

\subsection{Aligned and anti-aligned cases}
\label{ssec:Alignes}

\paragraph{Aligned case.}

As the aligned case, corresponding to $\rchi=0^{\circ}$, is a static field, it was treated separately from the other cases. In addition, even if in reality there are open field lines due to the plasma surrounding the neutron star, the Deutsch field does not have such magnetic field lines. Regarding the results of the simulations, we also found that in this case, all the protons crash onto the surface of the neutron star. 

\begin{figure*}[h]
\begin{subfigure}{.5\textwidth}
  \centering
  \includegraphics[width=\textwidth]{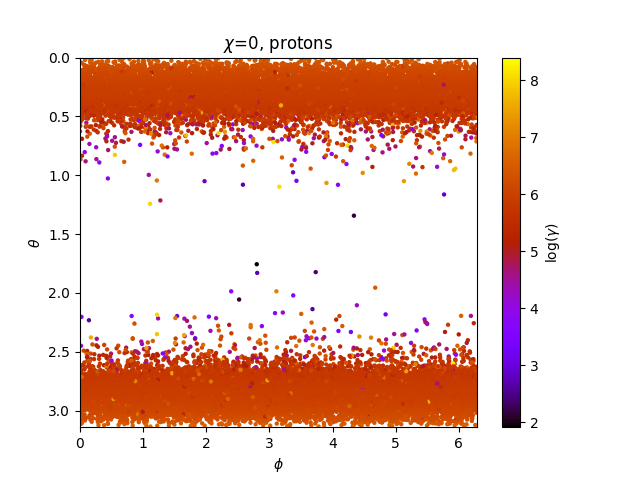}
        \put(-150,110) {\tiny (a)}
\end{subfigure}
\begin{subfigure}{.5\textwidth}
  \centering
  \includegraphics[width=\textwidth]{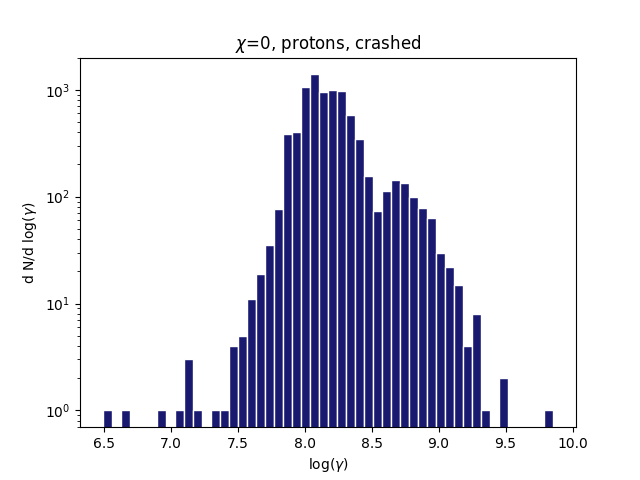}
        \put(-150,110) {\tiny (b)}
\end{subfigure}
\begin{subfigure}{.5\textwidth}
  \centering
  \includegraphics[width=\textwidth]{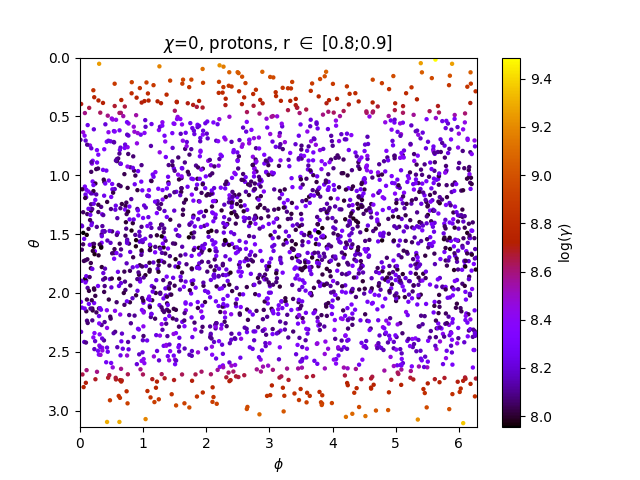}
        \put(-150,110) {\colorbox{white}{\tiny (c)}}
\end{subfigure}
\begin{subfigure}{.5\textwidth}
  \centering
  \includegraphics[width=\textwidth]{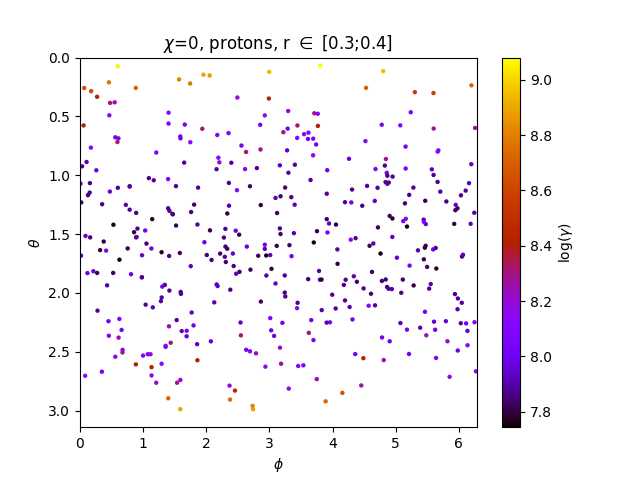}
        \put(-150,110) {\colorbox{white}{\tiny (d)}}
\end{subfigure}
\caption{Proton dynamics for an aligned rotator ($\rchi=0^{\circ}$). Panel~(a): impact map on the neutron star. Panel~(b): Lorentz factor distribution of the protons impacting the surface. Panel~(c): initial colatitude and azimuth, and final Lorentz factor of the protons starting at $r \in [0.8;0.9]$. Panel~(d): same as panel~(c) but for protons starting at $r \in [0.3;0.4]$.}
\label{fig:aligned}
\end{figure*}

Inspecting Figure~\ref{fig:aligned}, we found that particles impact the neutron star on its poles. When analysing the initial positions of the particles and linking them to their final Lorentz factors, particles injected closer to the neutron star were found to reach a lower Lorentz factor than those injected farther away. This phenomenon is due to the potential drop being greater for particles farther away from the neutron star than for those close to the surface. 
For a given starting radius, protons injected closer to the equator reach a lower energy than those injected close to the poles. The most probable explanation for this is the energy loss due to the curvature radiation. For instance, for a given radius, as we get closer to the equator ($\theta=10^{\circ}$), the curvature radius gets smaller, meaning that a particle injected there will have more energy loss than other particles starting at the same radius. 

When looking at the Lorentz factor distribution, we noticed that protons reach up to $\gamma=10^{9.8}$ and never go lower than $\gamma=10^{6.5}$. Also, the peak of the distribution is located at $\gamma=10^{8.1},$ with $N \sim 1300$, but a local maximum was also found at $\gamma=10^{8.6}$, with $N \sim 130$.

\paragraph{Anti-aligned case.}

The anti-aligned case was also treated separately for reasons similar to the aligned case (i.e. a constant field and no open magnetic field line in the case of the Deutsch field). However, the same comment could be made that the plasma normally around the neutron star changes the fields and `opens' the magnetic field lines beyond the light cylinder. 

\begin{figure*}[h]
\begin{subfigure}{.5\textwidth}
  \centering
  \includegraphics[width=\textwidth]{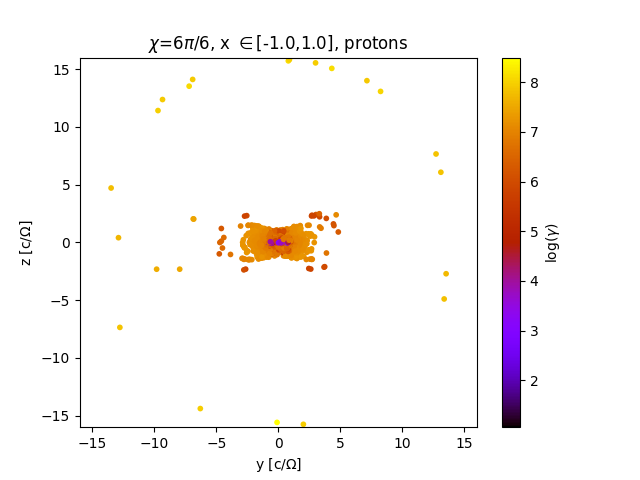}
        \put(-150,110) {\tiny (a)}
\end{subfigure}
\begin{subfigure}{.5\textwidth}
  \centering
  \includegraphics[width=\textwidth]{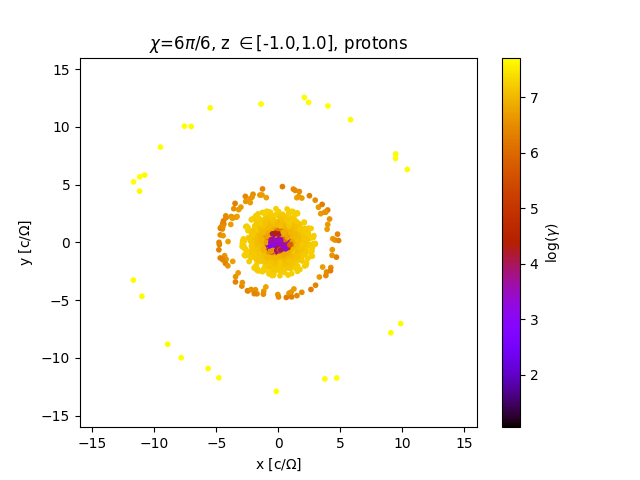}
        \put(-150,110) {\tiny (b)}
\end{subfigure}
\begin{subfigure}{.5\textwidth}
  \centering
  \includegraphics[width=\textwidth]{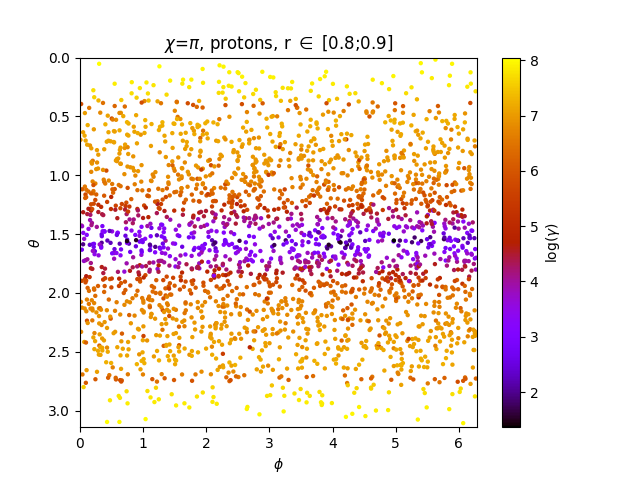}
        \put(-150,110) {\colorbox{white}{\tiny (c)}}
\end{subfigure}
\begin{subfigure}{.5\textwidth}
  \centering
  \includegraphics[width=\textwidth]{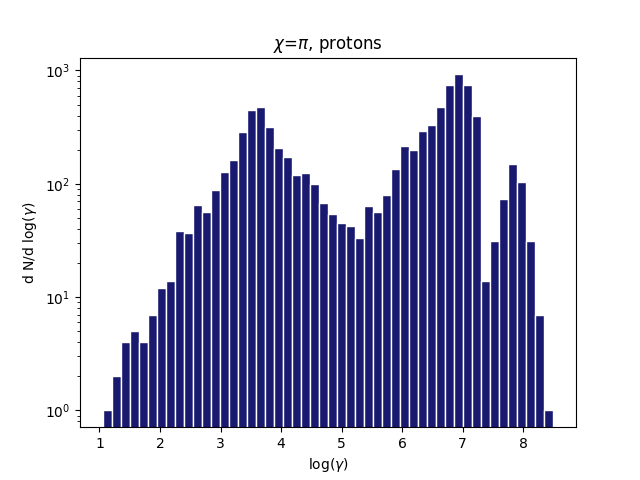}
        \put(-150,110) {\tiny (d)}
\end{subfigure}
\caption{Proton dynamics for the anti-aligned rotator ($\rchi=180^{\circ}$). Panel~(a): final position and Lorentz factor of protons in the $x\in[-1;1]$ slice. Panel~(b): the same in the $z\in[-1;1]$ slice. Panel~(c): initial colatitude and azimuth and final Lorentz factor of protons starting at $r\in[0.8;0.9]$. Panel~(d): Lorentz factor distribution of the protons around the neutron star.}
\label{fig:antialigned}
\end{figure*}

Looking at the starting positions of protons in Figure~\ref{fig:antialigned}, we observed that particles close to the poles are accelerated more efficiently than those close to the equator. We believe that, just like for the aligned case, the curvature radiation is more important for particles close to the equator than for particles close to the poles. 

Regarding the final positions of the protons, it appeared that a thick disc forms around the equator because of  radiation losses. Indeed, protons tend to oscillate between the north and south poles (as in a Van-Allen radiation belt), but as the radius of curvature of the field lines gets smaller, the radiation losses become more intense, and a particle that was previously oscillating between the poles finally becomes stuck in the equatorial plane. Inversely, farther away from the neutron star, the curvature radius of the magnetic field lines is larger, so protons continue to oscillate between the north and south poles and do not lose enough energy to become stuck in the equatorial plane. Particles starting close to the poles follow the field lines with curvature radii so large that by the end of the simulation they did not have time to reach the equatorial plane (forming a dome in each hemisphere), and since the energy loss is low because of the high curvature radius, these particles reach the highest energy. Additionally, Figure~\ref{fig:antialigned} shows not a disc but rather a disc and rings. The rings are in fact particles that are still oscillating between the north and south poles sufficiently so that they may reach coordinates out of the range $z \in [-1;1]$ at other times, and the holes between the rings are there since other particles have similar behaviours but were not in the range $z \in [-1;1]$ by the end of the simulation. 

Finally, regarding the spectral distribution of the Lorentz factors, we highlight three modes. First is a low energy mode with a maximum at $\gamma= 10^{3.5},$ with $N \sim 500,$ and ranging from $\gamma \sim 10$ to $\gamma=10^{5.3}$, with $N = 30$. 
The second mode is at $\gamma \in [10^{5.3};10^{7.2}]$ and has the overall maximum of the distribution at $\gamma=10^{6.9}$, at $N \sim 800$. 
The last mode is at $\gamma \in [10^{7.2};10^{8.4}]$, and it has its maximum at $\gamma=10^{7.9}$, with $N \sim 200$.

\begin{figure*}[h]
\begin{subfigure}{.5\textwidth}
  \centering
  \includegraphics[width=\textwidth]{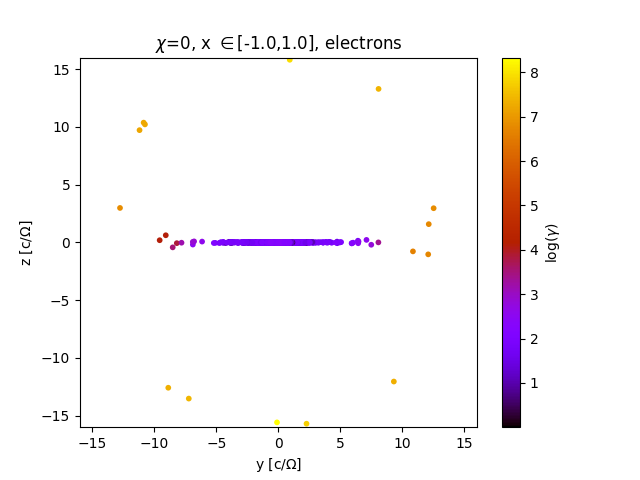}
        \put(-180,130) {\tiny (a)}
\end{subfigure}
\begin{subfigure}{.5\textwidth}
  \centering
  \includegraphics[width=\textwidth]{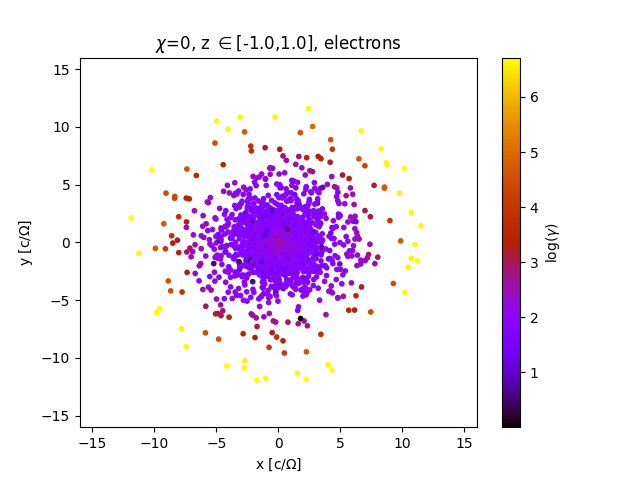}
        \put(-180,130) {\tiny (b)}
\end{subfigure}
\begin{subfigure}{.5\textwidth}
  \centering
  \includegraphics[width=\textwidth]{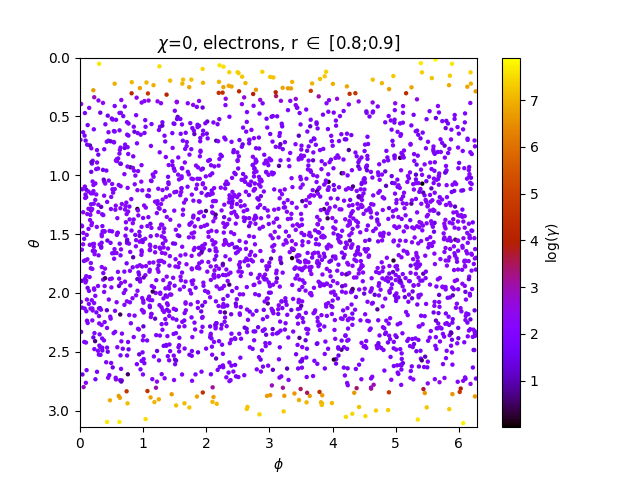}
        \put(-180,130) {\colorbox{white}{\tiny (c)}}
\end{subfigure}
\begin{subfigure}{.5\textwidth}
  \centering
  \includegraphics[width=\textwidth]{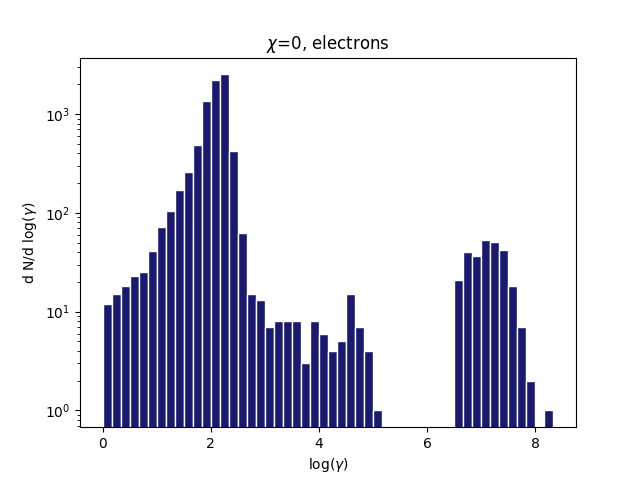}
        \put(-180,130) {\tiny (d)}
\end{subfigure}
\caption{Electron dynamics for the anti-aligned rotator ($\rchi=180^{\circ}$). Panel~(a): final position and Lorentz factor of electrons in the $x\in[-1;1]$ slice. Panel~(b): the same as (a) but in the $z\in[-1;1]$ slice. Panel~(c): initial colatitude and azimuth and final Lorentz factor of protons starting at $r\in[0.8;0.9]$. Panel~(d): Lorentz factor distribution of the protons around the neutron star.}
\label{fig:aligned_electrons}
\end{figure*}

For comparison, we took a look at the aligned case for electrons, too. Figure~\ref{fig:aligned_electrons} shows that electrons also form a disc by the end of the simulation, but due to their lower masses, radiation reaction makes them lose more energy than the protons, meaning that the disc formed by electrons is thinner than that of the protons. The Lorentz factor distribution is composed of a more populated low energy mode from $\gamma=1$ to $\gamma=10^{5}$ with the maximum at $\gamma=10^{2.5}$, while the high energy mode is between $\gamma=10^{6.5}$ and $\gamma=10^{8}$. It appears that the high energy electrons are those injected close to the poles (in terms of colatitude) because these particles follow magnetic field lines going farther away from the neutron star, meaning that these electrons spend less time in a strong magnetic field and thus lose less energy than those that remained close to the neutron star and ultimately formed the disc.

\section{Conclusions}
\label{sec:Conclusion}

In this paper, we studied the influence of radiation reaction on proton and electron dynamics near millisecond pulsars. We showed the drastic impact of radiative losses onto their trajectories and Lorentz factor. First of all, the evolution of the particles and their positions at the end of the simulations may still share some similarities when radiation reaction is enabled or disabled, but many particles have another behaviour, and almost all of them end the simulations with positions that are different from those in the simulations without radiation reaction. For example, the hot spots have different shapes and are located at different colatitudes regardless of whether radiation reaction is enabled. Regarding trapped particles, their radial position seems to be more impacted by radiation reaction than their colatitudes or azimuths. Moreover, the positions of particles are even more different when they are ejected away from the neutron star. Regarding the Lorentz factor distributions, it clearly appears that particles lose energy due to radiation reaction, as the Lorentz factors reached by the particles are more realistic than when radiation reaction is neglected. 
The interaction between particles is the next and last step to study the evolution of particles around a neutron star with realistic fields. We hope that the retroaction of the particles on the fields will help in obtaining results even closer to reality for the particle speeds and positions.

\section*{Acknowledgement}
We are grateful to the referee for helpful comments and suggestions. This work has been supported by CEFIPRA grant IFC/F5904-B/2018 and ANR-20-CE31-0010.


\begin{thebibliography}{26}
        \expandafter\ifx\csname natexlab\endcsname\relax\def\natexlab#1{#1}\fi
        
        \bibitem[{Arfken \& Weber(2005)}]{arfken_mathematical_2005}
        Arfken, G.~B. \& Weber, H.-J. 2005, Mathematical methods for physicists, 6th
        edn. (Boston: Elsevier)
        
        \bibitem[{Bogovalov(1999)}]{bogovalov_physics_1999}
        Bogovalov, S.~V. 1999, A\&A, 349, 1017
        
        \bibitem[{Boris(1970)}]{boris_relativistic_1970}
        Boris, J. 1970, Proceeding of Fourth Conference on Numerical Simulations of
        Plasmas
        
        \bibitem[{Brambilla {et~al.}(2018)Brambilla, Kalapotharakos, Timokhin, Harding,
                \& {Demosthenes Kazanas}}]{brambilla_electronpositron_2018}
        Brambilla, G., Kalapotharakos, C., Timokhin, A.~N., Harding, A.~K., \&
        {Demosthenes Kazanas}. 2018, ApJ, 858, 81
        
        \bibitem[{Cerutti {et~al.}(2015)Cerutti, Philippov, Parfrey, \&
                Spitkovsky}]{cerutti_particle_2015}
        Cerutti, B., Philippov, A., Parfrey, K., \& Spitkovsky, A. 2015, MNRAS, 448,
        606
        
        \bibitem[{Deutsch(1955)}]{deutsch_electromagnetic_1955}
        Deutsch, A.~J. 1955, Annales d'Astrophysique, 18, 1
        
        \bibitem[{Gordon \& Hafizi(2021)}]{gordon_special_2021}
        Gordon, D.~F. \& Hafizi, B. 2021, Comput. Phys. Commun, 258, 107628
        
        \bibitem[{Guépin {et~al.}(2020)Guépin, Cerutti, \&
                Kotera}]{guepin_proton_2020}
        Guépin, C., Cerutti, B., \& Kotera, K. 2020, A\&A, 635, A138
        
        \bibitem[{Hadad {et~al.}(2010)Hadad, Labun, Rafelski, Elkina, Klier, \&
                Ruhl}]{hadad_effects_2010}
        Hadad, Y., Labun, L., Rafelski, J., {et~al.} 2010, Phys. Rev. D, 82, 096012
        
        \bibitem[{Heintzmann \& Schrüfer(1973)}]{heintzmann_exact_1973}
        Heintzmann, H. \& Schrüfer, E. 1973, Physics Letters A, 43, 287
        
        \bibitem[{Kalapotharakos {et~al.}(2018)Kalapotharakos, Brambilla, Timokhin,
                Harding, \& Kazanas}]{kalapotharakos_three-dimensional_2018}
        Kalapotharakos, C., Brambilla, G., Timokhin, A., Harding, A.~K., \& Kazanas, D.
        2018, ApJ, 857, 44
        
        \bibitem[{Krause-Polstorff \&
                Michel(1985)}]{krause-polstorff_electrosphere_1985}
        Krause-Polstorff, J. \& Michel, F.~C. 1985, MNRAS, 213, 43P
        
        \bibitem[{Laue \& Thielheim(1986)}]{laue_acceleration_1986}
        Laue, H. \& Thielheim, K.~O. 1986, ApJS, 61, 465
        
        \bibitem[{Li {et~al.}(2021)Li, Decyk, Miller, Tableman, Tsung, Vranic, Fonseca,
                \& Mori}]{li_accurately_2021}
        Li, F., Decyk, V.~K., Miller, K.~G., {et~al.} 2021, Journal of Computational
        Physics, 438, 110367
        
        \bibitem[{Li {et~al.}(2012)Li, Spitkovsky, \& Tchekhovskoy}]{li_resistive_2012}
        Li, J., Spitkovsky, A., \& Tchekhovskoy, A. 2012, ApJ, 746, 60
        
        \bibitem[{Michel \& Li(1999)}]{michel_electrodynamics_1999}
        Michel, F. \& Li, H. 1999, Physics Reports, 318, 227
        
        \bibitem[{Nättilä {et~al.}(2017)Nättilä, Miller, Steiner, Kajava,
                Suleimanov, \& Poutanen}]{nattila_neutron_2017}
        Nättilä, J., Miller, M.~C., Steiner, A.~W., {et~al.} 2017, A\&A, 608, A31
        
        \bibitem[{Philippov \& Spitkovsky(2018)}]{philippov_ab-initio_2018}
        Philippov, A.~A. \& Spitkovsky, A. 2018, ApJ, 855, 94
        
        \bibitem[{Piazza(2008)}]{piazza_exact_2008}
        Piazza, A.~D. 2008, Lett Math Phys, 83, 305
        
        \bibitem[{Pétri(2020)}]{petri_relativistic_2020}
        Pétri, J. 2020, J. Plasma Phys., 86, 825860402
        
        \bibitem[{Pétri(2021)}]{petri_particle_2021}
        Pétri, J. 2021, MNRAS, 503, 2123
        
        \bibitem[{Pétri(2022)}]{petri_particle_2022-2}
        Pétri, J. 2022, A\&A, 666, A5
        
        \bibitem[{Pétri {et~al.}(2002)Pétri, Heyvaerts, \&
                Bonazzola}]{petri_global_2002}
        Pétri, J., Heyvaerts, J., \& Bonazzola, S. 2002, A\&A, 384, 414
        
        \bibitem[{Tomczak \& Pétri(2020)}]{tomczak_particle_2020}
        Tomczak, I. \& Pétri, J. 2020, J. Plasma Phys., 86, 825860401
        
        \bibitem[{Vay(2008)}]{vay_simulation_2008}
        Vay, J.-L. 2008, Physics of Plasmas (1994-present), 15, 056701
        
        \bibitem[{Vranic {et~al.}(2016)Vranic, Martins, Fonseca, \&
                Silva}]{vranic_classical_2016}
        Vranic, M., Martins, J.~L., Fonseca, R.~A., \& Silva, L.~O. 2016, Computer
        Physics Communications, 204, 141
        
\end{thebibliography}
\end{document}